\documentclass[prl,superscriptaddress,twocolumn,floatfix]{revtex4-2}

\usepackage[normalem]{ulem}
\usepackage{amsmath,mathtools,amsthm,amssymb,pifont}
\usepackage[utf8]{inputenc}
\usepackage[american]{babel}
\makeatletter
\let\l@en\l@american

\makeatother
\usepackage{graphicx,xcolor,bbold,titlesec}
\usepackage{braket}
\usepackage{comment}
\usepackage{MnSymbol}

\usepackage{xspace}
\usepackage{nameref}
\usepackage[colorlinks,
bookmarksopen,
bookmarksnumbered,
citecolor=red,
linkcolor=red,
pdfstartview=false,
urlcolor=red]{hyperref}
\usepackage{etoolbox}
\usepackage{tikz}

\newcommand{\Tr}{\operatorname{Tr}}

\graphicspath{{Images/}}

\newcommand{\lettersection}[1]{\textbf{#1}. ---}
\newcommand{\qnoise}{Noisy XX\xspace}

\usepackage{tikz}
\usetikzlibrary{calc,arrows.meta, positioning}

\begin{document}

\def\titleinfo{Emergence of universality in transport of noisy free fermions}
\title{\titleinfo} 
\author{João Costa}
\affiliation{CeFEMA-LaPMET, Departamento de Física, Instituto Superior Técnico, Universidade de Lisboa, Av. Rovisco Pais, 1049-001 Lisboa, Portugal}
\affiliation{Laboratoire de Physique Th\'eorique et Mod\'elisation, CY Cergy Paris Universit\'e, CNRS, F-95302 Cergy-Pontoise, France}

\author{Pedro Ribeiro}
\affiliation{CeFEMA-LaPMET, Departamento de Física, Instituto Superior Técnico, Universidade de Lisboa, Av. Rovisco Pais, 1049-001 Lisboa, Portugal}
\affiliation{Laboratoire de Physique Th\'eorique et Mod\'elisation, CY Cergy Paris Universit\'e, CNRS, F-95302 Cergy-Pontoise, France}
\affiliation{CY Advanced Studies, Maison internationale de la recherche, CY Cergy Paris Universit\'e, 95302 Cergy-Pontoise Cedex, France}

\author{Andrea De Luca}
\affiliation{Laboratoire de Physique Th\'eorique et Mod\'elisation, CY Cergy Paris Universit\'e, CNRS, F-95302 Cergy-Pontoise, France}
\begin{abstract}
We analyze the effects of various forms of noise on one-dimensional systems of non-interacting fermions. In the strong noise limit, we demonstrate, under mild assumptions, that the statistics of the fermionic correlation matrix in the thermodynamic limit follow a universal form described by the recently introduced quantum simple symmetric exclusion process (QSSEP). For charge transport, we show that QSSEP, along with all models in its universality class, shares the same large deviation function for the transferred charge as the classical SSEP model. The method we introduce to derive this result relies on a gauge-like invariance associated with the choice of the bond where the current is measured. This approach enables the explicit calculation of the cumulant generating function for both QSSEP and SSEP and establishes an exact correspondence between them. These analytical findings are validated by extensive numerical simulations. Our results establish that a wide range of noisy free-fermionic models share the same QSSEP universality class and show that their transport properties are essentially classical. 
\end{abstract}
\maketitle

\lettersection{Introduction}
Out-of-equilibrium quantum transport has become a central topic over the past two decades~\cite{eisert2015quantum,d2016quantum, abanin2019colloquium}, driven by major experimental advances in cold-atom systems~\cite{tang2018thermalization, Schneider2012, Lebrat2018, rauer2018recurrences, bloch2008many, Brown2015, Choi2016, boll2016spin, hild2014far} and solid-state platforms~\cite{scheie2021detection}, where one can now measure transferred-charge statistics across quantum point contacts~\cite{Gustavsson2006} and quantum dots~\cite{Fujisawa2006, Flindt2009}. On the theory side, generalized hydrodynamics (GHD)~\cite{bertini_transport_2016, PhysRevX.6.041065, PhysRevLett.121.160603, schemmer2019generalized, bastianello2020thermalization, doyon2025generalized} was a breakthrough for integrable dynamics, but general principles for non-equilibrium behaviour remain scarce beyond a few special cases~\cite{gallavotti1995dynamical, crooks1999entropy, maes1999fluctuation}. In one dimension, non-diffusive transport~\cite{krajnik2022exact, gopalakrishnan2024non, gopalakrishnan2019kinetic} and distinctive effects of dephasing and noise~\cite{gardiner2017quantum, znidaric_dephasing-induced_2010, PhysRevE.96.052118, knap2018entanglement} have further highlighted this challenge. A complementary line of attack is based on statistical sampling of dynamical processes, both for generic interacting systems~\cite{annurev:/content/journals/10.1146/annurev-conmatphys-031720-030658,10.21468/SciPostPhys.3.5.033, rowlands2018} and for noisy quantum dynamics~\cite{knap2018entanglement, christopoulos2023universal, swann2023spacetime}. In particular, random unitary circuits (RUCs)~\cite{nahum, nahum2018operator} provide a robust framework for quantum chaos~\cite{chan2018solution, chan2018spectral, friedman2019spectral, shivam2023many, chan2022many, bertini2018exact, bertini2019entanglement}, including the membrane picture for entanglement growth~\cite{mezei2018membrane, nahum, zhou2019emergent, zhou2020entanglement} and the butterfly effect in operator spreading~\cite{nahum2018operator, gopalakrishnan2018hydrodynamics, chan2019eigenstate, xu2019locality}. 

\begin{figure}[ht]
\includegraphics[width=1.\columnwidth]{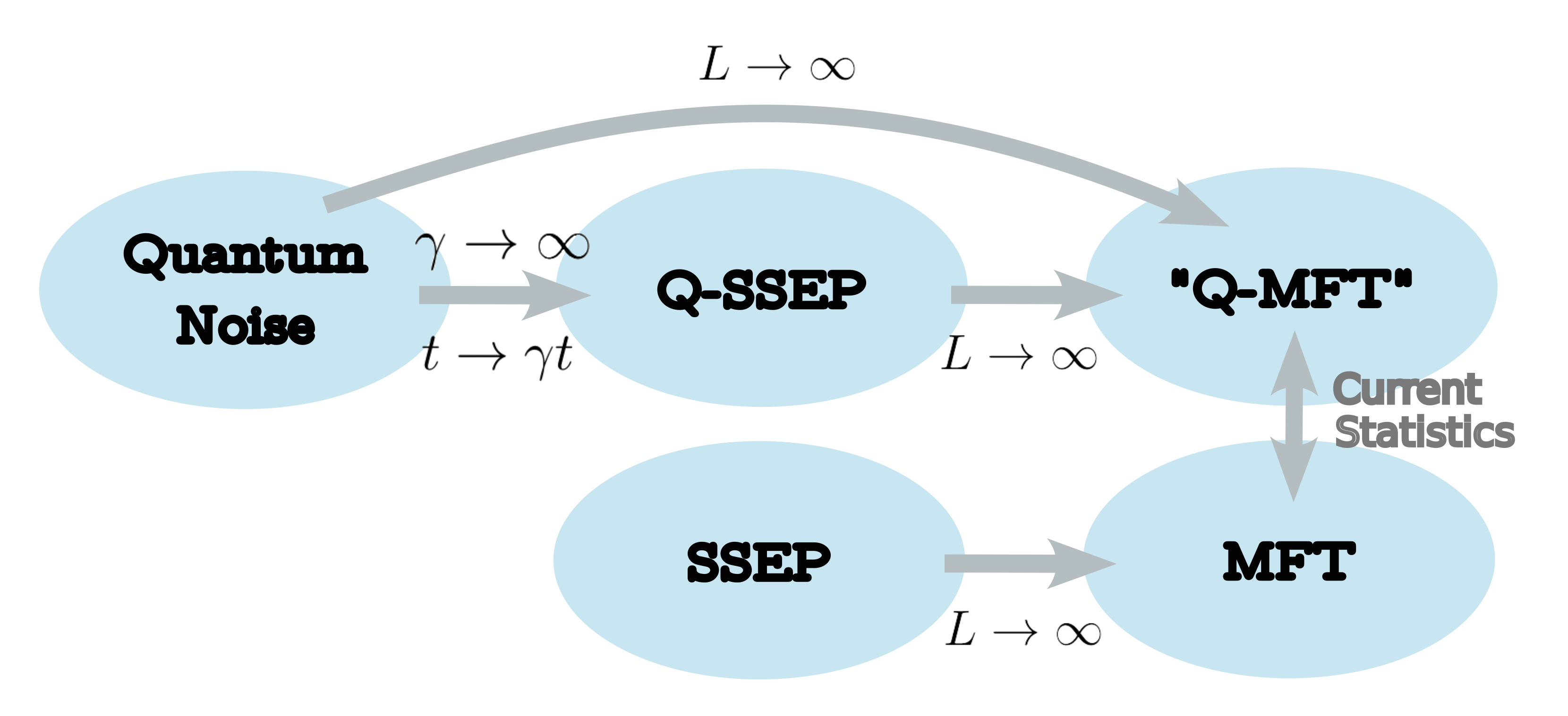}
\caption{
At large $L$ and $t$, QSSEP admits a diffusive scaling limit, which we refer to as Q-MFT~\cite{Bernard_2016, Bernard_2021}. Various noisy non-interacting fermion models in 1D reduce to QSSEP in the strong noise limit; since the effective noise grows with $L$, they fall into the Q-MFT universality class. For large charge transport deviations, Q-MFT reduces to usual MFT, describing SSEP in the scaling limit. }
\label{fig:schematic}
\end{figure}

Models of noisy free fermions have also been extensively studied~\cite{gullans2019entanglement, swann2023spacetime, christopoulos2023universal, DeLuca}. A key example is the quantum symmetric simple exclusion process (QSSEP), introduced by Bernard \textit{et al.}: a chain of spinless free fermions with nearest-neighbour hoppings drawn from independent white noise distributions~\cite{PhysRevLett.123.080601,bauer2019equilibrium,10.21468/SciPostPhys.3.5.033}. After noise averaging, QSSEP reduces to its classical counterpart~(see also~\cite{eisler_crossover_2011,Temme_2012}). On each noise realization, however, it
displays coherent hopping, still allowing exact analytical treatments~\cite{Bernard2021-vx}, with connections to combinatorics and free probability~\cite{PhysRevX.13.011045,biane2023combinatorics}, and explicitly bridging quantum and classical behaviours~\cite{PhysRevLett.123.080601, Bernard2021-vx, Bernard_2021, Derrida2009-nc}.


In this letter, we show the emergence of universality in the dynamics and transport of one-dimensional noisy free fermions, in close analogy with the emergence of classical macroscopic fluctuation theory (MFT) for lattice gases~\cite{Eyink1990,RevModPhys.87.593, kipnis2013scaling, spohn2012large, kipnis1989hydrodynamics, PhysRevLett.94.030601}. MFT describes diffusive non-equilibrium systems through coarse-grained density and current fields, with microscopic details entering only through the diffusion coefficient $D(\rho)$ and mobility $\sigma(\rho)$. 
Our analysis provides evidence of the existence of a quantum extension QMFT~\cite{Bernard_2016, Bernard_2021} that captures universal aspects of both quantum and noise fluctuations. Although its continuum formulation is still unknown~\cite{Bernard_2021}, QSSEP provides an explicit lattice representative with exact results~\cite{PhysRevLett.123.080601, PhysRevX.13.011045}. This universality class may even extend beyond free models~\cite{PhysRevLett.131.210402} and one dimension~\cite{Tony3dAnderson}.

To show the universality of noisy fermions beyond QSSEP, we consider homogeneous and static nearest-neighbour hopping, with noise only coupled to local densities. We refer to this as the \textit{ Quantum Dephasing Noise (\qnoise) model}~\cite{10.21468/SciPostPhys.3.5.033} (see Fig.~\ref{fig:schematic}). The system is connected at both ends to particle reservoirs modeled within the Lindblad framework. 
We argue that in the thermodynamic limit $L \to \infty$, the correlation-matrix distribution of \qnoise coincides with that of QSSEP. The key point is that increasing $L$ effectively produces a strong-noise limit, since particles accumulate more dephasing while traversing the system. In that limit, the exact mapping to QSSEP can be shown algebraically~\cite{10.21468/SciPostPhys.3.5.033} (see sec.~\ref{sec:snl} of \footnote{See supplemental material for extra details.} for an alternative proof). Related effective large-$L$ descriptions for noise-averaged correlations were discussed in~\cite{TonyQResistors}.

We then study charge transport through the full distribution of the charge transferred to one reservoir for a fixed noise realization. The transferred charge grows linearly in time and obeys a large-deviation principle; its cumulants become self-averaging at long times and scale diffusively as $O(1/L)$. Exploiting the gauge invariance associated with the bond where the current is measured, we prove that the leading cumulants of QSSEP, and hence of its universality class, coincide exactly with those of classical SSEP (see Fig.~\ref{fig:schematic}).

\lettersection{Model}\label{sec:Model}
\begin{figure}[ht]
\includegraphics[width=0.95\columnwidth]{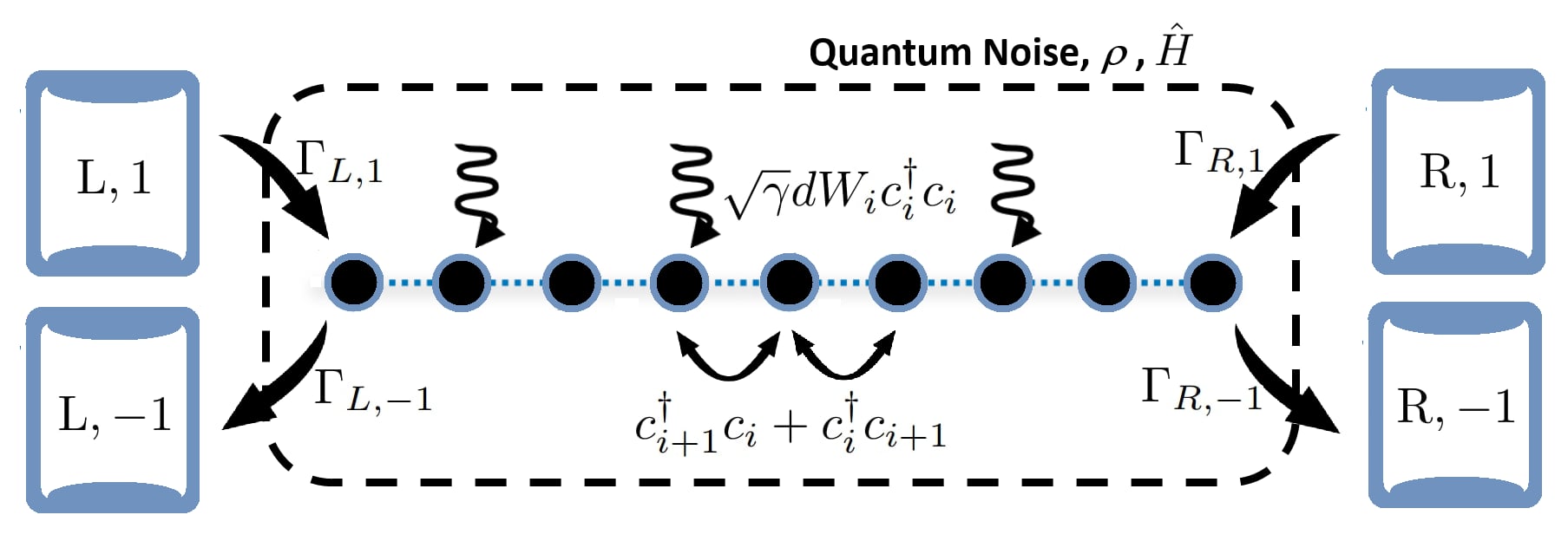}
\caption{\qnoise: 
a chain of spinless non-interacting fermions on $L$ sites with hopping Hamiltonian and subjected to stochastic noise coupled to the number operator at each site. 
At each boundary $\alpha \in \{\text{L}, \text{R}\}$, two reservoirs inject ($+1$) and remove ($-1$) particles with rates $\Gamma_{\alpha, \pm 1}$.}
\end{figure}
We introduce \qnoise for free fermions on a 1D lattice of $L$ sites. The evolution is controlled by a deterministic hopping Hamiltonian $\hat H_0$, which for reference is taken to be $\Hat{H}_0 = - \sum_{j=1}^{L-1} \left(c^{\dagger}_j c_{j+1} + c^{\dagger}_{j+1} c_{j}\right)$. However we discuss more general quadratic Hamiltonians when establishing the universality of our results. A stochastic noisy potential is coupled to the number density on each site setting the Hamiltonian increment $d\hat{H} = \hat{H}_0 dt + \sqrt{\gamma} \sum_{j=1}^L \hat{n}_j dW_j$. The $dW$'s denote Wiener processes, satisfying $\overline{dW_i}=0$, $\overline{dW_i dW_j}= \delta_{i,j} dt$. Using Ito calculus, we deduce the unitary evolution of the density matrix
\begin{equation}
    [d\rho]_{\rm uni} =  
    e^{-i d\hat H} \rho e^{i d\hat H} - \rho = 
    -i \left[d\hat{H}, \rho \right]   + \gamma \sum_{j=1}^L \mathcal{D}_{\Hat{n}_j}[\rho] dt,
    \label{Eq:TimeEvUni}
\end{equation}
where we denote as $\mathcal{D}_{\hat{O}}[\rho] = \hat{O} \rho \hat{O}^\dag - \frac{1}{2} \{ \hat{O}^{\dagger} \hat{O}, \rho \}$ the Lindblad superoperator associated with the jump operator $\hat{O}$. Additionally, the system exchanges particles on each of its edges with two reservoirs. The full setup of system+reservoirs, described by the total density matrix $\rho_T$, undergoes particle conserving quantum dynamics $\partial_t \rho_T = \mathcal{L}_T(\rho_T)$. 
For concreteness, we assume incoherent Markovian reservoirs, so that they can be traced out $\rho = \Tr_{\mathcal{R}}[\rho_T]$, eventually leading to the Lindblad description for the reduced density matrix $\rho$,
\begin{equation}
    d\rho =  [d\rho]_{\rm uni} + [d\rho]_{\rm bath} \;,  \quad [d\rho]_{\rm bath} := \sum_{\alpha, \sigma} \mathcal{D}_{\Hat{L}_{\alpha,\sigma}}[\rho] dt \;.
    \label{Eq:TimeEvGen}
\end{equation}
The index $\alpha \in \{\text{L}, \text{R}\}$ refers to the left/right reservoirs coupled with the sites $j_{\text{L}} = 1$ and $j_{\text{R}} = L$, respectively. Instead, $\sigma = \pm 1$  specifies the process of injection/removal of particles at each edge.
The jump operators read
\begin{equation}
\hat{L}_{(\alpha,1)} := \sqrt{\Gamma_{\alpha, 1}} c_{j_\alpha}^{\dag} \;, \quad
\hat{L}_{(\alpha,-1)} := \sqrt{\Gamma_{\alpha, -1}} c_{j_\alpha} \;,
\end{equation}
with $\Gamma_{\alpha, \sigma}$ the exchange rates at each boundary. 
The significant simplification involved in the Lindblad approach comes at the cost of losing direct access to the reservoirs' observables, such as the total number of particles on each reservoir $\Hat{N}_{\alpha, \sigma}$. Nevertheless, when studying transport properties, we will see how
these observables can still be retrieved from the full history of the system's evolution.

Considering the noise average $\overline{\rho}$, Eq.~\eqref{Eq:TimeEvUni} reduces to Lindblad dynamics with tight-binding hopping and on-site dephasing, which we call the \textit{dephasing XX model}~\cite{medvedyeva2016}. Its long-time behaviour was analysed in~\cite{znidaric2010,znidaric2010jphysa,LDdeph} and related at large $L$ to classical SSEP. Here, instead, we study the full statistics $\mathbb{P}_{t}(\rho)$ of the density matrix over noise realizations, formally governed by the Fokker-Planck equation associated with Eq.~\eqref{Eq:TimeEvGen}. Linear observables can be extracted from $\overline{\rho}_t$, but higher cumulants and other nonlinear functionals of $\rho$ cannot, and require new tools.
In the limit $t \to \infty$ at fixed $L$, $\mathbb{P}_{t}(\rho)$ converges to the stationary solution of the Fokker–Planck equation $\mathbb{P}_{\infty}(\rho)$, from which the steady-state statistics of any local observable can be extracted as one-time quantum expectation values.
Since we focus on quadratic models, Wick's theorem grants that these can be recovered entirely from the stationary distribution of the correlation matrix $G_{i,j} = \Tr\left(\rho c^{\dagger}_j c_i\right)$. 
More explicitly, we shall address the cumulants $\mathbb{E}_{\infty}^c\left[G^{\otimes n}\right]$, where $\mathbb{E}_{\infty}[\ldots]$ denotes noise-average in the steady state and the upperscript $c$ stands for connected part, defined in the usual way for a multivariate distribution (in this case, the distribution of the entries of $G$). 

\lettersection{Local observables and mapping to QSSEP}
We focus on the steady-state cumulants of the correlation matrix $G$ and then on the thermodynamic limit ($L \to \infty$), where universal behaviour emerges. Writing $x=i/L \in [0,1]$, the number of noisy sites crossed by a particle grows linearly with $L$, so the effective noise also scales with $L$. Equivalently, the diffusive timescale $t_{\rm Diff}\sim L^2$ eventually dominates the dephasing timescale $t_{\rm deph}\sim 1/\gamma$: taking $L \to \infty$ at fixed $\gamma$ or taking strong dephasing first leads to the same diffusion-dominated regime.

This heuristic argument suggests to look at the effective dynamics obtained at large $\gamma$. 
\begin{figure}[ht]
\includegraphics[width=1.\columnwidth]{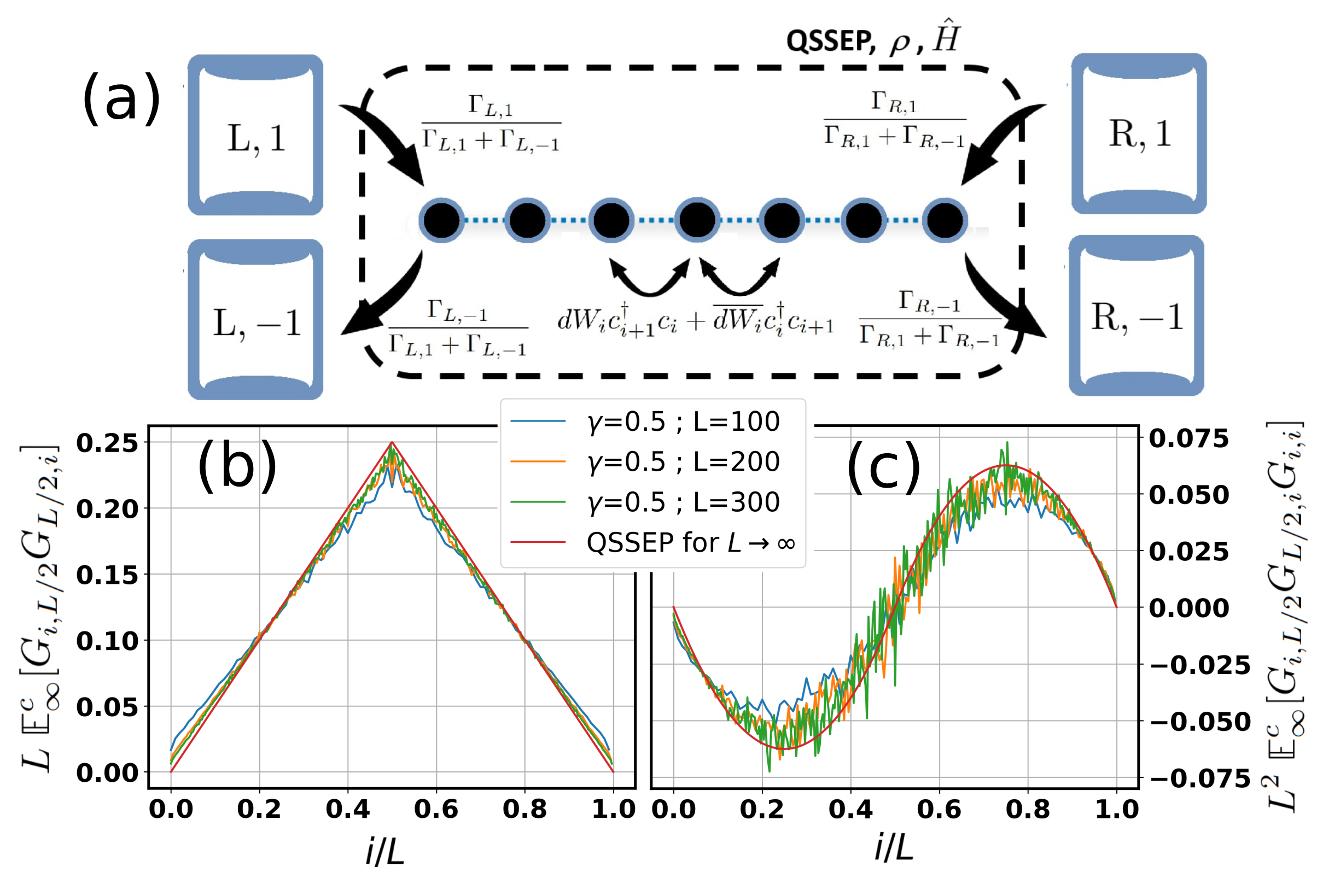}
\caption{In (a), a schematic representation of QSSEP is presented, with jumping rates obtained by taking $\gamma \to \infty$ in \qnoise, as described in appendix \ref{sec:snl} of~\cite{Note1}. In (b) and (c), steady-state cumulants of \qnoise's correlation matrix $G$ are plotted versus $i/L$. The convergence with $L$ of $\mathbb{E}_{\infty}^c[G_{i,L/2}G_{L/2,i}]$ and $\mathbb{E}_{\infty}^c[G_{i,i}G_{i,L/2}G_{L/2,i}]$ is checked against the analytical predictions for QSSEP. Each curve was obtained by averaging over $100$ realizations at sufficiently long times to reach the steady state. }
\label{fig:LargeL}
\end{figure}
Let us first consider the leading order, where the strong dephasing projects onto classical (diagonal) density matrices. However, upon rescaling time as $t \to \gamma t/2$, one can consider the effective residual dynamics. In~\cite{10.21468/SciPostPhys.3.5.033}, the authors showed that, for a closed chain, it coincides exactly with QSSEP \footnote{Even though in this paper the authors work with a spin chain, their results can be directly mapped to spinless fermions through a Jordan-Wigner transform}. QSSEP is a model of free fermions described by the Hamiltonian increment $d\Hat{H} = \sum_{j=1}^{L-1} \left(d\xi_j c^{\dagger}_{j+1} c_j + d\xi_j^* c^{\dagger}_{j} c_{j+1} \right)$, where the $d\xi_j$ are complex and independent Wiener increments with $\overline{d\xi_j d\xi_k^*} = dt \delta_{j,k}$. Here we consider QSSEP with open boundary conditions, where, just as in Eq.~\eqref{Eq:TimeEvGen}, two reservoirs exchange particles at the two boundaries. We shall denote the corresponding jumping rates with an additional tilde, $\tilde{\Gamma}_{\alpha, \sigma}$, to distinguish them from the \qnoise case. More explicitly, from Ito calculus we get the unitary part of the evolution of the density matrix
\begin{equation}
    [d\rho]_{\rm uni} =  -i [d\hat{H}, \rho ]  +  \sum_{j=1}^{L-1} \left(\mathcal{D}_{\Hat{L}_j}[\rho] + \mathcal{D}_{\Hat{L}_j^{\dagger}}[\rho]\right) dt\;,
    \label{Eq:TimeEvGenQSSEP}
\end{equation}
where $\Hat{L}_j = c^{\dagger}_{j+1} c_j$. In section~\ref{sec:snl} of \cite{Note1}, following a more algebraic approach, we show that the equivalence between QSSEP and \qnoise in the limit $\gamma \to \infty$ extends to the open chain, provided that we consider a QSSEP model with two less sites ($L-2$ sites) and the jumping rates $\tilde{\Gamma}_{\alpha,\sigma} = \Gamma_{\alpha, \sigma}/(\Gamma_{\alpha,-1} +\Gamma_{\alpha,1})$ (see Fig.~\ref{fig:LargeL}a). These modifications do not compromise our claim of equivalence of limits, as the slight change in the chain's length becomes irrelevant in the thermodynamic limit ($L \to \infty$). 
Moreover, under diffusive scaling, the reservoirs inject and remove particles on a much shorter timescale than the bulk dynamics, thereby fixing the boundary densities, $\rho_{\alpha}$, and making them the only parameters that remain relevant in the thermodynamic limit. Indeed, these agree for \qnoise and the corresponding QSSEP model, $\rho_{\alpha}=\Gamma_{\alpha,1}/\bigl(\Gamma_{\alpha,1}+\Gamma_{\alpha,-1}\bigr)=\tilde{\Gamma}_{\alpha,1}$ and $\rho_{\alpha}=\tilde{\Gamma}_{\alpha,1}/\bigl(\tilde{\Gamma}_{\alpha,1}+\tilde{\Gamma}_{\alpha,-1}\bigr)=\tilde{\Gamma}_{\alpha,1}$ respectively, thus estabishing that in the thermodynamic limit the boundaries of both models behave effectively the same.
The thermodynamic limit of QSSEP has been extensively studied in the literature and, in particular, in Ref.~\cite{PhysRevLett.123.080601} the authors showed that the leading order terms in $L$ of the steady state correlation matrix cumulants are described by simple formulas. In particular, $\mathbb{E}_{\infty}\left[ G_{i_1,j_1}\ldots G_{i_n,j_n} \right]$ is non-vanishing only when $j_{p} = i_{\sigma_p}$ (for some $n$-element permutation $\sigma$) and in these cases it is a polynomial function of the residual $i$'s away from the boundaries and contact points (i.e, when two indices approach each other).



Beyond leading order, Appendix \ref{sec:EquivalenceOfLimits} of \cite{Note1} shows that higher powers of $\gamma^{-1}$ are irrelevant in the diffusive scaling. This strongly supports the equivalence of the two models in the thermodynamic limit based on the full perturbative expansion at large $\gamma$.
%
%
%
%
To further support this equivalence, in Fig.~\ref{fig:LargeL}b and \ref{fig:LargeL}c we compare \qnoise numerics at fixed $\gamma$ and large $L$ with the analytical QSSEP predictions of \cite{PhysRevLett.123.080601}. The data are taken at times large enough to reach the steady state and show that the relevant cumulants of $G$ up to third order coincide in both models in the thermodynamic limit.

\lettersection{Transport Observables} 
Beyond system observables such as the site occupations, we also consider transport observables, which depend on the reservoirs' state (or at least on the history of the system's evolution). For the net charge transferred from the system to the right reservoirs, if these initially contain $N_0$ particles, measuring $\hat{N}_{\text{R}}$ at time $t$ gives
\begin{equation}
    P_t(\Delta N_{\text{R}}) = \sum_{N} \delta(N - N_0 - \Delta N_{\text{R}}) P_{\text{R}}(N,t).
\end{equation}
$P_{\text{R}}(N,t)$ is the Born-rule probability, $P_{\text{R}}(N,t)=\Tr\left( \rho_T \Hat{\Pi}_{\text{R},N}\right)$, with $\Hat{\Pi}_{\text{R},N}$ the projector onto the $N$-particle sector of the right reservoirs and $\rho_T$ the system+reservoirs density matrix. For Markovian reservoirs and finite $L$, the system has a finite correlation time $t_C$, so at long times $\Delta N_{\text{R}}$ obeys a large-deviation principle,
$P_t(\Delta N_{\text{R}}) \stackrel{t\gg t_C}{\sim} e^{-I(\Delta N_{\text{R}}/t) t}$.
The ratio $\Delta N_{\text{R}}/t=:J$ is the time-averaged current and $I(J)$ its rate function. Under mild assumptions, the Gartner-Ellis theorem applies~\cite{LDtouchette}: if the cumulant generating function (CGF), $\lambda(s) = \lim_{t \to \infty} \frac{1}{t}\log\left( \Tr\left[\rho_T e^{s \Delta \hat{N}_{\text{R}}(t)} \right] \right)$,
is well-defined and smooth, with $\Delta \hat{N}_{\text{R}}(t) = \hat{N}_{\text{R}}(t) - N_0$, then
$I(J) = \sup_{s \in \mathbb{R}} \left(s J - \lambda(s)\right)$. The parameter $s$ is known as the counting field.
Formally, $\lambda(s)$ and $I(J)$ depend on the noise realization, but they are self-averaging and independent of the initial condition (see Sec.~\ref{sec:proofSA} in \cite{Note1}). One can likewise define the CGF of the \textit{dephasing XX} model, $\lambda_{\rm Deph}(s) =\lim_{t \to \infty} \frac{1}{t}\log\left( \Tr\left[\overline{\rho_T} e^{s \Delta \hat{N}_{\text{R}}(t)} \right] \right)$. Derivatives at $s=0$ give the cumulants; for instance, $J_{\rm mp} := \lambda'(0)$ is the most probable current. While $\lambda'(0) = \lambda'_{\rm Deph}(0)$, $\lambda(s)$ captures large deviations for a fixed noise realization, whereas $\lambda_{\rm Deph}(s)$ also includes fluctuations between realizations. Consistently, $ \lambda(s) \leq \lambda_{\rm Deph}(s)$, by concavity of the logarithm. In the next section we show that equality is recovered at leading order in large $L$, where both models agree with SSEP~\cite{PhysRevLett.94.030601,RevModPhys.87.593,DerridaAdditivityPrinciple}.

\lettersection{CGF from gauge invariance}
To compute $\lambda(s)$, it is convenient to  introduce a pseudo density matrix $\rho_{T,s} := e^{\frac{s}{2} \Hat{N}_R} \rho_T e^{\frac{s}{2} \Hat{N}_R}$. 
In our assumptions of Markovianity, the reservoirs can be traced out for arbitrary $s$~(see appendix \ref{sec:countingfield_derivation}), defining $\rho_s = \Tr_{\mathcal{R}}(\rho_{T,s})$, which satisfies a modified stochastic Lindblad equation~\cite{Derrida_2007,derrida2011microscopic, LDtouchette}
\begin{equation}
    d\rho_s = [d\rho_s]_{\rm uni} + 
    [d\rho_s]_{\rm bath}+
    \sum_{\sigma} \left(e^{-\sigma s} - 1\right) \Hat{L}_{\text{R},\sigma} \rho_s \Hat{L}_{\text{R},\sigma}^{\dagger} dt \;.
    \label{Eq:TimeEvGenCGF}
\end{equation}
Due to the last term, this evolution is not trace preserving and
\begin{equation}
\lambda(s) = \lim_{t \to \infty} \frac{1}{t} \log \left[\Tr\left(\rho_s(t)\right)\right] = \lim_{t\to\infty} \frac{d}{dt}\overline{\log \left[\Tr\left(\rho_s(t)\right)\right]}.
\end{equation}
The second equality uses the self-averaging nature in time of $\lambda(s)$ at finite $L$.
Evaluating explicitly the time derivative in the previous expression yields
\begin{equation}
    \lambda(s) = \sum_{\sigma} \Gamma_{\text{R},\sigma} \left(e^{-\sigma s} - 1\right) \left(\delta_{\sigma,1} - \sigma \mathbb{E}_{\infty}[\left(G_s\right)_{L,L}]\right) \;,
    \label{Eq:TraceG}
\end{equation}
where $\left(G_s\right)_{i,j}=\Tr\left(\rho_s c_j^{\dagger} c_i\right)/\Tr(\rho_s)$ is the correlation matrix. Since the evolution of $\rho_s$ is quadratic, gaussianity is preserved and correlation functions of $\rho_s/\Tr(\rho_s)$ are fully determined by $G_s$~(see Sec.~\ref{sec:ProofGaussianity} in \cite{Note1}). Equation~\eqref{Eq:TraceG} therefore expresses $\lambda(s)$ through the diagonal element $\overline{(G_s)_{L,L}}$. However, the rightmost term in Eq.~\eqref{Eq:TimeEvGenCGF} makes $G_s$ obey a closed nonlinear SDE, so the averages of its tensor powers satisfy an infinite hierarchy,
\begin{equation}
\label{eq:Ghierarchy}
    \partial_t \overline{G_s^{\otimes n}} = \mathcal{F}^{(n)} [\{\overline{G_s^{\otimes m}}\}_{m=n-1}^{n+1}] \;.
\end{equation}
This form is general, though the functionals $\mathcal{F}^{(n)}(\ldots)$ are model-dependent. At finite $L$, determining $\lambda(s)$ requires the solution of the full hierarchy and is thus problematic. In contrast, for $s=0$, there is no dependence on higher moments $m > n$ of $G_s$ and for some models the quantities $\overline{G_{s=0}^{\otimes n}}$ can be systematically determined~\cite{PhysRevLett.123.080601}. 

In the large $L$ limit, we analyse $\tilde{\lambda}(s) = \lim_{L \to \infty} \frac{\gamma L}{2} \lambda(s)$. To study $\tilde{\lambda}(s)$, we first make use of the previous observation that in the steady state and for large system sizes, the statistical behaviour of system observables and of the transferred charge in \qnoise becomes identical to that of QSSEP (after appropriately rescaling time by $\gamma/2$).
Thus, in the following we focus on the latter. Secondly, we remark that, since only a finite amount of charge can be accumulated in any portion of the system, the transferred charge across any bond $j$, i.e. $\Delta \hat{N}_j := \Delta \hat{N}_{\text{R}} + \sum_{i > j}  \Delta \hat{n}_i$, has the same rate function in large $t$ as $\Delta \hat{N}_R$. Even more generally, we can collect the transferred charge across all bonds, setting $\Delta \hat{N}[f] := \frac{1}{L+1}\sum_{j=0}^L f_j \Delta \hat{N}_j$ for a set of coefficients $\{f_j\}$ satisfying $\sum_{j=0}^L f_j = L+1$, and, once again, $\Delta \hat{N}[f]$ has the same rate function as $\Delta \hat{N}_R$~(see Sec.~\ref{sec:proofGT} in \cite{Note1} for a rigorous proof). 
In the large $L$ limit, we choose the weights $f_j$ to converge to a smooth normalised function as $L^{-1} \sum_j f_j \to \int_{0}^1 dx \; f(x)$. This choice of $f(x)$ represents a huge gauge freedom that we exploit for QSSEP to distribute the counting field across the chain, turning Eq.~(\ref{Eq:TraceG}) into (see appendix \ref{sec:QSSEP_CGF_derivation})
\begin{equation}
    \Tilde{\lambda}(s) = -  s \int_0^1 dx \; f^2(x) \left(\frac{\partial_x g_s(x)}{f(x)}-sg_s(x) \left(1-g_s(x)\right)\right),  
    \label{Eq:CGFv1Main}
\end{equation}
where $g_s(x) = \lim_{L \to \infty} \mathbb{E}_{\infty}[\left(G_s\right)_{xL,xL}]$. As explained above, $\Tilde{\lambda}(s)$ must be independent of $f(x)$. This is true because  Eq.~\eqref{eq:Ghierarchy}, which, for $n=1$,  determines $g_s(x)$, is also modified when the weights $f(x)$ are included (see Eq.~\eqref{eq:eqforgs}). Crucially, if the function $f(x)$ satisfies
\begin{equation}
 \partial_x f_s(x) = s f_s^2(x) \left(2 g_s(x) -1 \right),
 \label{eq:fconstr}
\end{equation}
then $g_s(x)$ decouples from the higher moments and satisfies a closed differential equation that can be easily solved. From this, we derive
\begin{equation}
 \Tilde{\lambda}(s) = \Theta(w_s-1)\,\left(\text{arccosh}(w_s)\right)^2 -  \Theta(1-w_s)\,\left(\text{arccos}(w_s)\right)^2
\label{eq:CGFSSEP}
\end{equation}
where $w_s = \sqrt{\left(1 + \left(e^s - 1\right) \tilde{\Gamma}_{L,1}\right) \left(1 + \left(e^{-s} - 1\right) \tilde{\Gamma}_{R,1}\right) }$ and $\Theta(x)$ is the Heaviside function. This result matches perfectly
with the one derived for SSEP~\cite{Derrida2004,Derrida_2007, derrida2011microscopic, mallick2015exclusion}. Indeed, the use of the gauge freedom can be applied
to SSEP itself, providing a new derivation of this well-known result.

As claimed in the introduction, classical SSEP behavior emerges very generally, extending to any 1D quadratic model with quasi-local hopping $\Hat{H}_0 = - \sum_{j<k} \left(J_{j,k} c^{\dagger}_j c_{k} + J_{j,k}^* c^{\dagger}_{k} c_{j}\right)$, where $J_{j,k}\to 0$ sufficiently fast with $|j-k| \to \infty$. The large-$\gamma$ expansion then leads to a mapping to a quasi-local extension of QSSEP. In the large-$L$ limit, gauge invariance again gives $\lambda(s) \stackrel{L\to\infty}{\simeq} \frac{2}{\gamma L \mathcal{J}} \Tilde{\lambda}(s)$, with $\mathcal{J}$ set by the spatial integral of the $J_{j,k}$'s. Since the argument relies on quasi-locality and gauge invariance rather than on specifically one-dimensional features, we expect the same equivalence to extend to arbitrary higher-dimensional geometries, as discussed in Sec.~\ref{sec:higherdimensions} of \cite{Note1}.
\begin{figure}[ht]
\includegraphics[width=0.9\columnwidth]{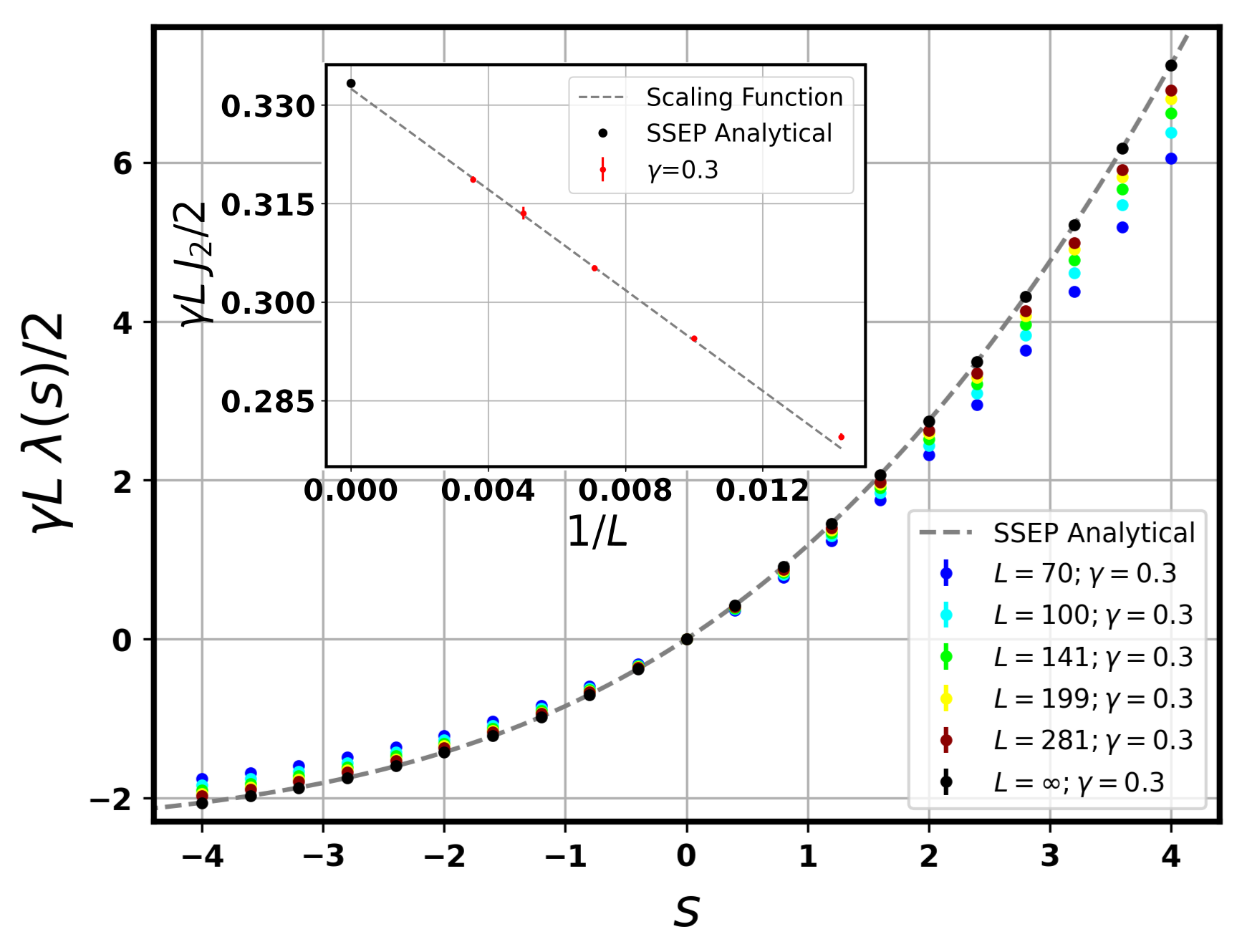}
\caption{CGF of the current for different system sizes of \qnoise, compared with the large-$L$ SSEP prediction rescaled by $L$. The reservoir rates are $\Gamma_{\text{L},1}=\Gamma_{\text{R},-1}=2$ and $\Gamma_{\text{L},-1}=\Gamma_{\text{R},1}=0$. Inset: rescaled second cumulant $\gamma L J_2/2$ versus $L$, obtained from the discrete second derivative around $s=0$ of each CGF curve.}
\label{fig:CGF}
\end{figure}

In order to support this analysis, we show in Fig.~\ref{fig:CGF} the CGF of the current for different system sizes of \qnoise at a finite fixed $\gamma$, against our analytical predictions given in Eq.~\eqref{eq:CGFSSEP}~(a more detailed analysis of the numerical data can be found in Sec.~\ref{sec:ComplementNum} of \cite{Note1}.). The convergence with $L$ is clear, which corroborates the findings of the last two sections, namely in relating \qnoise to QSSEP in the thermodynamic limit and now the current statistics of QSSEP and SSEP.

\lettersection{Conclusions}
In this letter, we analysed out-of-equilibrium dynamics of noisy free fermions on a $1D$ lattice exchanging particles with reservoirs at both edges. We considered \qnoise as a simple model for a quantum transport setup and focused on the total transported charge, which obeys a large deviation principle. We argue for universality by reducing, through coarse-graining, the model to QSSEP. Then, we introduced a new technique that exploits the arbitrariness on where to measure the current to calculate exactly the cumulants of charge transferred in QSSEP in the thermodynamic limit and show they coincide with SSEP. Extending this method to other stochastic models of transport in $1D$~\cite{derrida1993exact} is an interesting perspective.

More generally, our analysis applies to local one-dimensional free-fermion systems with a well-defined thermodynamic limit. We conclude that current fluctuations in spinless $U(1)$-conserving free fermions with unitary noise are universal and match those of SSEP. Natural open questions concern the role of interactions in noisy boundary-driven systems~\cite{PhysRevLett.131.210402}, and the effects of long-range hopping or higher dimensions.

\lettersection{Acknowledgements}
ADL warmly thanks Denis Bernard, Tony Jin and Jacopo De Nardis for discussions. ADL acknowledges support from the ANR JCJC grant ANR-21-CE47-0003 (TamEnt). 
This work was supported by Fundação para a Ciência e Tecnologia (FCT) through grants No.2022.11940.BD (JC), UID/04540/2025 (DOI: 10.54499/UID/04540/2025) to the I\&D unit Centro de Física e Engenharia de Materiais Avançados (CeFEMA) (JC, PR), and through project SCALE-QLT (DOI: 10.54499/2024.16192.PEX) (PR).
This work was also supported by FCT I.P. under Project 2023.10864.CPCA.A1 at OBLIVION supercomputer. Finally, this work received funding by the CY Initiative of Excellence (grant “Investissements d’Avenir” ANR-16-IDEX-0008) and part of it was developed during PR stay at CY Advanced Studies whose support is gratefully acknowledged.

\onecolumngrid
\newpage 

\appendix
\setcounter{equation}{0}
\setcounter{figure}{0}
\renewcommand{\thetable}{S\arabic{table}}
\renewcommand{\theequation}{S\thesection.\arabic{equation}}
\renewcommand{\thefigure}{S\arabic{figure}}
\setcounter{secnumdepth}{2}

\begin{center}
{\Large End matter \\ 
\vspace{0.22cm}
\titleinfo
}
\end{center}

\section{Derivation of QSSEP's CGF}
\label{sec:QSSEP_CGF_derivation}
In this section, we shall prove that, in the thermodynamic limit and under the appropriate rescaling, the QSSEP's CGF agrees with the well-known result established for SSEP in \cite{Derrida2004}. The derivation presented here can be adjusted, as we shall describe, to provide a rederivation of the SSEP's CGF expression.

We begin by considering the time evolution of QSSEP's pseudo-density matrix $\rho_s$, obtained by substituting Eq.~(\ref{Eq:TimeEvGenQSSEP}) into Eq.~(\ref{Eq:TimeEvGenCGF}). For brevity, we denote this evolution by $\partial_t \rho_s = \mathcal{L}_s(\rho_s; t)$. At this stage, we make use of the gauge invariance discussed in the main text to redistribute the counting field smoothly across the chain. As explained in Sec.~\ref{sec:proofGT} of \cite{Note1}, this is achieved by considering the transformed (tilted) density matrix $\Tilde{\rho}_s = \Hat{U} \rho_s \Hat{U}$, where $\Hat{U} = e^{\frac{\tilde{s}}{2} \sum_{j=1}^{L} F_j c^{\dagger}_j c_j}$ and $\tilde{s} = \frac{s}{L+1}$. Choosing $F_j$ such that $F_0 = 0$ and $F_{L+1} = L + 1$, and for notational simplicity dropping the tilde on the transformed density matrix, we arrive at the modified evolution equation:
\begin{multline}
    d\rho_s = \sum_{0 < j < L} \left( e^{\tilde{s} f_j} \hat{L}_{j} \rho_s \hat{L}^{\dagger}_{j} + e^{-\tilde{s} f_j} \hat{L}^{\dagger}_{j} \rho_s \hat{L}_{j} - \frac{1}{2} \{\hat{L}^{\dagger}_{j} \hat{L}_{j} + \hat{L}_{j} \hat{L}^{\dagger}_{j}, \rho_s \} \right)  + \sum_{\sigma} \left( e^{\sigma \Tilde{s} f_0} \Hat{L}_{\text{L},\sigma} \rho_s \Hat{L}_{\text{L},\sigma}^{\dagger} + e^{-\sigma \Tilde{s} f_L} \Hat{L}_{\text{R},\sigma} \rho_s \Hat{L}_{\text{R},\sigma}^{\dagger} \right) -  \\ - \frac{1}{2} \sum_{\sigma} \{\Hat{L}_{\text{L},\sigma}^{\dagger} \Hat{L}_{\text{L},\sigma}+\Hat{L}_{\text{R},\sigma}^{\dagger} \Hat{L}_{\text{R},\sigma}, \rho_s\}  
    - i \sum_{0 < j < L}\left( e^{\frac{\tilde{s} f_j}{2}}\hat{L}_{j} \rho_s d\xi_j + e^{-\frac{\tilde{s} f_j}{2}}\hat{L}^{\dagger}_{j} \rho_s d\xi^*_j - e^{-\frac{\tilde{s} f_j}{2}} d\xi_j  \rho_s \hat{L}_{j} - e^{\frac{\tilde{s} f_j}{2}} d\xi^*_j \rho_s \hat{L}^{\dagger}_{j}\right), 
\label{eq:BulkCFmain}
\end{multline} 
where, for convenience, we denote the discrete derivative of this function by $f_j = F_{j+1}-F_{j}$. The function $f_j$ is only constrained to satisfy $\sum_{j=0}^{L} f_j = L+1$ and it corresponds exactly to the function $f_j$ introduced in the main text.

At the same time, this transformation also changes relation \eqref{Eq:TraceG}, which becomes (see Sec.~\ref{sec:BigEqs} in \cite{Note1})
\begin{equation}
        \lambda(s) =  \sum_{j=1}^{L-1} \left( \left(e^{\tilde{s} f_{j}}-1\right) \mathbb{E}_{\infty}\left[\left(G_s\right)_{j,j}\left(1-\left(G_s\right)_{j+1,j+1} \right)\right]+  \left(e^{-\tilde{s} f_{j}}-1\right) \mathbb{E}_{\infty}\left[\left(G_s\right)_{j+1,j+1}\left(1-\left(G_s\right)_{j,j}\right)\right] \right).
        \label{eq:cgfexact}
\end{equation}
In writing this equation, we chose $f_0=f_L=0$, made use of the expression $\overline{d\log\left(\Tr\left(\rho_s\right)\right)}=\overline{\frac{d \Tr\left(\rho_s\right)}{\Tr\left(\rho_s\right)}} - \frac{1}{2}\overline{\left(\frac{d \Tr\left(\rho_s\right)}{\Tr\left(\rho_s\right)}\right)^2}$ and applied Wick's theorem (see Sec.~\ref{sec:ProofGaussianity} in \cite{Note1}).
Resorting to the same set of identities, one can rewrite Eq.~\ref{eq:BulkCFmain} for $\overline{d\left(G_s\right)_{j,j}}$ instead, and realize that the RHS depends now on terms of the form $\overline{\left(G_s\right)_{j,k}\left(G_s\right)_{k,j}}$ and $\overline{\left(G_s\right)_{j,k}\left(G_s\right)_{k,j}\left(G_s\right)_{k\pm1,k\pm1}}$ (see Eq.~(\ref{Eq:Gevs}) of Sec.~\ref{sec:BigEqs} in \cite{Note1}).


At this point, we introduce the key assumption that the scaling with $L$ of the cumulants $\mathbb{E}_{\infty}^c\left(G_s^{\otimes N}\right)$ is independent of $s$ and thus reduces to the scaling observed at $s=0$, which was first described in~\cite{PhysRevLett.123.080601} for the QSSEP without a counting field. More concretely, we use that 
\begin{equation}
    \lim_{L \to \infty} L \hspace{1mm} \mathbb{E}_{\infty}^c\left[\left(G_s\right)_{x L, y L}\left(G_s\right)_{y L,x L}\right]= H_s\left(x,y\right), \hspace{2mm} \lim_{L \to \infty} \frac{\mathbb{E}_{\infty}^c\left[\left(G_s\right)_{x L,x L}\mathcal{F}\left(G_s\right)\right]}{\mathbb{E}_{\infty}\left[\left(G_s\right)_{x L,x L}\right] \hspace{1mm}\mathbb{E}_{\infty}\left[\mathcal{F}\left(G_s\right)\right]} = 0,
    \label{eq:assumptions}
\end{equation}
where $\mathcal{F}\left(G_s\right)$ represents a power of the entries of $G_s$ for which $\mathbb{E}_{\infty}\left[\mathcal{F}\left(G_s\right)\right] \neq 0$ and $H_{s}(x,y)$ is a well-behaved function of $x$ and $y$ that may depend parametrically on $s$. Even though we do not have a rigorous proof of this, one can argue heuristically by taylor expanding $G_s$ around $s=0$ and proceeding by induction at all orders, see Sec.~\ref{sec:CGFassumptions} in \cite{Note1}. Equipped with the relations in Eq.~(\ref{eq:assumptions}) and expanding in large $L$, one derives Eq.~(\ref{Eq:CGFv1Main}) from the continuum limit of Eq.~(\ref{eq:cgfexact}). Additionally, from the equation for $\overline{\left(G_s\right)_{j,j}}$ (Eq.~(\ref{Eq:Gevs}) of Sec.~\ref{sec:BigEqs} in \cite{Note1}) evaluated in the steady state, one obtains
\begin{multline}\label{eq:eqforgs}
    \partial_x^2 g_s(x) - s (1- 2 g_s(x)) \left(2 \partial_x\left(f(x) g_s(x)\right) - \partial_x f(x)  g_s(x)\right) - s \partial_x f(x) g_s^2(x) + s^2 f^2(x) g_s(x) \left( 1 - g_s(x)\right) \left( 1 - 2 g_s(x)\right) =\\ = -  \int_0^1 dy\; \left(  s^2 f^2(y) (2 g_s(y)-1) - s \partial_y f(y) \right) H_s(x,y).
\end{multline}
It is now clear that choosing $f(x)$ according to Eq.~(\ref{eq:fconstr}) eliminates the term containing $H_s(x,y)$ and we are left with two closed coupled differential equations for $g_s(x)$ and $f_s(x)$. $g_s$ satisfies $g_s(0)= \rho_{\text{L}} = \frac{\Gamma_{\text{L},1}}{\Gamma_{\text{L},1}+\Gamma_{\text{L},-1}}$ and $g_s(1)= \rho_{\text{R}} = \frac{\Gamma_{\text{R},1}}{\Gamma_{\text{R},1}+\Gamma_{\text{R},-1}}$ on the boundaries and $f_s(x)$ satisfies $\int_0^1 dx \; f_s(x) = 1$. Rewriting this system of equations for $h_s(x) = f_s(x) g_s(x)$ instead of $g_s(x)$ and rescaling $x \to x/s$, we summarize all the previous results as 
\begin{equation}
\begin{cases}
    \partial_x^2 h_s(x) - 2 h_s(x) \partial_x h_s(x) = 0, \hspace{1mm}  h_s(0) = \rho_{\text{L}} f_s(0), \hspace{1mm} h_s(s) = \rho_{\text{R}} f_s(s),\\ 
    \partial_x f_s(x) =  f_s(x) \left(2 h_s(x) - f_s(x) \right), \hspace{1mm} \int_0^s dx \; f_s(x) = s.
\end{cases} \Longrightarrow \tilde{\lambda}(s) =   s \int_0^s dx \; \left( h_s^2(x) - \partial_x h_s(x)\right).
\label{eq:finalsystem}
\end{equation}
The solution of these equations yields the result that was already known for SSEP, 
\begin{equation}
 \tilde{\lambda}(s) =  - \left(\arccos{(w_s)} \right)^2 \Theta(1-w_s) +  \left( \text{arccosh}(w_s)\right)^2 \Theta(w_s-1) , \hspace{1mm} \text{ where } w_s = \sqrt{\left(1 + \left(e^s - 1\right) \rho_{\text{L}}\right) \left(1 + \left(e^{-s} - 1\right) \rho_{\text{R}}\right) },
\end{equation}
where $\Theta(x)$ is the Heaviside function. The exact same argument can be used to rederive the CGF formula for SSEP, by assuming instead that the correlators $\langle \hat{n}_{i_1} ... \hat{n}_{i_N} \rangle_s^c$ have the same large-$L$ scaling limit for all $s$. 


\section{Counting Field}
\label{sec:countingfield_derivation}
In order to keep the discussion self-contained, in this appendix we show that the CGF of the current (as defined in the introduction) can be obtained through Eq.~\eqref{Eq:TimeEvGenCGF} and Eq.~\eqref{Eq:TraceG}.

We shall consider a general (unidimensional) system interacting with reservoirs on its left and right. Of course if the whole setup is isolated, it undergoes unitary evolution; however, for generality, we shall consider that the time evolution is generated by a Lindbladian $\partial_t \rho_T = \mathcal{L}_T\left(\rho_T\right)$ that conserves the total particle number. Since we are ultimately interested in studying the total amount of charge transferred to the right reservoirs, we write the full density matrix of the System + Reservoirs as $\rho_T = \sum_{M \alpha, N \beta} \rho^{(M\alpha,N\beta)}_{L,S} \otimes \ket{R; M \alpha} \bra{R; N \beta}$, where $\ket{R;M\alpha}$ represents a state of the right reservoirs on which they contain $M$ particles in total ($\alpha$ is a label for other degrees of freedom).

Anticipating the rest of the argument, we write the superoperator $\mathcal{L}_T$ as a sum over components that induce a specific number of particle jumps from the system to the right reservoirs, $\mathcal{L}_T = \sum_{n,m} \left(\mathcal{L}_T\right)_{(n,m)}$, where  $\left(\mathcal{L}_T\right)_{(n,m)} [\rho_T] = \sum_{M, N, \alpha, \alpha', \beta,\beta'} C_{M, N, \alpha, \alpha', \beta,\beta'} \ket{R; M+m ,  \alpha} \bra{R; M \alpha'} \rho_T \ket{R; N \beta'} \bra{R; N+n , \beta}$, for some appropriate coefficients $C$.

Ultimately, we shall be interested in computing the CGF of the current, which, as described in the main text, is defined by $\lambda(s) = \lim_{t \to \infty} \frac{1}{t} \log \left(\sum_{n} P_{\text{R}}(n;t) e^{s \left(n-n_0\right)} \right)$. Using Born's rule, this expression can be rewritten as
\begin{equation}
    \lambda(s) = \lim_{t \to \infty} \frac{1}{t} \log \left(\Tr \left( \rho_T e^{s \Hat{N}_R}\right)\right) = \lim_{t \to \infty} \frac{1}{t} \log \left(\Tr \left( e^{\frac{s}{2} \Hat{N}_R} \rho_T e^{\frac{s}{2} \Hat{N}_R}\right)\right) = \lim_{t \to \infty} \frac{1}{t} \log \left(\Tr \left(  \rho_{T,s} \right)\right),
    \label{eq:cgfrhoTmain}
\end{equation}
where $\rho_{T,s} = e^{\frac{s}{2} \Hat{N}_R} \rho_T e^{\frac{s}{2} \Hat{N}_R}$. $\lambda(s)$ can thus be determined from the time evolution of $\rho_{T,s}$, which reads
\begin{equation}
    \frac{d}{dt} \rho_T = \mathcal{L}_T(\rho_T) \Longrightarrow \frac{d}{dt} \rho_{T,s} = e^{\frac{s}{2} \Hat{N}_R} \mathcal{L}_T(\rho_T) e^{\frac{s}{2} \Hat{N}_R} = \mathcal{L}_{T,s}(\rho_{T,s}),\hspace{2mm}
    \mathcal{L}_{T,s} =\sum_{n,m} e^{\frac{s}{2}\left(n+m\right)} \left(\mathcal{L}_T\right)_{(n,m)}.
    \label{eq:EEcfTmain}
\end{equation}
In second quatized notation, $\mathcal{L}_{T,s}$ is obtained from  $\mathcal{L}_{T}$ by attaching a $e^s$ ($e^{-s}$) factor to every creation (annihilation) operator of a right reservoirs' mode that acts to the left of $\rho_T$ and a $e^{-s}$ ($e^{s}$) factor if it acts to the right of $\rho_T$. 

In this article, we are concerned with the case of Hamiltonian evolution in the total setup of System ($S$) + Reservoirs ($\mathcal{B}$), $\Hat{H}_T = \Hat{H}_S + \Hat{H}_{\mathcal{B}} + \Hat{H}_{int}$, but assuming that the latter are Markovian. This means that a closed time evolution of Lindblad form can be obtained for the system's density matrix $\rho_S = \Tr_{\mathcal{B}}\left(\rho_T\right)$ after a sequence of appropriate approximations, $\frac{d}{dt} \rho_S = \mathcal{L}_S(\rho_S)$. The same sequence of approximations can still be employed to Eq.~(\ref{eq:EEcfTmain}) to write a closed equation for $\rho_{S,s} = \Tr_{\mathcal{B}}\left(\rho_{T,s}\right)$, i.e, $\frac{d}{dt} \rho_{S,s} = \mathcal{L}_{S,s}(\rho_{S,s})$. Note that, from Eq.~(\ref{eq:cgfrhoTmain}), one obtains $\lambda(s) = \lim_{t \to \infty} \frac{1}{t} \log \left(\Tr \left(  \rho_{S,s} \right)\right)$.
From our previous analysis, $\mathcal{L}_{S,s}$ and $\mathcal{L}_{S}$ differ only in the terms that contain creation/annihilation operators that come from $H_{int}$ and express some particle exchange between the system and right reservoirs. Each such creation/annihilation operator carries a $e^s$ or $e^{-s}$ factor depending on whether they act to the left or right of $\rho_S$. Consequently, in general, in order to compute $\mathcal{L}_{S,s}$, one needs to track back these operators to the original equation through the sequence of approximations performed. In the specific case considered in this article, Eq.~(\ref{Eq:TimeEvGen}), the origin of each term is clear and, using the notation introduced in the main text, we conclude that
\begin{equation}
    \Hat{L}_{(R,\sigma)} \rho_{S} \Hat{L}_{(R,\sigma)}^{\dagger} \Longrightarrow e^{- s \sigma} \Hat{L}_{(R,\sigma)} \rho_{S,s} \Hat{L}_{(R,\sigma)}^{\dagger}
\end{equation}
and that all other terms remain the same as they conserve the number of particles on the System + Left reservoirs.

\newpage

\begin{center}
{\Large Supplementary Material \\ 
\vspace{0.22cm}
\titleinfo
}
\end{center}
In this supplementary material we collect the technical details supporting the main text. We discuss the strong-noise reduction to QSSEP in Appendix~\ref{sec:snl}, the mathematical basis of the gauge trick and the self-averaging nature of the current CGF in Appendices~\ref{sec:proofGT} and~\ref{sec:proofSA}, respectively, and the preservation of gaussianity under time evolution in Appendix~\ref{sec:ProofGaussianity}. We then provide further computational details complementing some equations of the main text in Appendix~\ref{sec:BigEqs}, more detailed numerical checks in Appendix~\ref{sec:ComplementNum}, and, in Appendix~\ref{sec:CGFassumptions}, a sketch of a proof of the assumptions made in the derivation of QSSEP's CGF in Appendix~\ref{sec:QSSEP_CGF_derivation}. Finally, we give a more thorough and quantitative discussion of the validity of the equivalence of limits $L\to\infty$ and $\gamma\to\infty$ followed by $L\to\infty$ in Appendix~\ref{sec:EquivalenceOfLimits}, and extend the gauge trick to QSSEP in higher dimensions in Appendix~\ref{sec:higherdimensions}.

\section{Strong noise limit}
\label{sec:snl}

In this section, we are interested in characterizing the dynamics of a system that obeys Eq.~\eqref{Eq:TimeEvGen} in the limit of very strong dissipation, i.e $\gamma \to \infty$. We shall make no assumption about the form of the Hamiltonian (besides being quadratic and having a zero diagonal, $H_{i,i}=0$) and so this model can be taken to represent a quantum system on a generic graph $\mathcal{G}$ of $L$ vertices with two of them, labeled by $1$ and $L$, connected to Markovian reservoirs. The generalization to a more general setup where the system interacts with Markovian reservoirs at $M$ different vertices is straightforward from our analysis, as we shall briefly specify at the end of the section. 
Since we are ultimately concerned with transport properties, we start directly with the evolution equation for the correlation matrix $G_s$ with a counting field. 
Despite one only needing the large time behavior of $\overline{G_s}$ to determine $\lambda_N(s)$ (see Eq.~\eqref{Eq:TraceG}), due to the non-linearity of its steady state equation
, computing the time evolution of $\overline{G_s}$ implies knowledge of $\overline{G_s^{\otimes n}}$ for all $n$ as well. As a consequence, our analysis of the $\gamma \to \infty$ limit must include all the latter n-point correlators. 

In order to simplify the notation, we shall start by re-expressing the time evolution in the infinite dimensional vector space $\mathcal{V}=\oplus_{n=0}^{\infty} V^{\otimes n}$, for $V=\mathbb{C}^{2L}$ a $L^2$-dimensional vector space on which the correlation matrix $G_s$ is represented by $\ket{G_s^{(1)}} = \sum_{i,j} \left(G_s\right)_{i,j} \ket{i;j}$. In a similar fashion, generic tensor products $ \left(G_s^{(n)}\right) = G_s^{\otimes n}$  can be written as vectors $\ket{G_s^{(n)}} = \sum_{\mathbf{i_n}; \mathbf{j_n}} \left(\prod_{k=1}^L\left(G_s\right)_{i_k,j_k}\right) \ket{\mathbf{i_n}; \mathbf{j_n}}$ that belong to $V^{\otimes n}$ ($\mathbf{i}_n=\{i_1,...i_n\}$). Note that, even though of course $\ket{G_s^{(n)}}=\ket{G_s^{(1)}}^{\otimes n}$, we will be interested in the average over noise of these quantities, for which $\overline{\ket{G_s^{(n)}}} \neq \left(\overline{\ket{G_s^{(1)}}}\right)^{\otimes n}$ generically. Nevertheless, there is still a redundancy under permutations of $G_s$ that we could fix by considering the symmetrized sector of $V^{\otimes n}$ (i.e identifying $\ket{\mathbf{i}_n,\mathbf{j}_n} \sim \ket{\sigma(\mathbf{i_n}),\sigma(\mathbf{j_n})})$, for some permutation $\sigma$). However, for simplicity we shall not do so, and the time evolution equation will be such that the identity $ \braket{\sigma(\mathbf{i_n});\sigma(\mathbf{j_n})|G_s^{(n)}} =\braket{\mathbf{i_n};\mathbf{j_n}|G_s^{(n)}}$ is preserved. From now on, we shall refer to the indices  $\mathbf{i_n}$ ($\mathbf{j_n}$) in $\ket{\mathbf{i_n},\mathbf{j_n}}$ as $+$ ($-$) indices.

To fix the notation and simplify the presentation of the following results, we shall introduce the operators $\ket{k}\bra{k'}_{a,+}$ and $\ket{l}\bra{l'}_{a,-}$ which act on the basis vector as $ \left(\ket{k}\bra{k'}_{a,+}\right) \left(\ket{l}\bra{l'}_{b,-} \right)\ket{\mathbf{i}_n;\mathbf{j}_n} = \delta_{i_a,k'_a}\delta_{j_b,l'_b}\ket{i_1...i_{a-1} k_a i_{a+1}...i_n;j_1...j_{b-1} l_b j_{b+1}...j_n}$. Essentially, $\ket{k}\bra{l}_{a,+}$ maps the vector $\ket{l}$ on the copy $(+,a)$ to $\ket{k}$ and annihilates all states orthogonal to $\ket{l}$. For simplicity, we shall drop the index $a$ when $n=1$ (which implies $a=1$).

With this notation, the averaged time evolution equation of $G_s$ reads 
\begin{multline}
      d\ket{G_{s}^{(1)}} = \left(-i \sum_{i,j}\left(H_{i,j} \ket{i}\bra{j}_{+} - H_{i,j}^T \ket{i}\bra{j}_{-}\right)\ket{G_{s}^{(1)}}+ \Gamma_{\text{L},1}\ket{1;1} + e^{-s}\Gamma_{\text{R},1}\ket{L;L}-\gamma(\mathbb{1}-\sum_k \ket{k}\bra{k}_{+}
      \ket{k}\bra{k}_{-})\ket{G_{s}^{(1)}}\right)dt +\\ +i\sqrt{\gamma}\left(\Hat{dW}_{+} - \Hat{dW}_{-} \right)\ket{G_{s}^{(1)}} -\left( \left(\Gamma_{\text{L}} \Hat{Q}_1 + \Gamma_{\text{R}}^{(s)} \Hat{Q}_L\right)\ket{G_{s}^{(1)}}+ \Gamma_s \sum_{k,l}\ket{k;l}\braket{k,L;L,l|G_{s}^{(2)}} \right) dt,\\
      \text{for} \hspace{1mm} \Hat{dW}_{+/-}=\sum_{k=1}^L dW_k \ket{k}\bra{k}_{+/-}, \hspace{1mm}   \Hat{Q}_{k} = \ket{k}\bra{k}_+ + \ket{k}\bra{k}_-\hspace{1mm} \text{and} \hspace{1mm} \Gamma_s = \sum_{\sigma\in \{-1,1\}} \sigma \Gamma_{\text{R},-\sigma}  \left(e^{\sigma s} - 1\right).
\label{Eq:TevGket}
\end{multline}

The time evolution equation for $\ket{G_s^{(n)}}$ can be directly obtained from the previous one by direct application of Itô's rule,
\begin{equation}
    d \ket{G_s^{(n)}} = \sum_{i=1}^n \ket{G_s^{(i-1)}} \otimes d\ket{G_s} \otimes \ket{G_s^{(n-i)}} + \sum_{i<j} \ket{G_s^{(i-1)}} \otimes d\ket{G_s} \otimes \ket{G_s^{(j-i-1)}} \otimes d\ket{G_s} \otimes \ket{G_s^{(n-j)}}.
    \label{Eq:NG}
\end{equation}

Before proceeding, we shall perform a change of variables that turns out to be more convenient for the following perturbative expansion, namely
\begin{equation}
    \ket{\Tilde{G}_s} = \ket{G_s} - \frac{\Gamma_{\text{L},1}}{\Gamma_{\text{L},1}+\Gamma_{\text{L},-1}} \ket{1;1}  - \frac{\Gamma_{\text{R},1}}{\Gamma_{\text{R},1}+e^s\Gamma_{\text{R},-1}} \ket{L;L}.
\end{equation}

Plugging this in Eq.~\eqref{Eq:TevGket}, one can show that

\begin{multline}
      d\ket{\Tilde{G}_{s}^{(1)}} = \left(-i \sum_{i,j}\left(H_{i,j} \ket{i}\bra{j}_{+} - H_{i,j}^T \ket{i}\bra{j}_{-}\right) \ket{\Tilde{G}_{s}^{(1)}}+ \ket{\mathcal{H}}-\gamma(\mathbb{1}-\sum_k \ket{k}\bra{k}_{+} \ket{k}\bra{k}_{-})\ket{\Tilde{G}_{s}^{(1)}}\right)dt +\\ +i\sqrt{\gamma}\left(\Hat{dW}_+ - \Hat{dW}_-\right)\ket{\Tilde{G}_{s}^{(1)}}- \left(\left(\Gamma_{\text{L}} \Hat{Q}_1 + \Gamma_{\text{R}}^{(0)} \Hat{Q}_L\right)\ket{\Tilde{G}_{s}^{(1)}} +  \Gamma_s \sum_{k,l}\ket{k;l}\braket{k,L;L,l|\Tilde{G}_{s}^{(2)}} \right) dt,\\
      \text{for} \hspace{1mm} \ket{\mathcal{H}} = -i \sum_k \left( \rho_{\text{L}}\left( H_{k,1} \ket{k;1}- H_{1,k} \ket{1;k}\right) + \rho_{\text{R}}^{(s)} \left( H_{k,L} \ket{k;L}- H_{L,k} \ket{L;k}\right)\right), 
\label{Eq:TevGtket}
\end{multline}
where now the inhomogeneous term depends on the Hamiltonian, on $\rho_{\text{L}} = \frac{\Gamma_{\text{L},1}}{\Gamma_{\text{L},1} + \Gamma_{\text{L},-1}}$ and on $ \rho_{\text{R}}^{(s)}=\frac{\Gamma_{\text{R},1}}{\Gamma_{\text{R},1} + e^s \Gamma_{\text{R},-1}} $. The evolution equation for $\ket{\Tilde{G}_s^{(n)}}$ is given by replacing all $G_s$ in Eq.~\eqref{Eq:NG} by $\Tilde{G}_s$.

Since we are ultimately interested in studying the time evolution of the system perturbatively in $\gamma^{-1}$, we start by characterizing the action of the operator proportional to $\gamma$ on the space $V^{\otimes n}$, that we denote by $\Hat{F}_n$. $\Hat{F}_n$ is a block component of an operator $\Hat{F}$ defined on the whole space $\mathcal{V}$, i.e $\Hat{F}_n = \mathbf{\Hat{P}}_{n} \Hat{F} \mathbf{\Hat{P}}_{n} $, for $\mathbf{\Hat{P}}_{n}$ a projection operator on the $V^{\otimes n}$ subspace of $\mathcal{V}$. From Eq.~\eqref{Eq:NG} and Eq.~\eqref{Eq:TevGtket}, one can show that $\Hat{F}_n$ is of the form 
\begin{multline}
          \Hat{F}_n = \sum_{a=1}^n \left(\mathbb{1} - \sum_{k=1}^L  \ket{k}\bra{k}_{a,-}\ket{k}\bra{k}_{a,+} \right) +  \sum_{a<b} \sum_{k}\left( \ket{k}\bra{k}_{a,+} - \ket{k}\bra{k}_{a,-}\right) \left( \ket{k}\bra{k}_{b,+} - \ket{k}\bra{k}_{b,-}\right) \Longleftrightarrow\\ \Longleftrightarrow \Hat{F}_n = \frac{1}{2} \sum_{k=1}^L \left(\sum_{a=1}^n \left(\ket{k}\bra{k}_{a,+} - \ket{k}\bra{k}_{a,-}\right)  \right)^2.
      \label{Eq:Dn}
\end{multline}

We conclude that $\Hat{F}_n$ is a diagonal operator on this basis, $\Hat{F}_n = \sum_{\mathbf{i},\mathbf{j}}  \left(f_n\right)_{\mathbf{i_n}\mathbf{j_n}} \ket{\mathbf{i_n}; \mathbf{j_n}} \bra{\mathbf{i_n};\mathbf{j_n}}$, with eigenvalues given by
\begin{equation}
    \left(f_n\right)_{\mathbf{i}\mathbf{j}} = \frac{1}{2}\lVert\Vec{\mathbf{n}}^{(i)}-\Vec{\mathbf{n}}^{(j)}\rVert^2 \implies \Hat{F}_n \ket{\mathbf{i_n};\sigma(\mathbf{i_n})} = 0,
    \label{Eq:NoiseEig}
\end{equation}
where $\sigma$ is a permutation and $\Vec{\mathbf{n}}^{(i)}$ is a $L$-dimensional vector defined by $\Vec{\mathbf{n}}^{(i)}_m = \sum_{k=1}^n \delta_{m,i_k}$ (the definition of $\Vec{\mathbf{n}}^{(j)}$ is the same but with $i$ replaced by $j$). As specified in Eq.~\eqref{Eq:NoiseEig}, the kernel of the operator $\Hat{F}_n$ is composed of all vectors with the same set of left and right indices, $\ket{\mathbf{i_n};\sigma(\mathbf{i_n})}$. As a consequence, when studying the evolution of the averaged $n$-tensor product of $\Tilde{G}_s$'s in the limit of large $\gamma$, one expects that all components on the orthogonal subspace to the kernel of $\Hat{F}_n$ are highly suppressed and the dynamics mostly occurs on the latter. This suggests splitting the whole space $\mathcal{V}$ into the direct sum of the Kernel of $\Hat{F}$ and its orthogonal complement, that we denote by $\perp$. For reasons that become clear after the following analysis, it is convenient to also split the kernel of $\Hat{F}$ into the direct sum of two orthogonal subspaces: the span of the vectors $\ket{\mathbf{i},\sigma(\mathbf{i})}$ that contain at least one index on the boundary (i.e $i_k\in \{1,L\}$ for some $k$), which denote by $\partial$, and its orthogonal complement, represented by $\parallel$. We write the projector onto these subspaces as $\mathbf{\Hat{P}}_{\alpha}$ ($\alpha \in \{\perp, \parallel,\partial\}$) and $\mathbf{\Hat{P}}_{n,\alpha} = \mathbf{\Hat{P}}_{\alpha} \mathbf{\Hat{P}}_{n}$.

Having introduced this notation, let us write explicitly the form of the averaged time evolution equation projected onto a subspace with fixed $n$, 

\begin{equation}
        \frac{d}{dt} \begin{bmatrix}
        \ket{g_{s,\parallel}^{(n)}}\\
        \ket{g_{s,\partial}^{(n)}}\\
        \ket{g_{s,\perp}^{(n)}}
    \end{bmatrix} = - \begin{bmatrix}
        0 & 0 &  \Hat{B}_{\parallel,\perp} \\
        0 &  \Hat{A} &  \Hat{B}_{\partial,\perp}\\
        \Hat{B}_{\perp,\parallel} &  \Hat{B}_{\perp,\partial} &  \gamma \Hat{F}
    \end{bmatrix} \begin{bmatrix}
        \ket{g_{s,\parallel}^{(n)}}\\
        \ket{g_{s,\partial}^{(n)}}\\
        \ket{g_{s,\perp}^{(n)}}
    \end{bmatrix} - \begin{bmatrix}
        0\\
        \Hat{\mathcal{U}}_{\partial,\perp}\ket{g_{s,\perp}^{(n-1)}}\\
        \Hat{\mathcal{U}}_{\perp,\parallel} \ket{g_{s,\parallel}^{(n-1)}} + \Hat{\mathcal{U}}_{\perp,\partial} \ket{g_{s,\partial}^{(n-1)}} +  \Hat{\mathcal{U}}_{\perp,\perp} \ket{g_{s,\perp}^{(n-1)}} 
    \end{bmatrix} +  \begin{bmatrix}
         \Hat{\mathcal{D}}_{\parallel,\partial}  \ket{g_{s,\partial}^{(n+1)}}\\
         \Hat{\mathcal{D}}_{\partial,\partial}  \ket{g_{s,\partial}^{(n+1)}}\\
         \Hat{\mathcal{D}}_{\perp,\partial}  \ket{g_{s,\perp}^{(n+1)}}
    \end{bmatrix}, 
    \label{eq:NtensorEQ}
\end{equation}
where $\ket{g_s^{(n)}} = \overline{\ket{G_s^{(n)}}}$ and $\mathbf{\Hat{P}}_{n,\alpha} \ket{g_s^{(n')}} = \delta_{n,n'} \ket{g_{s,\alpha}^{(n)}}$. For clarity, all the operators in Eq.~\eqref{eq:NtensorEQ} are defined in $\mathcal{V}$ and can be directly obtained from Eq.~\eqref{Eq:TevGtket} and Eq.~\eqref{Eq:NG}. In particular, $\Hat{F}$, $\Hat{B}_{\alpha,\beta}$ and $\Hat{A}$ are block diagonal in the sense that they obey the following set of equalities written for $\Hat{M}$: $\Hat{M}_n = \mathbf{\Hat{P}}_{n}  \Hat{M} =  \Hat{M}  \mathbf{\Hat{P}}_{n} = \mathbf{\Hat{P}}_{n}  \Hat{M}  \mathbf{\Hat{P}}_{n}$, where $\Hat{M} \in \{\Hat{F},\Hat{A},\Hat{B}_{\alpha,\beta}\}$.  

The operator $\Hat{F}_n$ has already been specified in Eq.~\eqref{Eq:Dn}. $\Hat{B}_{\alpha,\beta}$ and $\Hat{A}$ are of the form 
\begin{align}
    &\left[\Hat{B}_n\right]_{\alpha,\beta} = i \sum_{a=1}^n \sum_{kl} \mathbf{\Hat{P}}_{n,\alpha}  \left( H_{k,l}   \ket{k}\bra{l}_{a,+} -  H_{k,l}^T \ket{k}\bra{l}_{a,-} \right)    \mathbf{\Hat{P}}_{n,\beta}  \;, \\
    &\Hat{A}_n = \sum_{a=1}^n  \mathbf{\Hat{P}}_{\partial} \left(\Gamma_{\text{L}}  \left( \ket{1}\bra{1}_{a,+}  + \ket{1}\bra{1}_{a,-}\right) + \Gamma_{\text{R}}^{(0)}  \left( \ket{L}\bra{L}_{a,+} + \ket{L}\bra{L}_{a,-}\right)   \right) \mathbf{\Hat{P}}_{\partial} \;. 
\label{Eq:ABdef}
\end{align}
The operators $\Hat{\mathcal{U}}_{\alpha,\beta}$ and $\Hat{\mathcal{D}}_{\alpha,\beta}$ connect states that belong to different sectors, they are just slightly off block diagonal: $\mathbf{\Hat{P}}_{n} \Hat{\mathcal{U}}_{\alpha,\beta}\mathbf{\Hat{P}}_{n'} = \delta_{n,n'+1} [\Hat{\mathcal{U}}_n]_{\alpha,\beta}$ and $\mathbf{\Hat{P}}_{n} \Hat{\mathcal{D}}_{\alpha,\beta} \mathbf{\Hat{P}}_{n'} = \delta_{n,n'-1} [\Hat{\mathcal{D}}_n]_{\alpha,\beta}$. With these definitions, we obtain
\begin{multline}
    \Hat{\mathcal{U}}_n = i \sum_{a=1}^n \sum_{k,l} \sum_{\mathbf{i_{n-1}},\mathbf{j_{n-1}}}  \left(\rho_{\text{L}} \left(H_{k,1} \delta_{l,1} - H_{1,l}  \delta_{k,1} \right)+ \rho_{\text{R}}^{(s)} \left(H_{k,L} \delta_{l,L} - H_{L,l}  \delta_{k,L}\right)\right) \ket{i_1...i_{a-1} k i_{a} ... i_{n-1};j_1...j_{a-1} l j_{a} ... j_{n-1}} \bra{\mathbf{i_{n-1}};\mathbf{j_{n-1}}}   ,  \\
    \Hat{\mathcal{D}}_n =  \Gamma_s  \sum_{a=1}^n  \sum_{\mathbf{i_n},\mathbf{j_n}}   \ket{\mathbf{i_n};\mathbf{j_n}} \bra{i_1...i_{a} L i_{a+1} ... i_n; j_1...j_{a-1} L j_{a} ...j_n}     .
\label{Eq:RLdef}
\end{multline}
It is implicitly assumed that if, for instance, $a=1$, the string $i_1...i_{a-1}$ is to be removed (and the same applies in analogous cases). 

The form of Eq.~\eqref{eq:NtensorEQ} suggests that we perform a time rescaling $t \to \gamma t$ and, accordingly, $\ket{g_{s,\perp}^{(n)}}\to \gamma^{-1} \ket{g_{s,\perp}^{(n)}}$, $\ket{g_{s,\partial}^{(n)}} \to \gamma^{-1} \ket{g_{s,\partial}^{(n)}}$ for all $n$, which yields

\begin{equation}
        \frac{d}{dt} \begin{bmatrix}
        \ket{g_{s,\parallel}^{(n)}}\\
        \ket{g_{s,\partial}^{(n)}}\\
        \ket{g_{s,\perp}^{(n)}}
    \end{bmatrix} = - \begin{bmatrix}
        0 & 0 &  \Hat{B}_{\parallel,\perp} \\
        0 &  \gamma \Hat{A} &  \gamma \Hat{B}_{\partial,\perp}\\
        \gamma^2 B_{\perp,\parallel} &  \gamma \Hat{B}_{\perp,\partial} &  \gamma^2 \Hat{F}
    \end{bmatrix} \begin{bmatrix}
        \ket{g_{s,\parallel}^{(n)}}\\
        \ket{g_{s,\partial}^{(n)}}\\
        \ket{g_{s,\perp}^{(n)}}
    \end{bmatrix} - \begin{bmatrix}
        0\\
        \gamma \Hat{\mathcal{U}}_{\partial,\perp}\ket{g_{s,\perp}^{(n-1)}}\\
        \gamma^2 \Hat{\mathcal{U}}_{\perp,\parallel} \ket{g_{s,\parallel}^{(n-1)}} + \gamma \Hat{\mathcal{U}}_{\perp,\partial} \ket{g_{s,\partial}^{(n-1)}} + \gamma \Hat{\mathcal{U}}_{\perp,\perp} \ket{g_{s,\perp}^{(n-1)}} 
    \end{bmatrix} +  \begin{bmatrix}
         \Hat{\mathcal{D}}_{\parallel,\partial}  \ket{g_{s,\partial}^{(n+1)}}\\
         \gamma \Hat{\mathcal{D}}_{\partial,\partial}  \ket{g_{s,\partial}^{(n+1)}}\\
         \gamma \Hat{\mathcal{D}}_{\perp,\partial}  \ket{g_{s,\perp}^{(n+1)}}
    \end{bmatrix}.
    \label{eq:NtensorEQresc}
\end{equation}

The presence of a factor of $\gamma$ in front of every term in the evolution equations for $\ket{g_{s,\partial}^{(n)}}$ and $\ket{g_{s,\perp}^{(n)}}$ implies that, at this time scale, these are fast modes and thus they quickly relax to their steady state solution, given by

\begin{equation}
    \Hat{F} \ket{g_{s,\perp}^{(n)}} = - \Hat{B}_{\perp,\parallel} \ket{g_{s,\parallel}^{(n)}} - \Hat{\mathcal{U}}_{\perp,\parallel} \ket{g_{s,\parallel}^{(n-1)}} \hspace{1mm} \text{ and }
    \label{Eq:Largegamma1}
\end{equation}
\begin{equation}
    \Hat{A} \ket{g_{s,\partial}^{(n)}} = \Hat{B}_{\partial,\perp} \Hat{F}^{-1}  \Hat{B}_{\perp,\parallel} \ket{g_{s,\parallel}^{(n)}} + \left(\Hat{B}_{\partial,\perp}  \Hat{\mathcal{U}}_{\perp,\parallel} + \Hat{\mathcal{U}}_{\partial,\perp}] \Hat{B}_{\perp,\partial} \right) \ket{g_{s,\parallel}^{(n-1)}} + \Hat{\mathcal{U}}_{\partial,\perp} \Hat{\mathcal{U}}_{\perp,\parallel} \ket{g_{s,\parallel}^{(n-2)}} + \Hat{\mathcal{D}}_{\partial,\partial}  \ket{g_{s,\partial}^{(n+1)}},
    \label{Eq:Largegamma2_0}
\end{equation}
where we have used the first equation to simplify the latter. 

We now note that in Eq.~\eqref{Eq:Largegamma2_0}, the terms that belong to the $\parallel$-subspace couple to those in $\partial$ that contain at most one pair of indices in the boundary. As a consequence, denoting the subspace of $\partial$ generated by vectors with two or more pair of indices on the boundary by $\partial \partial$, we obtain that
\begin{equation}
    \mathbf{\Hat{P}}_{\partial \partial} \hat{A}  \ket{g_{s,\partial}^{(n)}} = \mathbf{\Hat{P}}_{\partial \partial} \Hat{\mathcal{D}}_{\partial,\partial}  \ket{g_{s,\partial}^{(n+1)}} \Longleftrightarrow   \hat{A}  \ket{g_{s,\partial \partial}^{(n)}} = \mathbf{\Hat{P}}_{\partial \partial} \Hat{\mathcal{D}}_{\partial,\partial}  \ket{g_{s,\partial \partial}^{(n+1)}},
\end{equation}
since $[\mathbf{\Hat{P}}_{\partial \partial}, \hat{A}]=0$ and $\mathbf{\Hat{P}}_{\partial} \Hat{\mathcal{D}} \mathbf{\Hat{P}}_{\partial} = \mathbf{\Hat{P}}_{\partial} \Hat{\mathcal{D}}\mathbf{\Hat{P}}_{\partial \partial}$. As a consequence, these equations are consistently solved by $\ket{g_{s,\partial \partial}^{(n)}}=0$ and thus $\Hat{\mathcal{D}}_{\partial,\partial} \ket{g_{s,\partial}^{(n+1)}} = \Hat{\mathcal{D}}_{\partial,\partial} \ket{g_{s,\partial \partial}^{(n+1)}} = 0$. This simplifies Eq.~\eqref{Eq:Largegamma2_0} to
\begin{equation}
    \Hat{A} \ket{g_{s,\partial}^{(n)}} = \Hat{B}_{\partial,\perp} \Hat{F}^{-1}  \Hat{B}_{\perp,\parallel} \ket{g_{s,\parallel}^{(n)}} + \left(\Hat{B}_{\partial,\perp}  \Hat{\mathcal{U}}_{\perp,\parallel} + \Hat{\mathcal{U}}_{\partial,\perp} \Hat{B}_{\perp,\parallel} \right) \ket{g_{s,\parallel}^{(n-1)}} + \Hat{\mathcal{U}}_{\partial,\perp} \Hat{\mathcal{U}}_{\perp,\parallel} \ket{g_{s,\parallel}^{(n-2)}}.
    \label{Eq:Largegamma2}
\end{equation}
Plugging Eq.~\eqref{Eq:Largegamma1} and Eq.~\eqref{Eq:Largegamma2} in the top equation of Eq.~\eqref{eq:NtensorEQresc}, we obtain
\begin{multline}
    \frac{d}{dt} \ket{g_{s,\parallel}^{(n)}} = \Hat{B}_{\parallel,\perp} \Hat{F}^{-1} \Hat{B}_{\perp,\parallel} \ket{g_{s,\parallel}^{(n)}} + \Hat{B}_{\parallel,\perp} \Hat{\mathcal{U}}_{\perp,\parallel} \ket{g_{s,\parallel}^{(n-1)}} + \\+ \Hat{\mathcal{D}}_{\parallel,\partial} \Hat{A}^{-1} \left(\Hat{B}_{\partial,\perp} \Hat{F}^{-1}  \Hat{B}_{\perp,\parallel} \ket{g_{s,\parallel}^{(n+1)}} + \left(\Hat{B}_{\partial,\perp}  \Hat{\mathcal{U}}_{\perp,\parallel} + \Hat{\mathcal{U}}_{\partial,\perp} \Hat{B}_{\perp,\parallel} \right) \ket{g_{s,\parallel}^{(n)}} + \Hat{\mathcal{U}}_{\partial,\perp} \Hat{\mathcal{U}}_{\perp,\parallel} \ket{g_{s,\parallel}^{(n-1)}}\right) 
\label{Eq:BKEgeneral}
\end{multline}

We now make a few remarks that will substantially simplify the previous equation. First, since the operators $\Hat{B}_{\perp,\parallel}$ can only change one of the indices in $\ket{\mathbf{i_n},\sigma(\mathbf{i_n})}$, $\Hat{F} \Hat{B}_{\perp,\parallel} \ket{\mathbf{i_n},\sigma(\mathbf{i_n})} = \ket{\mathbf{i_n},\sigma(\mathbf{i_n})}$ and thus $\Hat{F}$ can be removed from Eq.~\eqref{Eq:BKEgeneral} without affecting the results. By the same reason (but applied now to $B_{\partial,\perp}$, $\Hat{\mathcal{U}}_{\partial,\perp}$ and $\Hat{\mathcal{U}}_{\perp,\parallel} $) and the fact that $\Hat{\mathcal{D}}_{\parallel,\partial}$ is only non-vanishing when it acts on a vector containing one left and one right index equal to $L$, $\Hat{A}$ can also be safely replaced by $2\Gamma_{\text{R}}^{(0)}= \Gamma_{\text{R},1} + \Gamma_{\text{R},-1}$ in the previous equation: $\Hat{P}^{(L,a)}_L \Hat{P}^{(R,a-1)}_L \Hat{A} \Hat{M}_{\partial,\perp} \Hat{M}_{\perp,\parallel} = \left( \Gamma_{\text{R},1} + \Gamma_{\text{R},-1}\right) \Hat{P}^{(L,a)}_L \Hat{P}^{(R,a-1)}_L \Hat{M}_{\partial,\perp} \Hat{M}_{\perp,\parallel}$, where $\Hat{M}$ can be any of the aforementioned operators. 

In fact, using these properties and the definitions in Eq.~\eqref{Eq:ABdef}, one can show that
\begin{multline}
    \Hat{B}_{\parallel,\perp} \Hat{F}^{-1} \Hat{B}_{\perp,\parallel} = 2 \sum_{a,a'=1}^n \sum_{k=2}^{L-1} \sum_{\mathbf{i_n},\sigma} |H_{i_a,k}|^2    \delta_{i_a,i_{\sigma(a')}} \ket{i_1...i_{a-1} k i_{a+1}...i_n;i_{\sigma(1)}...i_{\sigma(a'-1)} k i_{\sigma(a'+1)}...i_{\sigma(n)}}\bra{\mathbf{i_n};\mathbf{j_n}} -\\
    - 2 \sum_{a=1}^n \sum_{k=1}^L \sum_{\mathbf{i_n},\sigma} |H_{i_a,k}|^2 \ket{\mathbf{i_n};\sigma(\mathbf{i_n})}\bra{\mathbf{i_n};\sigma(\mathbf{i_n})}
\end{multline}
Furthermore, by direct application of the definitions in Eq.~\eqref{Eq:RLdef} and Eq.~\eqref{Eq:ABdef}, one can show that
\begin{equation}
\begin{split}
     \mathbf{\Hat{P}}_{n} \Hat{B}_{\parallel,\perp} \Hat{\mathcal{U}}_{\perp,\parallel} \mathbf{\Hat{P}}_{n-1}= 2  \sum_{a=1}^n \sum_{k=2}^{L-1} \sum_{\mathbf{i_{n-1}},\sigma}  \left( \rho_{\text{L}}  | H_{1,k}|^2   + \rho_{\text{R}}^{(s)} |H_{k,L}|^2   \right)  \ket{i_1...i_{a-1} k i_{a} ... i_{n-1};i_{\sigma(1)} ...i_{\sigma(a-1)} k i_{\sigma(a)} ...i_{\sigma(n-1)}} \bra{\mathbf{i_{n-1}};\sigma(\mathbf{i_{n-1}})} ,\\
      \mathbf{\Hat{P}}_{n} \Hat{\mathcal{D}}_{\parallel,\partial} \Hat{A}^{-1} \Hat{\mathcal{U}}_{\partial,\perp} \Hat{\mathcal{U}}_{\perp,\parallel} \mathbf{\Hat{P}}_{n-1}= 2  \Tilde{\Gamma}_s  \left(\rho_{\text{R}}^{(s)}\right)^2 \sum_{a=1}^n \sum_{k=2}^{L-1} \sum_{\mathbf{i_{n-1}},\sigma}  |H_{k,L}|^2   \ket{i_1...i_{a-1} k i_{a} ... i_{n-1}; i_{\sigma(1)}...i_{\sigma(a-1)} k i_{\sigma(a)} ...i_{\sigma(n-1)}} \bra{\mathbf{i_{n-1}};\sigma(\mathbf{i_{n-1}})} \\
       \Longrightarrow  \Hat{B}_{\parallel,\perp} \Hat{\mathcal{U}}_{\perp,\parallel}  + \Hat{\mathcal{D}}_{\parallel,\partial} \Hat{A}^{-1} \Hat{\mathcal{U}}_{\partial,\perp} \Hat{\mathcal{U}}_{\perp,\parallel}  = \Hat{\mathcal{R}}_{\parallel,\parallel}, \hspace{1mm} \text{ for } \hspace{1mm} \Hat{\mathcal{R}}_{\parallel,\parallel} = \sum_n [\Hat{\mathcal{R}}_n]_{\parallel,\parallel} \text{ and }
\end{split}
\end{equation}
\begin{equation}
    [\Hat{\mathcal{R}}_n]_{\parallel,\parallel} = 2  \sum_{a=1}^n \sum_{k=2}^{L-1} \sum_{\mathbf{i_{n-1}},\sigma}  \left( \rho_{\text{L}} |H_{k,1}|^2   + e^{-s} \rho_{\text{R}}^{(0)} |H_{k,L}|^2  \right) \ket{i_1...i_{a-1} k i_{a} ... i_{n-1};i_{\sigma(1)}. ...i_{\sigma(a-1)} k i_{\sigma(a)} ... i_{\sigma(n-1)}} \bra{\mathbf{i_{n-1}};\sigma(\mathbf{i_{n-1}})} .
\end{equation}
Note that along these steps we used the definition $ \Tilde{\Gamma}_s = \sum_{\epsilon \in \{-1,1\}} \epsilon \frac{\Gamma_{\text{R},-\epsilon}}{\Gamma_{\text{R},-1}+\Gamma_{\text{R},1}} \left(e^{\epsilon s} - 1\right)$  and the identity $\rho_{\text{R}}^{(s)} \left( 1 +  \Tilde{\Gamma}_s  \rho_{\text{R}}^{(s)}\right)= e^{-s} \rho_{\text{R}}^{(0)}$.

Last, by a similar reasoning, one also obtains that
\begin{multline}
    \Hat{\mathcal{D}}_{\parallel, \partial} \Hat{A}^{-1} \left(\Hat{B}_{\partial,\perp} \Hat{\mathcal{U}}_{\perp,\parallel} + \Hat{\mathcal{U}}_{\partial,\perp} \Hat{B}_{\perp,\parallel} \right)= 4 \left(e^{-s}-1\right) \rho_{\text{R}}^{(0)} \sum_{a=1}^n \sum_{k} \sum_{\mathbf{i_n},\sigma}  |H_{k,L}|^2 \delta_{i_a,k}\ket{\mathbf{i_n};\sigma(\mathbf{i_n})} \bra{\mathbf{i_n};\sigma(\mathbf{i_n})} , \hspace{2mm} \Tilde{\Gamma}_s\rho_{\text{R}}^{(s)} =\left(e^{-s}-1\right) \rho_{\text{R}}^{(0)},\\
    \Hat{\mathcal{D}}_{\parallel,\partial} \Hat{A}^{-1} \Hat{B}_{\partial,\perp} \Hat{F}^{-1}  \Hat{B}_{\perp,\parallel}= 2 \Tilde{\Gamma}_s \sum_{a=1}^n \sum_{k} \sum_{\mathbf{i_n},\sigma}  |H_{k,L}|^2 \ket{\mathbf{i_n};\sigma(\mathbf{i_n})} \bra{i_1 ...i_a k i_{a+1} ... i_n; i_{\sigma(1)} ... i_{\sigma(a-1)} k i_{\sigma(a)} i_{\sigma(n)}}.
\end{multline}

Let us denote by $\mathcal{G}'$ the graph that is obtained from $\mathcal{G}$ by removing the vertices $1$ and $L$ (and the associated edges). Then, by direct comparison with Eq.~\eqref{Eq:BKEgeneral} and all the subsequent equations specifying each of its terms, one concludes that the averaged time evolution of $\ket{G_s^{(n)}} $ (when generated by Eq.~\eqref{Eq:TevGket}, restricted to vertices $\{2...L-1\}$ and after the time rescaling by $\gamma$) is the same as the one induced by the following equation defined on $\mathcal{G}'$:
\begin{multline}
      d\ket{G_{s}^{(1)}} = -i \sum_{k,l}\left(H_{k,l} d\xi_{k,l}  \ket{k}\bra{l}_+ - H_{k,l}^T  d\xi_{k,l}^* \ket{k}\bra{l}_- \right)\ket{G_{s}^{(1)}} + \sum_{k,l}  |H_{k,l}|^2 \left( 2 \ket{k}\bra{l}_+ \ket{k}\bra{l}_- -  \left(\ket{l}\bra{l}_+ + \ket{l}\bra{l}_- \right)  \right) \ket{G_{s}^{(1)}}  dt  +\\ + 2 \sum_k \left( \rho_{\text{L}} |H_{k,1}|^2  + e^{-s} \rho_{\text{R}} |H_{k,L}|^2 \right) \ket{k;k}dt  -\sum_k \left(  \Tilde{\Gamma}_{k}^{(s)}  \Hat{Q}_k  \ket{G_{s}^{(1)}}+ 2 \Tilde{\Gamma}_s |H_{k,L}|^2 \sum_{l,m} \ket{l;m}\braket{l,k;k,m|G_{s}^{(2)}} \right) dt,
\label{Eq:QTevGket}
\end{multline}
where, for brevity, we used $\rho_{\text{R}} = \rho_{\text{R}}^{(0)}$, $\Tilde{\Gamma}_{k}^{(s)} = |H_{k,1}|^2 + \left( 2 \left(e^s - 1\right) \rho_{\text{R}} +1\right) |H_{k,L}|^2  $, $\overline{d\xi_{k,l} d\xi_{k',l'}^*} = 2 \delta_{k,k'}\delta_{l,l'} dt$ and $d\xi_{k,l} = d\xi_{l,k}^*$. We remark that the phases of the matrix elements $H_{k,l}$ of the original Hamiltonian become irrelevant in the limit $\gamma \to \infty$ as they can be absorbed in a redefinition of $d\xi_{k,l}$. 

\subsection{Equivalence between models}

 At this point, we observe that Eq.~\eqref{Eq:QTevGket} represents the time evolution of a quantum system on $\mathcal{G}'$ modelled by a stochastic Hamiltonian $\Hat{H} = \sum_{k,l} |H_{k,l}| dW_{k,l} c^{\dagger}_k c_l$ and interacting at possibly each vertex $k$ with four different Markovian reservoirs, two providing particles at rates $\alpha = 2 |H_{k,1}|^2 \rho_{\text{L}}$ and $\delta = 2 |H_{k,L}|^2 \rho_{\text{R}}$ and two removing them at rates $\gamma = 2 |H_{k,1}|^2 \left(1 - \rho_{\text{L}}\right)$ and $\beta = 2 |H_{k,L}|^2 \left(1-\rho_{\text{R}}\right)$.  

The generalization of this result to a quantum system on the graph $\mathcal{G}$ but connected to $M$ pairs or reservoirs at $M$ vertices $\{b_1,...,b_M\}$, such that on $M'<M$ of which the net flow of particles is monitored, is straightforward: the counting field is inserted in the time evolution equation as described in the introduction and the jumping rates to each pair of reservoirs from the vertex $k$ are given by $2 |H_{k,b_m}|^2 \rho_m$ and $2 |H_{k,b_m}|^2 \left(1-\rho_m\right)$. It is also easy to see that our conclusions find a natural extension to the case where one monitors the activity instead, which amounts to counting the total number of jumps to and from each pair of reservoirs. Thus, the only difference is that a factor of $e^s$ is assigned to both reservoirs and so every step of our derivation can be repeated with all $e^{-s}$ replaced by $e^s$.

We have proved that the average of the $n$-tensor product of $G_s$ is the same at all times $t$ for both stochastic processes considered (Eq.~\eqref{Eq:TevGket} in the limit $\gamma \to \infty$ (model $\mathbf{1}$) and Eq.~\eqref{Eq:QTevGket} (model $\mathbf{2}$)), given that their initial value is the same. This implies that, for any initial probability distribution $d\mu_0(G)$ over the space of correlation matrices, the moments of the distribution $d\mu_{\mathbf{k},t}(G)$, obtained by solving the Fokker-Planck equation of model $\mathbf{k}$ up to time $t$, are the same in both models. For any reasonably well-behaved probability distribution this also implies that $d\mu_{\mathbf{1},t}(G)=d\mu_{\mathbf{2},t}(G)$. In this appendix, we further showed that the current (and activity) statistics also matches in both cases.

In this article, we are, however, mostly interested in the specific case of the model we introduced (i.e \qnoise), i.e, $H_{i,j}= \left(\delta_{i+1,j} + \delta_{i,j+1}\right)$. The associated model in the limit $\gamma \to \infty$ under the correspondence established in Eq.~\eqref{Eq:QTevGket} is known in the literature as QSSEP (Quantum Symmetric Simple Exclusive Process).  

\section{Proof of Gauge Trick}\label{sec:proofGT}
In this section, we provide a rigorous derivation of the validity of the Gauge Trick. As described in the main text, the cumulant generating function is given by $\lambda(s) = \lim_{t \to \infty} t^{-1} \log{\left(\Tr{\left(\rho_s(t)\right)}\right)}$, where $\rho_s(t)$ is the time-evolved state under the tilted Liouvillian $\mathcal{L}_s$, starting from an arbitrary initial condition $\rho_s(0)$:
$\rho_s(t) = \mathcal{T}e^{\int_0^t dt' \mathcal{L}_s(t')}\left(\rho_s(0)\right)$, as defined in Eq.~(\ref{Eq:TimeEvGenCGF}). The essence of the Gauge Trick lies in the observation that $\lambda(s)$ can be equivalently obtained by evolving a (possibly different) initial state under a modified tilted Liouvillian. Indeed, we can consider the evolution $\tilde{\rho}_s(t) = \mathcal{T}e^{\int_0^t dt' \mathcal{L}'_s(t')}\left(\tilde{\rho}_s(0)\right)$, where the transformed Liouvillian is defined as $\mathcal{L}'_s(\cdot; t) = \Hat{U} \mathcal{L}_s\left(\Hat{U}^{-1} \cdot \Hat{U}^{-1}; t\right) \Hat{U}$ and $\Hat{U}$ is any operator of the form 
\begin{equation}\label{eq:gaugetransf}
    \Hat{U} = e^{\frac{\tilde{s}}{2} \sum{j=1}^{L} F_j c^{\dagger}_j c_j}, \text{ with   } \tilde{s} = \frac{s}{L+1}.
\end{equation}
This leads to the central statement of the Gauge Trick:
\begin{equation}\label{eq:GaugeTrick}
    \lambda(s) \coloneqq \lim_{t \to \infty} t^{-1} \log{\left(\Tr{\left(\rho_s(t)\right)}\right)} = \lim_{t \to \infty} t^{-1} \log{\left(\Tr{\left(\Tilde{\rho}_s(t)\right)}\right)}.
\end{equation}
To establish this equality, we consider the initial condition of the transformed system as $\tilde{\rho}_s(0) = \Hat{U} \rho_s(0) \Hat{U}$. Then,
\begin{equation}\label{eq:rhoT=rho}
\frac{\log \left(\Tr\left( \Tilde{\rho}_s(t)\right)\right)}{t} = \frac{\log \left(\Tr\left( \mathcal{T}e^{\int_0^t dt' \mathcal{L}'_s(t')}\left(\Hat{U} \rho_s(0) \Hat{U} \right) \right)\right)}{t} = \frac{\log \left(\Tr\left( \Hat{U} \left(\mathcal{T}e^{\int_0^t dt' \mathcal{L}_s(t')}\left( \rho_s(0) \right) \right)\Hat{U}\right)\right)}{t} = \frac{\log \left(\Tr\left( \Hat{U} \rho_s(t) \Hat{U}\right)\right)}{t},
\end{equation}
where we have used the definition of $\mathcal{L}'_s$. Here, $\mathcal{T}$ denotes time-ordering.

We now argue that the density matrix $\rho_s(t)$ remains positive for all $t$. In general, the Liouvillian evolution preserves positivity, and this property continues to hold even after the introduction of a counting field in the tilted Liouvillian $\mathcal{L}_s$. To make this explicit, consider the part of the Liouvillian associated with the coupling to the reservoirs and modified by the counting field. By introducing an auxiliary Itô increment $d\xi$, these contributions can be recast as:
\begin{equation}
    \left(e^s \Hat{L}_\alpha \rho_s \Hat{L}_\alpha^{\dagger} - \frac{1}{2} \{\Hat{L}_\alpha^\dagger \Hat{L}_\alpha, \rho_s\}\right)dt =  \mathbb{E}_{\xi}\left[\left(\mathbb{1} + e^{s/2} d\xi \Hat{L}_\alpha\right) \left(\mathbb{1} - \frac{1}{2} \Hat{L}_\alpha^\dagger \Hat{L}_\alpha dt\right) \rho_s \left(\mathbb{1} - \frac{1}{2} \Hat{L}_\alpha^\dagger \Hat{L}_\alpha dt\right) \left(\mathbb{1} + e^{s/2} d\xi \Hat{L}_\alpha^\dagger\right)\right],
\end{equation}
where $\Hat{L}_\alpha$ denotes a jump operator representing a specific reservoir. The expression inside the expectation value $\mathbb{E}_{\xi}$ is manifestly positive, and therefore, the average also preserves positivity. This confirms that the tilted Liouvillian $\mathcal{L}_s$ indeed maintains the positivity of $\rho_s(t)$ throughout the evolution.

Now, consider the occupation number basis states $\ket{\mathbf{n}} \coloneqq \ket{n_1, ..., n_L}$. The positivity of $\rho_s(t)$ implies $\bra{\mathbf{n}} \rho_s(t) \ket{\mathbf{n}} \geq 0$, from which we deduce the following bounds:
\begin{equation}
\frac{\mathbf{F}_{\text{min}}+\log\left(\Tr\left( \rho_s(t) \right)\right)}{t} \leq \frac{\log \Tr\left(\sum_{\mathbf{n}} e^{\tilde{s} \sum_j F_j n_j} \bra{\mathbf{n}} \rho_s(t) \ket{\mathbf{n}}\right)}{t} = \frac{\log\left( \Tr\left( \Hat{U} \rho_s(t) \Hat{U}\right) \right)}{t} \leq \frac{ \mathbf{F}_{\text{max}}+\log\left( \Tr\left( \rho_s(t) \right) \right)}{t},
\end{equation}
where we defined $\mathbf{F}_{\text{min}} = \min_{\mathbf{n}} \left( \tilde{s} \sum_j F_j n_j \right)$ and $\mathbf{F}_{\text{max}} = \max_{\mathbf{n}} \left( \tilde{s} \sum_j F_j n_j \right)$.
Taking the long-time limit and using Eq.(\ref{eq:rhoT=rho}), we obtain
\begin{equation}
\lambda(s) = \lim_{t \to \infty} \frac{1}{t} \log \left(\Tr\left( \rho_s(t)\right)\right) = \lim_{t \to \infty} \frac{1}{t} \log\left( \Tr\left( \Hat{U} \rho_s(t) \Hat{U}\right) \right) = \lim_{t \to \infty} \frac{1}{t} \log \left(\Tr\left( \Tilde{\rho}_s(t)\right)\right),
\end{equation}
which finally completes the proof of the Gauge Trick's validity.

\section{Proof that the CGF of the current is self-averaging}\label{sec:proofSA}
In this appendix, we prove that, at finite system size $L$, the cumulant generating function of the current is self-averaging, i.e that the limits $\lim_{t \to \infty} \frac{1}{t} \overline{\log \left(\Tr\left( \rho_s(t)\right)\right)}$ and $\lim_{t \to \infty} \frac{1}{t} \log \left(\Tr\left( \rho_s(t)\right)\right)$ are well-defined almost everywhere and that 
\begin{equation}
    \lim_{t \to \infty} \frac{1}{t} \overline{\log \left(\Tr\left( \rho_s(t)\right)\right)} = \lim_{t \to \infty} \frac{1}{t} \log \left(\Tr\left( \rho_s(t)\right)\right) \; a.s.
\end{equation}

Intuitively, as discussed in the main text, the presence of noise introduces a finite correlation time into the system dynamics. Consequently, contributions to  $\log \left(\Tr\left( \rho_s(t)\right)\right)$
from sufficiently separated time intervals are effectively independent. By the law of large numbers, one can then naively conclude that $\lim_{t \to \infty} \log \left(\Tr\left( \rho_s(t)\right)\right)$ converges to the same value almost everywhere, thereby coinciding with the average over noise realizations.

To prove this more rigurously, we start by considering a vectorized (tilted) density matrix $\ket{\rho_s(t)} \in \mathbb{C}^{4^L}$, so that now $\mathcal{T}e^{\int_0^t d\mathcal{L}_s(t')} \in \text{GL}\left(4^L,\mathbb{C}\right)$. At this point, we define the stochastic process
\begin{equation}\label{eq:ourSP}
   \{X_n\}_{n \in \mathbb{N}_0}, \;\text{ with } X_n = \mathcal{T}e^{\int_n^{n+1} d\mathcal{L}_s(t')}.
\end{equation}
Since the charge-coupled noise in the model we consider is delta-correlated in time, the random matrices $X_n$ are independent and identically distributed on $\text{GL}\left(4^L,\mathbb{C}\right)$.

Furstenberg-Kesten theorem, or more generally Oseledets' Multiplicative Ergodic Theorem~\cite{Furstenberg-Kesten,oseledets1968multiplicative}, state that, provided the $X_n$ are measurable, the stochastic process $\{X_n\}_n$ is metrically transitive (or ergodic) and $\overline{\log^+\left\lVert X_n\right\rVert}<\infty$, for some operator norm $\lVert \rVert$ induced by a norm on $\mathbb{C}^{4^L}$ (that we also denote by $\lVert \rVert$), there exists a splitting
\begin{equation}
    \mathbb{C}^{4^L} = E_1 \oplus ... \oplus E_k,
\end{equation}
such that
\begin{equation}\label{eq:Oseledets}
    \lim_{n \to \infty} \frac{1}{n} \log \left(\lVert Y_n v_k\rVert\right) = \lambda_k \;a.s, \text{ for all non-zero } v_k \in E_k \text{ and } Y_n=X_n X_{n-1} ... X_0.
\end{equation}
Additionally,
\begin{equation}
      \lim_{n \to \infty} \frac{1}{n} \overline{\log \left(\lVert Y_n\rVert\right)} = \lim_{n \to \infty} \frac{1}{n} \log \left(\lVert Y_n\rVert\right) = \lambda_M \;a.s, \; \text{ for } \lambda_M=\max\left(\{\lambda_k\}_k\right).
\end{equation}
Note that Eq.~(\ref{eq:Oseledets}) also implies that, for an initial random vector $v \in \mathbb{C}^{4^L}$, 
\begin{equation}
    \lim_{n \to \infty} \frac{1}{n} \log \left(\lVert Y_n v\rVert\right) = \lambda_M \;a.s.
\end{equation}

Note also that since all norms in a finite dimensional vector space are equivalent, all the previous results hold for any induced operator norm.

In order to show that the stochastic process we defined in Eq.~(\ref{eq:ourSP}) satisfies the assumptions of Oseledet's theorem, we note that, if we define $Z_n(t)$ to be the solution of the following linear SDE,
\begin{equation}\label{eq:SDEZn}
    dZ_n(t) = d\mathcal{M}_s(n+t) \; Z_n(t), \; Z_n(0) = \mathbb{1},\;d\mathcal{M}_s(n+t)=d\mathcal{L}_s(n+t)+\frac{1}{2} \left(d\mathcal{L}_s(n+t)\right)^2,
\end{equation}
then $X_n =Z_n(1)$. Note that for this to be true, we needed to include $\left(d\mathcal{L}_s\right)^2$ in the expression for the generator $d\mathcal{M}_s$, because of Ito calculus, $dW^2=dt$.

This implies immediately that the $X_n$ are measurable. Additionally, from  Eq.~(\ref{eq:SDEZn}), one can derive the inequality
\begin{equation}
    d \log \left(\lVert Z_n(t)\rVert\right) \leq \lVert d\mathcal{M}_s(n+t)\rVert \Longrightarrow  d \log^+ \left(\lVert Z_n(t)\rVert\right) \leq \lVert d\mathcal{M}_s(n+t)\rVert.
\end{equation}
It is now straightforward to conclude that
\begin{equation}
    \overline{\log^+ \left(\lVert X_n\rVert\right)} = \overline{\log^+ \left(\lVert Z_n(1)\rVert\right)} \leq \int_0^1 \overline{ \lVert d\mathcal{M}_s(n+t)\rVert} < \infty.
\end{equation}
Finally, since the random matrices $X_n$ are i.i.d, they form trivially a metrically transitive stochastic process and thus we have established the validity of all assumptions to Oseledet's theorem for the stochastic process in Eq.~(\ref{eq:ourSP}). 

Noting that since, as established in appendix~\ref{sec:proofGT}, the (tilted) density matrix $\rho_s(t)$ remains positive at all times, we obtain that
\begin{equation}\label{eq:bounds}
     t^{-1}\log \left(\lVert \rho_s(t) \rVert_2 \right) \leq t^{-1}\log \left(\Tr \left(\rho_s(t) \right) \right) \leq t^{-1}\log \left( 2^L \lVert \rho_s(t) \rVert_2 \right).
\end{equation}
where $\lVert \cdot \rVert_2$ denotes the operator norm acting on the space of $2^L \times 2^L$ complex matrices, which is induced by the $\ell^2$ norm on the corresponding $2^L$-dimensional complex vector space. After vectorization of the density matrix, this norm naturally induces an operator norm on the space of $4^L \times 4^L$ complex matrices, which can then be readily employed in Oseledets’ theorem applied to the sequence Eq.~(\ref{eq:ourSP}) to establish that
\begin{equation}\label{eq:oseledetapplied}
    \lim_{n \to \infty} n^{-1}\log \left(\lVert \rho_s(n) \rVert_2 \right) =  \lim_{n \to \infty} n^{-1}\log \left(2^L \lVert \rho_s(n) \rVert_2 \right) = \lambda_M \; a.s.
\end{equation}
Applying Eq.~(\ref{eq:oseledetapplied}) to the bounds derived in Eq.~(\ref{eq:bounds}), we finally obtain that
\begin{equation}
    \lim_{t \to \infty} t^{-1}\log \left(\Tr \left(\rho_s(t) \right) \right) = \lambda_M \;a.s.
\end{equation}
This concludes the proof that the CGF of the current is self-averaging.

\section{Proof of gaussianity (in the presence of counting field)}\label{sec:ProofGaussianity}
Due to the structure of the time evolution equation Eq.~(\ref{Eq:TimeEvGen}), it is convenient to recast the density matrix as a vector in an enlarged Hilbert space—a procedure known as vectorization. More explicitly, considering the occupation number basis states $\ket{\mathbf{n}} \coloneqq \ket{n_1, ..., n_L}$, we perform the mapping
\begin{equation}
    \ket{\mathbf{n}} \bra{\mathbf{n}'}  \to  \ket{\mathbf{n}} \otimes   \bra{\mathbf{n}'}^T, \; \; \; \hat{O}_1\ket{\mathbf{n}} \bra{\mathbf{n}'} \hat{O}^{\dagger}_2  \to
     \left(\hat{O}_1 \otimes \hat{O}_2^{\dagger^T} \right) \ket{\mathbf{n}} \otimes   \bra{\mathbf{n}'}^T,
\end{equation}
for arbitrary operators $\hat{O}_1$ and $\hat{O}_2$. In the vectorized space, we define the following vector of creation and annihilation operators:
\begin{equation}\label{vect_cao_bar}
 \bar{a} = \Big[
 c_{1}\otimes e^{i\pi\hat{N}^T},\dots,
 c_{L}\otimes e^{i\pi\hat{N}^T},
 \mathbb{1}\otimes c_{1}^{\dagger^{T}}e^{i\pi\hat{N}^T},\dots,
 \mathbb{1}\otimes c_{L}^{\dagger^{T}}e^{i\pi\hat{N}^T}, 
  c_{1}^{\dagger}\otimes e^{i\pi\hat{N}^T},\dots,
 c_{L}^{\dagger}\otimes e^{i\pi\hat{N}^T},
 \mathbb{1}\otimes e^{i\pi\hat{N}^T}c_{1}^{T},\dots,
 \mathbb{1}\otimes e^{i\pi\hat{N}^T}c_{L}^{T}\Big].
\end{equation}

From this definition, $\bar{a}$ is related to $\bar{a}^{\dagger}$ via $\bar{a}^{\dagger}_i = \sum_j \bar{S}_{i,j} \bar{a}_j$, with $\bar{S}_{i,j} = \delta_{i,j+2L} + \delta_{i+2L,j}$—a relation we refer to as particle-hole symmetry. As a consequence, the canonical anti-commutation relations $\{\bar{a}_i,\bar{a}^{\dagger}_j\}=\delta_{i,j}$ imply $\{\bar{a}^{\dagger}_i,\bar{a}^{\dagger}_j\}=\bar{S}_{i,j}$ and $\{\bar{a}_i,\bar{a}_j\}=\bar{S}_{i,j}$.

Before proceeding, we establish some properties of quadratic forms—operators acting on the vectorized space of the form $\hat{A} = \frac{1}{2} \sum_{i,j} \bar{a}^{\dagger}_i A_{i,j} \bar{a}_j$. Each quadratic form $\hat{A}$ is uniquely specified by a $4L \times 4L$ matrix $A$. However, this correspondence is not bijective: due to particle-hole symmetry, multiple matrices $A$ may represent the same quadratic form $\hat{A}$. Nonetheless, imposing the condition $\bar{S} A^T \bar{S} = -A$ ensures that the mapping becomes injective. Henceforth, we assume that any matrix $A$ appearing in a quadratic form satisfies this condition. 

Considering the Lie algebra formed by all complex matrices satisfying this constraint (with the Lie bracket given by the matrix commutator), one finds that it is isomorphic to the Lie algebra of quadratic forms (with Lie bracket given by the operator commutator), since $[\frac{1}{2} \sum_{i,j} \bar{a}^{\dagger}_i A_{i,j} \bar{a}_j, \frac{1}{2} \sum_{i,j} \bar{a}^{\dagger}_i B_{i,j} \bar{a}_j] = \frac{1}{2}  \sum_{i,j} \bar{a}^{\dagger}_i [A,B]_{i,j} \bar{a}_j$. Via the Baker–Campbell–Hausdorff formula, this implies that
\begin{equation}
    e^{\frac{1}{2} \sum_{i,j} \bar{a}^{\dagger}_i A_{i,j} \bar{a}_j} e^{\frac{1}{2} \sum_{i,j} \bar{a}^{\dagger}_i B_{i,j} \bar{a}_j} = e^{\frac{1}{2} \sum_{i,j} \bar{a}^{\dagger}_i C_{i,j} \bar{a}_j}, \text{ for  } e^C = e^A e^B.
\end{equation}      
Note that, in exponential form, particle-hole symmetry implies $\bar{S} e^{A^T} \bar{S} = e^{-A}$, a property clearly satisfied by $C$ as defined above. 

An additional identity that will be important later is
\begin{equation}\label{eq:identityqduadlinear}
    e^{-\frac{1}{2} \sum_{i,j} \bar{a}^{\dagger}_i A_{i,j} \bar{a}_j} \bar{a}_k e^{\frac{1}{2} \sum_{i,j} \bar{a}^{\dagger}_i A_{i,j} \bar{a}_j} = \sum_j \left(e^{A}\right)_{k,j} \bar{a}_j.
\end{equation}

Having established these properties of quadratic forms, we note that, using the definition in Eq.~(\ref{vect_cao_bar}) and discarding normalization factors, the time evolution generator—even in the presence of a counting field (see Eq.~(\ref{Eq:TimeEvGenCGF}))—can be written as a quadratic form: $d\hat{\mathcal{L}}_t = \frac{1}{2} \sum_{i,j} \bar{a}^{\dagger}_i \left(dL_t\right)_{i,j} \bar{a}_j$. Provided the initial state is Gaussian, i.e., a state obtained by acting with the exponential of a quadratic form on a vacuum state,
\begin{equation}
    \ket{\rho_0}\propto e^{\frac{1}{2}\sum_{i,j}\bar{a}^{\dagger}_i\left(\Omega_0\right)_{i,j}\bar{a}_j} \ket{\mathbf{0}},
\end{equation}
where $\ket{\mathbf{0}} = \ket{0} \otimes \ket{0}$ denotes the vacuum of the vectorized space and $\ket{0}$ the Fock vacuum in which all sites are empty. Since the dependence on the counting field plays no role in this section, we omit it for simplicity.
Using the properties of quadratic forms discussed earlier, we find that the time-evolved vectorized density matrix at time $t$ retains its Gaussian form and can be written as
\begin{equation}
    \ket{\rho_t} \propto \mathcal{T}e^{\int_0^t dt'\; d\hat{\mathcal{L}}_{t'} }\ket{\rho_0} = e^{\frac{1}{2}\sum_{i,j}\bar{a}^{\dagger}_i\left(\Omega_t\right)_{i,j}\bar{a}_j} \ket{\mathbf{0}}, \text{  with   } e^{\Omega_t} = \mathcal{T} e^{\int_0^t dt'\; dL_{t'}} e^{\Omega_0}.
\end{equation}
Here $\mathcal{T}$ denotes time-ordering.

So far, we have shown that Gaussianity is preserved under time evolution. Each Gaussian state $\ket{\rho_t}$ is uniquely determined by a $4L \times 4L$ matrix $\Omega_t$, which, due to particle-hole symmetry, admits the parametrization
\begin{equation} 
    e^{\Omega_t} = \begin{bmatrix}
    e^{\omega_t} + \eta_t e^{-\omega_t^T} \phi_t  & \eta_t e^{-\omega_t^T}\\
    e^{-\omega_t^T} \phi_t &e^{-\omega_t^T} \\
    \end{bmatrix}=\exp{\left(\Omega_t^{(\eta)}\right)} \exp{\left( \Omega_t^{(\omega)}\right)}\exp{\left( \Omega_t^{(\phi)}\right)}, \; \; \Omega_t^{(\eta)} =  \begin{bmatrix}
    0   & \eta_t \\
    0  & 0 \\
    \end{bmatrix},\;\; \Omega_t^{(\omega)}=\begin{bmatrix}
    \omega_t   & 0\\
     0 & -\omega_t^T  \\
    \end{bmatrix},\;\; \Omega_t^{(\phi)}=\begin{bmatrix}
    0   & 0\\
     \phi_t & 0  \\
    \end{bmatrix},
    \label{step1Omeg}
\end{equation}
for arbitrary $\omega_t$ and antisymmetric matrices $\eta_t$ and $\phi_t$. Since each of the exponentials on the right-hand side of Eq.~\eqref{step1Omeg} individually satisfies particle-hole symmetry, we may apply the Baker-Campbell-Hausdorff (BCH) formula in reverse to write:
\begin{equation}\label{eq:BCHreverse}
    \ket{\rho_t}\propto e^{\frac{1}{2}\sum_{i,j} \bar{a}_i^{\dagger} \left(\Omega_t\right)_{i,j} \bar{a}_j} \ket{\mathbf{0}} = e^{\frac{1}{2}\sum_{i,j} \bar{a}_i^{\dagger} \left(\Omega_t^{(\eta)}\right)_{i,j} \bar{a}_j}  e^{\frac{1}{2}\sum_{i,j} \bar{a}_i^{\dagger} \left(\Omega_t^{(\omega)}\right)_{i,j} \bar{a}_j}
     e^{\frac{1}{2}\sum_{i,j} \bar{a}_i^{\dagger} \left(\Omega_t^{(\phi)}\right)_{i,j} \bar{a}_j} \ket{\mathbf{0}} \propto e^{\frac{1}{2}\sum_{i,j} \bar{a}_i^{\dagger} \left(\Omega_t^{(\eta)}\right)_{i,j} \bar{a}_j} \ket{\mathbf{0}}\mathrel{=\vcentcolon}\ket{\eta_t}.
\end{equation}
To compute the normalization factor $N_t$ such that $\ket{\rho_t} = N_t^{-1} \ket{\eta_t}$, we note that the trace of an operator $\hat{O}$ is given by
\begin{equation}
    \Tr{\left(\hat{O}\right)}=\braket{\mathbb{1}|\hat{O}}, \text{ with} \;\; \bra{\mathbb{1}} = \bra{\mathbf{0}} e^{\frac{1}{2} \sum_{i,j} \bar{a}^{\dagger}_i \mathcal{I}_{i,j} \bar{a}_j},\;\;\mathcal{I}=\begin{bmatrix}
     0 & 0 \\
     S J & 0  \\
\end{bmatrix},\;\;S=\begin{bmatrix}
     0 & \mathbb{1}_{L\times L} \\
     \mathbb{1}_{L\times L} & 0  \\
\end{bmatrix}\text{   and   }
J=\begin{bmatrix}
     \mathbb{1}_{L\times L}  & 0\\
     0 & -\mathbb{1}_{L\times L}  \\
\end{bmatrix}.
\end{equation}
It then follows, using the properties of Gaussian quadratic forms, that the normalization factor is $N_t = \braket{\mathbb{1}|\eta_t} = \sqrt{\det\left(\mathbb{1} + S J \eta_t \right)}$.

From the previous discussion, we conclude that $\eta_t$ fully determines the normalized density matrix $\ket{\rho_t}$. In fact, one can also show that the correlation matrix of $\ket{\rho_t}$ is directly obtained from $\eta_t$ via $G_{\alpha,\beta} = \bra{\mathbb{1}}\bar{a}_\beta^{\dagger}\bar{a}_\alpha \ket{\rho_t}= \delta_{\alpha,\beta}-\left(\left( \mathbb{1}+ SJ \eta_t\right)^{-1}\right)_{\beta,\alpha}$, for $\alpha,\beta\leq L$. We shall not prove this result here, as it is not required in the main text. Instead, we conclude this section by proving that Wick's theorem holds for the Gaussian states discussed.

For notational simplicity, we explicitly address the case of 4-point functions, noting that generalization to arbitrary $n$-point functions is straightforward. In this case, Wick's theorem states:
\begin{equation}\label{eq:wick4ptfunc}
    \bra{\mathbb{1}}\bar{a}_i \bar{a}_j \bar{a}_k \bar{a}_l\ket{\rho_t} = \bra{\mathbb{1}}\bar{a}_i \bar{a}_j \ket{\rho_t}  \bra{\mathbb{1}} \bar{a}_k \bar{a}_l\ket{\rho_t} - \bra{\mathbb{1}}\bar{a}_i \bar{a}_k \ket{\rho_t}  \bra{\mathbb{1}} \bar{a}_k \bar{a}_l\ket{\rho_t} + \bra{\mathbb{1}}\bar{a}_i \bar{a}_l \ket{\rho_t}  \bra{\mathbb{1}} \bar{a}_j \bar{a}_k\ket{\rho_t}.
\end{equation}
We now note that, just as in Eq.~(\ref{eq:BCHreverse}), we may write:
\begin{equation}
    \ket{\eta_t}=e^{\frac{1}{2}\sum_{i,j} \bar{a}_i^{\dagger} \left(\Omega_t^{(\eta)}\right)_{i,j} \bar{a}_j} \ket{\mathbf{0}}=e^{\frac{1}{2}\sum_{i,j} \bar{a}_i^{\dagger} \left(\Omega_t^{(\eta)}\right)_{i,j} \bar{a}_j} e^{\frac{1}{2}\sum_{i,j} \bar{a}_i^{\dagger} \left(\varphi_t\right)_{i,j} \bar{a}_j} \ket{\mathbf{0}}, \text{ with   } \; \varphi_t=\begin{bmatrix}
    0  & 0\\
    -\left(\mathbb{1}+SJ \eta_t\right)^{-1} S J &  0  \\
\end{bmatrix},
\end{equation}
where we used the fact that the terms in the exponential of $\varphi_t$ annihilate the vacuum $\ket{\mathbf{0}}$.
We now insert the identity $\mathbb{1}=e^{\hat{\Omega}_t^{(\eta)}}e^{\hat{\varphi}_t}e^{-\hat{\varphi}_t}e^{-\hat{\Omega}_t^{(\eta)}}$ between each pair of fermionic operators $\bar{a}$ in the correlation function, where $\hat{\Omega}_t^{(\eta)}$ and $\hat{\varphi}_t$ denote the quadratic operators explicitly written in the previous equation. Using the identity from Eq.~(\ref{eq:identityqduadlinear}), we obtain:
\begin{equation}
    e^{-\hat{\varphi}_t} e^{-\hat{\Omega}_t^{(\eta)}} \bar{a}_i e^{\hat{\Omega}_t^{(\eta)}} e^{\hat{\varphi}_t}  =  \sum_{j,k} \left(e^{\Omega_t^{(\eta)}}\right)_{i,j} \left(e^{\varphi_t}\right)_{j,k} \bar{a}_k \mathrel{=\vcentcolon} \bar{b}_i \Longrightarrow  \bra{\mathbb{1}}\bar{a}_i \bar{a}_j \bar{a}_k \bar{a}_l\ket{\rho_t} = \frac{\bra{\mathbb{1}} e^{\hat{\Omega}_t^{(\eta)}} e^{\hat{\varphi}_t} \bar{b}_i \bar{b}_j \bar{b}_k \bar{b}_l\ket{\mathbf{0}}}{\braket{\mathbb{1}|\eta_t} } = \bra{\mathbf{0}}  \bar{b}_i \bar{b}_j \bar{b}_k \bar{b}_l\ket{\mathbf{0}},
\end{equation}
where we used that $\bra{\mathbf{0}} e^{\hat{\mathcal{I}}} e^{\hat{\Omega}_t^{(\eta)}} e^{\hat{\varphi}_t} = \braket{\mathbb{1}|\eta_t} \bra{\mathbf{0}}$. Since the operators $\bar{b}$ are linear combinations of the original $\bar{a}$, we may now apply the standard Wick’s theorem for vacuum expectation values:
\begin{equation}
    \bra{\mathbf{0}}  \bar{b}_i \bar{b}_j \bar{b}_k \bar{b}_l\ket{\mathbf{0}} =  \bra{\mathbf{0}}  \bar{b}_i \bar{b}_j \ket{\mathbf{0}}  \bra{\mathbf{0}} \bar{b}_k \bar{b}_l\ket{\mathbf{0}} - \bra{\mathbf{0}}  \bar{b}_i \bar{b}_k \ket{\mathbf{0}}  \bra{\mathbf{0}} \bar{b}_j \bar{b}_l\ket{\mathbf{0}} + \bra{\mathbf{0}}  \bar{b}_i \bar{b}_l \ket{\mathbf{0}}  \bra{\mathbf{0}} \bar{b}_j \bar{b}_k\ket{\mathbf{0}}.
\end{equation}
At no point in the derivation was the number of fermionic operators fixed, and thus the above procedure generalizes to arbitrary $n$-point functions. Moreover, since the transformation from $\bar{a}$ to $\bar{b}$ is invertible, we can revert the computation inside each 2-point correlator to recover Eq.~(\ref{eq:wick4ptfunc}), thus completing the proof and concluding this section.

\section{Explicit additional computations}\label{sec:BigEqs}
In this section, we explicitly derive and justify several of the key equations referenced in the main text, with particular emphasis on the time evolution equations.   

We begin by considering the evolution of the (tilted) density matrix as described in Eq.~(\ref{Eq:TimeEvGenCGF}), which is applicable to both \qnoise and QSSEP models, with the counting field located at the right boundary. As a reminder, the scaled cumulant generating function is defined as
$\lambda(s)=\lim_{t \to \infty} t^{-1}\overline{\log\left( \Tr{\left(\rho_s(t)\right)}\right)}$. In order to derive this quantity, we first compute the increment of $\overline{\log\left( \Tr{\left(\rho_s(t)\right)}\right)}$. This is given by
\begin{equation}\label{eq:dlog}
    \overline{d\log\left(\Tr\left(\rho_s\right)\right)}=\overline{\frac{d \Tr\left(\rho_s\right)}{\Tr\left(\rho_s\right)}} - \frac{1}{2}\overline{\left(\frac{d \Tr\left(\rho_s\right)}{\Tr\left(\rho_s\right)}\right)^2}.
\end{equation}
The expansion to second order in $d\Tr(\rho_s)$ is required due to the presence of the Itô increment $dW$ in the stochastic evolution of $\rho_s$. Substituting Eq.~(\ref{Eq:TimeEvGenCGF}) into the expression above, we obtain
\begin{equation}
    \overline{d\log\left(\Tr\left(\rho_s\right)\right)}= \sum_\sigma \Gamma_{\text{R},\sigma} \left(e^{-\sigma s} - 1\right)  \Tr{\left(\frac{\rho_s}{\Tr{\rho_s}}\hat{L}_{\text{R},\sigma}^{\dagger}\hat{L}_{\text{R},\sigma} \right)} dt = \sum_{\sigma} \Gamma_{\text{R},\sigma} \left(e^{-\sigma s} - 1\right) \left(\delta_{\sigma,1} - \sigma \overline{\left(G_s\right)}_{L,L}\right)dt \;,
\end{equation}
where we used the definition of the normalized correlation matrix introduced earlier, namely, $\left(G_s\right)_{i,j} = \Tr{\left(\frac{\rho_s}{\Tr{\rho_s}} c^{\dagger}_j c_i \right)}$.
Taking the long-time limit $t \to \infty$, and assuming that $\overline{\left(G_s\right)}_{L,L}$ admits a well-defined stationary value, we finally obtain
\begin{equation}
    \lambda(s) = \sum_{\sigma} \Gamma_{\text{R},\sigma} \left(e^{-\sigma s} - 1\right) \left(\delta_{\sigma,1} - \sigma \mathbb{E}_{\infty}[\left(G_s\right)_{L,L}]\right) \;.
\end{equation}
From this discussion, and as already emphasized in the main text, we conclude that computing $\lambda(s)$ reduces to determining the value of $\left(G_s\right)_{L,L}$. To obtain it, one needs the evolution equation for the correlation matrix $G_s$, which can be derived directly from Eq.~(\ref{Eq:TimeEvGenCGF}) by inserting the operator $c_j^{\dagger} c_i$ and taking the trace on both sides.
In the case of \qnoise, this procedure leads to Eq.~(\ref{Eq:TevGket}), while for QSSEP it yields Eq.~(\ref{Eq:QTevGket}). Both expressions are written in the vectorized notation introduced in Appendix~\ref{sec:snl}. It is evident in both cases that the evolution depends on the second moment, $\overline{G \otimes G}$, which introduces a hierarchy of equations and significantly complicates their resolution. For QSSEP, as discussed in the main text, a useful strategy is to split the counting field across the system, which leads to Eq.~(\ref{eq:BulkCFmain}). We now turn to how $\lambda(s)$ can be extracted from the correlation matrix $G_s$ associated with the (tilted) density matrix $\rho_s$ that solves Eq.~(\ref{eq:BulkCFmain}), and we write down the full time-evolution equation for this $G_s$.

As before, we start by evaluating the increment of $\overline{\log\left( \Tr{\left(\rho_s(t)\right)}\right)}$, as given in Eq.~(\ref{eq:dlog}). To that end, we compute the quantity $\frac{d \Tr{\rho_s}}{\Tr{\rho_s}}$, which can be directly obtained from Eq.~(\ref{eq:BulkCFmain}). This yields
\begin{multline}
\frac{d \Tr{\rho_s}}{\Tr{\rho_s}} = \sum_\sigma  \Gamma_{\text{R},\sigma} \left(e^{-\sigma \tilde{s}f_L} - 1\right) \left(\delta_{\sigma,1} - \sigma \left(G_s\right)_{L,L}\right) dt + \sum_\sigma  \Gamma_{\text{L},\sigma} \left(e^{\sigma \tilde{s} f_0} - 1\right) \left(\delta_{\sigma,1} - \sigma \left(G_s\right)_{1,1}\right) dt + \\ + \sum_{0<j<L} \left(\left(e^{\tilde{s} f_j} - 1\right) \left(G_s\right)_{j,j} \left(1-\left(G_s\right)_{j+1,j+1}\right) + \left(e^{-\tilde{s} f_j} - 1\right) \left(G_s\right)_{j+1,j+1} \left(1-\left(G_s\right)_{j,j}\right) + \right. \\ \left. + \left(e^{\tilde{s} f_j} + e^{-\tilde{s} f_j} - 2\right) \left(G_s\right)_{j,j+1} \left(G_s\right)_{j+1,j} \right) dt -i \sum_{0<j<L} \left(\left(e^{\frac{\tilde{s} f_j}{2}} - e^{-\frac{\tilde{s} f_j}{2}} \right)  \left( \left(G_s\right)_{j,j+1} d\xi_j - \left(G_s\right)_{j+1,j} d\xi_j^*\right) \right).
\end{multline}
In the derivation above, we have used Wick’s theorem (see Appendix~\ref{sec:ProofGaussianity}) to express four-point correlations in terms of products of two-point functions.
Substituting this result into Eq.~(\ref{eq:dlog}) and applying the same reasoning as before, we find
\begin{multline}
\lambda(s) = \sum_\sigma  \Gamma_{\text{R},\sigma} \left(e^{-\sigma \tilde{s}f_L} - 1\right) \left(\delta_{\sigma,1} - \sigma \mathbb{E}_{\infty}[\left(G_s\right)_{L,L}]\right) dt + \sum_\sigma  \Gamma_{\text{L},\sigma} \left(e^{\sigma \tilde{s} f_0} - 1\right) \left(\delta_{\sigma,1} - \sigma \mathbb{E}_{\infty}\left[\left(G_s\right)_{1,1}\right]\right)  + \\ + \sum_{j=1}^{L-1} \left( \left(e^{\tilde{s} f_{j}}-1\right) \mathbb{E}_{\infty}\left[\left(G_s\right)_{j,j}\left(1-\left(G_s\right)_{j+1,j+1} \right)\right] +  \left(e^{-\tilde{s} f_{j}}-1\right) \mathbb{E}_{\infty}\left[\left(G_s\right)_{j+1,j+1}\left(1-\left(G_s\right)_{j,j}\right)\right] \right).
\end{multline}
Setting $f_0 = f_L = 0$, we recover Eq.~(\ref{eq:cgfexact}). Under this same condition, we can proceed as before: we insert the operator $c_j^{\dagger} c_i$ into Eq.~(\ref{eq:BulkCFmain}), take the trace on both sides, yielding
\begin{equation}
     \left(dG_s\right)_{j,k} = \frac{\Tr\left(d\rho_s\, c^{\dagger}_k c_j\right)}{\Tr\left(\rho_s \right)}- \frac{\Tr\left(d\rho_s\, c^{\dagger}_k c_j\right)}{\Tr\left(\rho_s \right)} \frac{d\Tr\left(\rho_s\right)}{\Tr\left(\rho_s \right)}-\left(G_s\right)_{j,k}\left(\frac{d\Tr\left(\rho_s\right)}{\Tr\left(\rho_s \right)}-\left(\frac{d\Tr\left(\rho_s\right)}{\Tr\left(\rho_s \right)}\right)^2\right),
\end{equation}
and finally apply Wick’s theorem to express all this in terms of the two-point function $G_s$,
\begin{multline}\label{eq:completeGs}
    d\left(G_s\right)_{j,k} = \left(\delta_{j,k} \left(\left(G_s\right)_{j+1,k+1}+\left(G_s\right)_{j-1,k-1}\right)- 2\left(G_s\right)_{j,k} +   \sum_{\alpha\in\{\text{L},\text{R}\}}\left(\delta_{j,j_\alpha}\delta_{k,j_\alpha} \Gamma_{\alpha,1}  -\sum_\sigma \left(\delta_{j,j_\alpha}+\delta_{k,j_\alpha}\right)\Gamma_{\alpha,\sigma}\left(G_s\right)_{j,k}\right)\right)dt\\
    - \sum_{l=1}^{L-1} \left(e^{\tilde{s} f_l}-1\right)\left( \left(G_s\right)_{j,l}\left(G_s\right)_{l,k} \left(1-\left(G_s\right)_{l+1,l+1}\right) - \left(G_s\right)_{l,l} \left(\delta_{j,l+1}-\left(G_s\right)_{j,l+1}\right) \left(\delta_{k,l+1}-\left(G_s\right)_{l+1,k}\right)\right)dt\\
    - \sum_{l=1}^{L-1} \left(e^{-\tilde{s} f_l}-1\right)\left(\left(G_s\right)_{j,l+1} \left(G_s\right)_{l+1,k} \left(1-\left(G_s\right)_{l,l}\right) - \left(G_s\right)_{l+1,l+1} \left(\delta_{j,l}-\left(G_s\right)_{j,l}\right) \left(\delta_{k,l}-\left(G_s\right)_{l,k}\right)\right)dt\\
    -i\left(e^{\tilde{s}f_{j-1}/2} d\xi_{j-1} \left(G_s\right)_{j-1,k}-e^{-\tilde{s}f_{k}/2} d\xi_{k} \left(G_s\right)_{j,k+1}-e^{\tilde{s}f_{k-1}/2} d\xi_{k-1}^*\left(G_s\right)_{j,k-1}+e^{-\tilde{s}f_{j}/2} d\xi_{j}^* \left(G_s\right)_{j+1,k}\right)\\
    + i \sum_{l=1}^{L-1} \left(e^{\tilde{s} f_l/2}-e^{-\tilde{s} f_l/2}\right)\left(\left(G_s\right)_{j,l+1}\left(G_s\right)_{l,k}d\xi_l-\left(G_s\right)_{j,l}\left(G_s\right)_{l+1,k}d\xi_l^*\right),
\end{multline}
where we recall that $j_{\text{L}}=1$ and $j_{\text{R}}=L$.
Since the case $i = j$ is particularly relevant for the discussion in the main text, we write explicitly the evolution equation for $\overline{\left(G_s\right)}_{i,i}$ by taking the average over the noise on both sides of Eq.~(\ref{eq:completeGs}), yielding
\begin{multline}\label{Eq:Gevs}
    \frac{d}{dt} \overline{\left(G_s\right)}_{i,i}= \overline{\left(G_s\right)}_{i+1,i+1} - 2 \overline{\left(G_s\right)}_{i,i} + \overline{\left(G_s\right)}_{i-1,i-1} +   \delta_{1,i}\sum_\sigma \Gamma_{\text{L},\sigma} \left(\delta_{\sigma,1} -  \overline{\left(G_s\right)}_{1,1}\right) + \delta_{i,L} \sum_\sigma \Gamma_{\text{R},\sigma} \left(\delta_{\sigma,1} -  \overline{\left(G_s\right)}_{L,L}\right) \\
    - \sum_{k=1}^{L-1} \left(e^{\tilde{s} f_k}-1\right)\left( \overline{\left(G_s\right)_{k,i}\left(G_s\right)_{i,k} \left(1-\left(G_s\right)_{k+1,k+1}\right)} - \overline{\left(G_s\right)_{k,k} \left(\delta_{i,k+1}-\left(G_s\right)_{i,k+1}\right) \left(\delta_{i,k+1}-\left(G_s\right)_{k+1,i}\right)}\right)\\
    - \sum_{k=1}^{L-1} \left(e^{-\tilde{s} f_k}-1\right)\left(\overline{\left(G_s\right)_{i,k+1} \left(G_s\right)_{k+1,i} \left(1-\left(G_s\right)_{k,k}\right)} - \overline{\left(G_s\right)_{k+1,k+1} \left(\delta_{i,k}-\left(G_s\right)_{i,k}\right) \left(\delta_{i,k}-\left(G_s\right)_{k,i}\right)}\right).
\end{multline}
Applying the assumptions detailed in Appendix~\ref{sec:QSSEP_CGF_derivation}—in particular Eq.(\ref{eq:assumptions})—and solving for the stationary solution, one can simplify Eq.(\ref{Eq:Gevs}) and take its continuous limit, thereby arriving at Eq.~(\ref{eq:eqforgs}).

\section{Complement to the numerics}\label{sec:ComplementNum}
\begin{figure}[ht]
 \centering
  \includegraphics[width=0.47\linewidth]{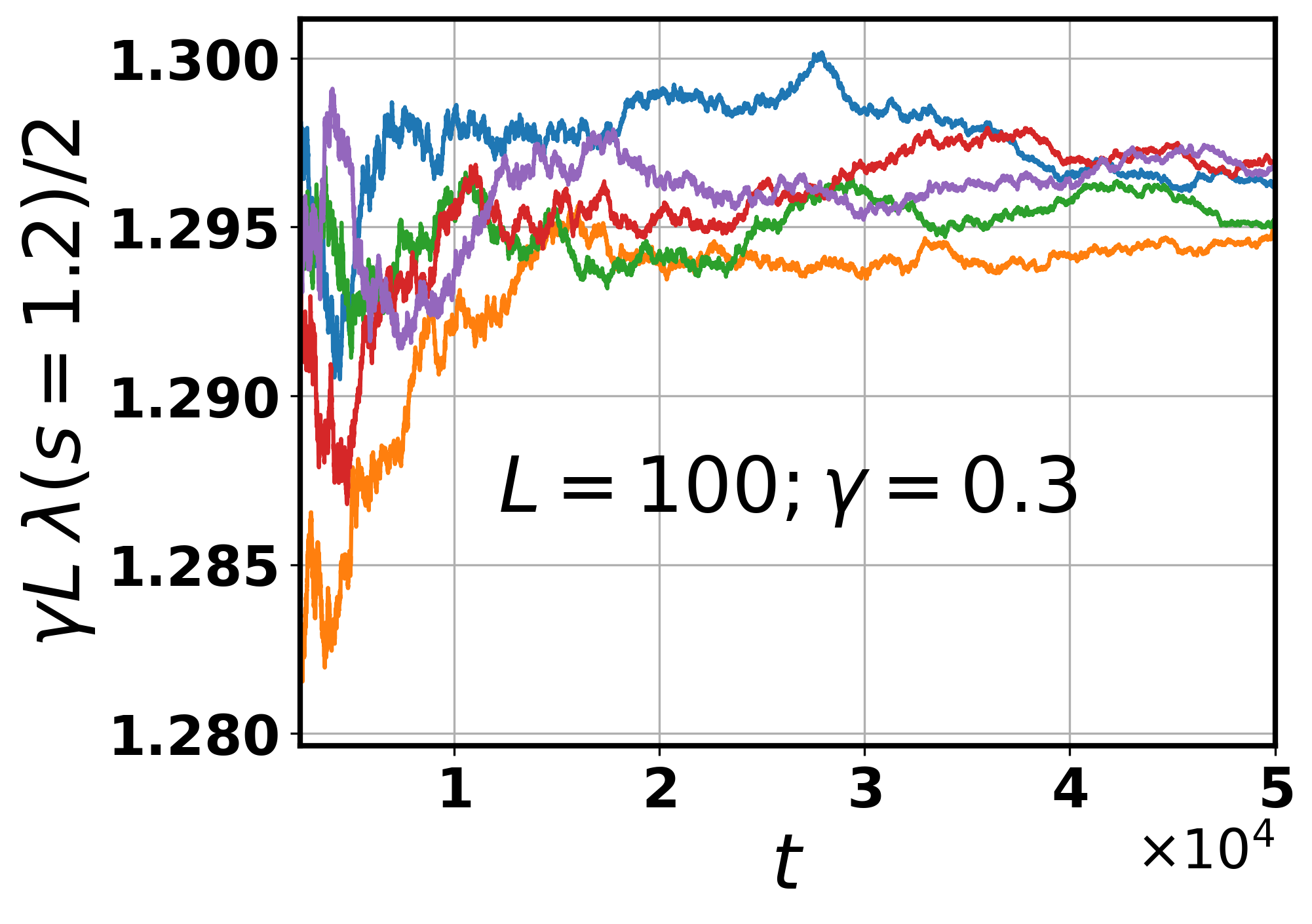}
  \includegraphics[width=0.45\linewidth]{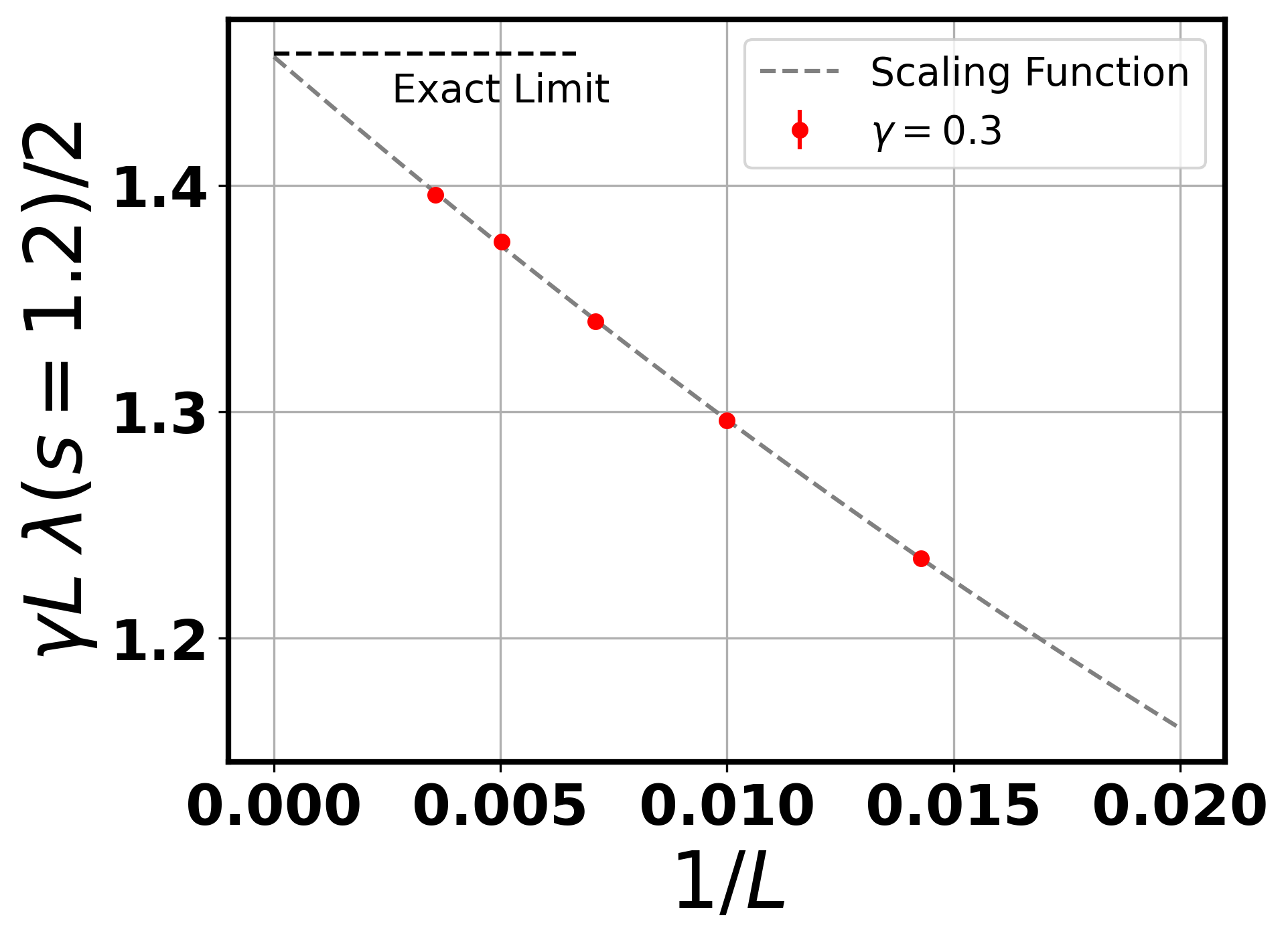}
\caption{In this figure, the left panel displays the time evolution of $\frac{\gamma L}{2}\log\left(\Tr\left(\rho_s(t)\right)\right)/t$ at $s=1.2$, $L=100$, and $\gamma=0.3$, for $n=5$ different noise realizations. The right panel shows the estimator of the CGF at $s=1.2$ and $\gamma=0.3$ as a function of the system size $L$, illustrating the agreement of the finite-size scaling with the SSEP result in the limit $L \to \infty$.}
\label{fig:Specific1}
\end{figure}

In this appendix, we provide some complementary plots to fig.~(\ref{fig:CGF}) from the main text. In fig.~(\ref{fig:CGF}), we plot the CGF of the current for \qnoise at different system sizes. These plots were obtained by simulating the dynamics described in Eq.~(\ref{Eq:TimeEvGenCGF}) up to very long times ($t_{\text{max}} \sim 10^4$) and for $n=5$ independent noise realizations. For each realization, we analyze the time evolution of $\frac{\gamma L}{2}\log\left(\Tr\left(\rho_s(t)\right)\right)/t$ and estimate the CGF for each value of the counting field and system size by averaging this quantity over a time window taken at the end of the evolution and over the different noise realizations. The associated error is estimated from the variance across the noise realizations. Although fig.~(\ref{fig:CGF}) appears to display no error bars, they are in fact included; their apparent absence is due to the very small numerical uncertainty. This is more clearly illustrated in fig.~(\ref{fig:Specific1}), where both the fluctuations along individual trajectories and the variations between different trajectories are relatively small.

In fig.~(\ref{fig:CGF}), the $L=\infty$ curve is obtained by performing a finite-size scaling analysis of the remaining numerical data. This procedure is illustrated in the right panel of fig.~(\ref{fig:Specific1}) for the representative case $s=1.2$ (the behaviour is analogous for other values of the counting field or system size), where we observe excellent agreement with the expected $L \to \infty$ result, namely the SSEP CGF. As before, error bars are included in the plot, although they are barely visible due to their small magnitude. 

For the inset of fig.~(\ref{fig:CGF}), we follow the same procedure as before, but now focus on the second cumulant of the current. To this end, we evaluate the discrete second derivative of the CGF around $s=0$ and consider the time evolution of the estimator
\begin{equation}\label{eq:discreteJ2}
J_2(t) = \frac{\log\left(\Tr\left(\rho_{s=h}(t)\right)\right)+\log\left(\Tr\left(\rho_{s=-h}(t)\right)\right)}{h^2\;t},
\end{equation}
where in our numerics we take $h=0.4$.

\begin{figure}[ht]
 \centering
  \includegraphics[width=0.47\linewidth]{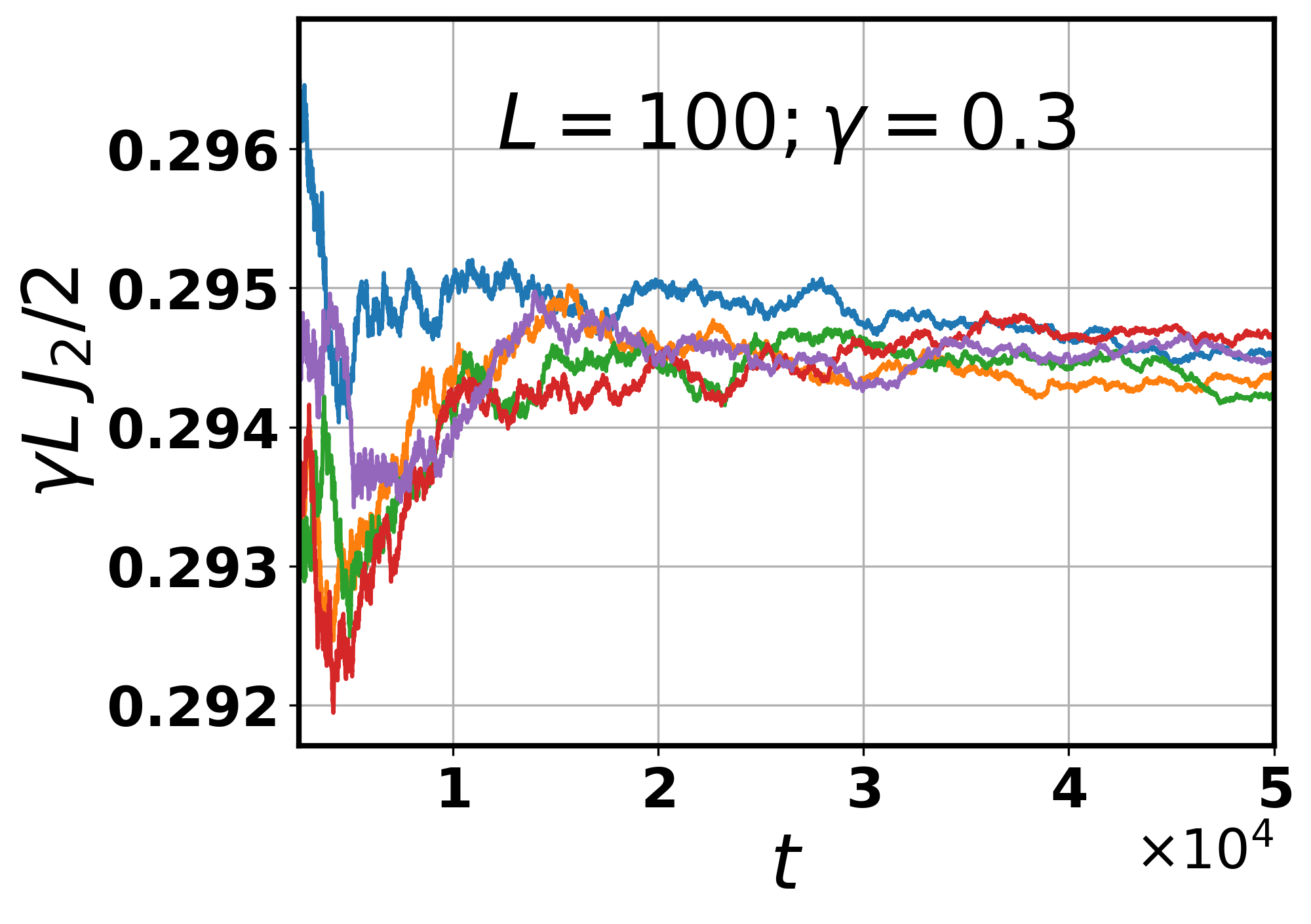}
  \includegraphics[width=0.45\linewidth]{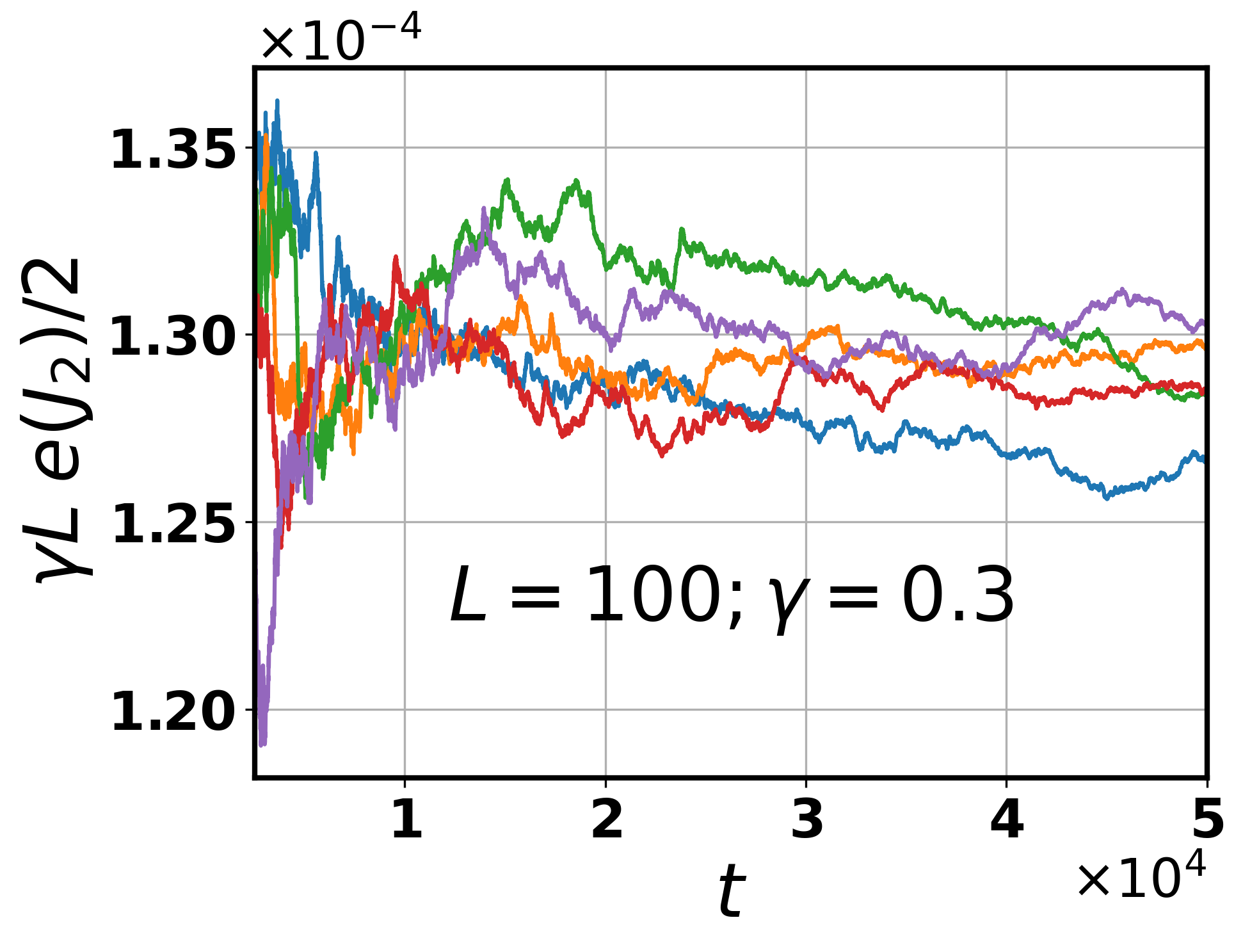}
\caption{In this figure, we show the estimator of the second cumulant, computed using Eq.~(\ref{eq:discreteJ2}), as a function of time for $L=100$ and $\gamma=0.3$ (left panel), together with the corresponding numerical error arising from the discrete derivative, evaluated via Eq.~(\ref{eq:discreteJ2error}) (right panel).}
\label{fig:Specific2}
\end{figure}

There are two sources of error in this procedure. The first arises from temporal fluctuations of this quantity and from variations between different noise realizations; these effects are small, as illustrated in the left panel of fig.~(\ref{fig:Specific2}). The second source of error originates from the use of a discrete derivative. An estimate of this latter contribution is given by
\begin{equation}\label{eq:discreteJ2error}
e(J_2) = \left|\frac{\lambda^{(4)}(0)}{12}\right| h^2,
\end{equation}
where the fourth derivative of $\lambda(s)$ is itself evaluated discretely, in direct analogy with Eq.~(\ref{eq:discreteJ2}). We plot this estimate in fig.~(\ref{fig:Specific2}); it is again found to be small. This is expected, since higher-order derivatives of the CGF are numerically much smaller than the second derivative. Indeed, for the SSEP—whose CGF is recovered in the limit $L \to \infty$ of \qnoise's CGF—one has $\lambda_{\text{SSEP}}^{(4)}(0) \sim -\frac{1}{105 L}$, which is numerically negligible compared to $\lambda_{\text{SSEP}}^{(2)}(0) \sim \frac{1}{3 L}$.

We emphasize again that error bars are included in all the plots discussed here, although they are barely visible due to their small magnitude.

\section{On the assumptions in the CGF derivation}\label{sec:CGFassumptions}
In this appendix, we justify the assumptions in Eq.~(\ref{eq:assumptions}), which underlie the computation of the QSSEP's CGF in the thermodynamic limit. 
\subsection{Numerical Evidence}
\begin{figure}[ht]
 \centering
  \includegraphics[width=1.\linewidth]{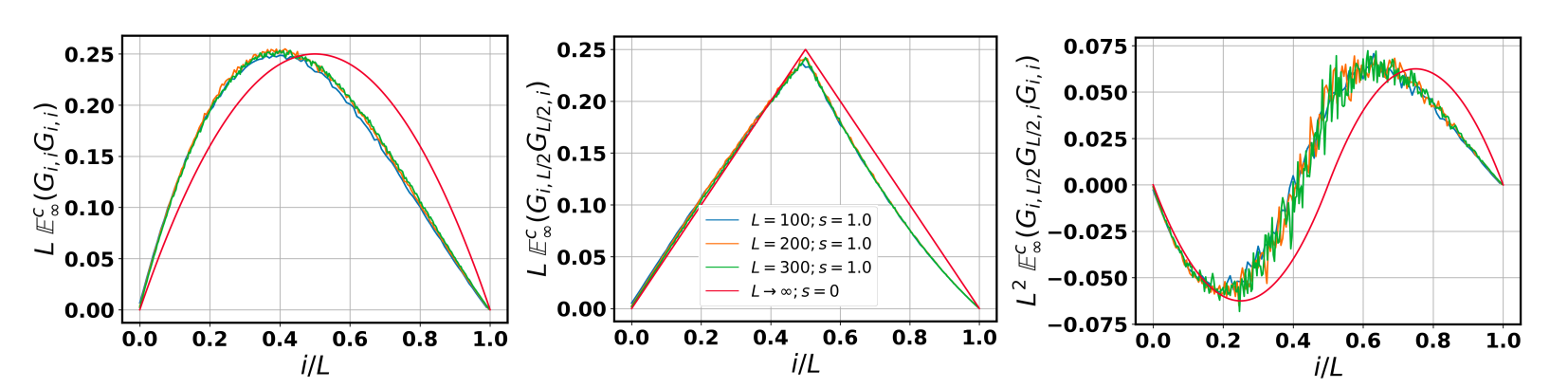}
\caption{In this figure we show the scaling with $L$ of relevant choices of QSSEP's second and third cumulants of $G_s$ for $s=1$. Each of the curves on these plots were obtained by averaging over $100$ different realizations that were simulated for sufficiently long times such that, effectively, $t \to \infty$.}
\label{fig:Gs1cumulants}
\end{figure}

Before turning to a more detailed analysis, we first present numerical evidence that strongly supports the validity of Eq.~(\ref{eq:assumptions}). In Fig.~(\ref{fig:Gs1cumulants}), we show the spatial dependence of the 2- and 3-point functions at finite $s$ (specifically $s=1.$), rescaled with $L$ according to the scaling predicted by Eq.~(\ref{eq:assumptions}), which coincides with the $s=0$ scaling derived in~\cite{PhysRevLett.123.080601}. We observe convergence of these functions with $L$, although to limits different from those at $s=0$, thereby demonstrating that the scaling proposed in Eq.~(\ref{eq:assumptions}) remains valid at finite $s$.
To obtain these plots, we performed the time evolution with the counting field applied at the right boundary, as in Eq.~(\ref{Eq:TimeEvGenCGF}). 

One might enquire whether the scaling with $L$ could in principle be affected by the choice of the gauge, i.e how the counting field is distributed along the bonds of the chain. In the next paragraph we provide an argument explaining that it is not the case: the scaling remains independent of the gauge. 

\paragraph{Independence of the scaling with $L$ from the gauge. -- } To establish this, we argue that if the $L$-scaling of the cumulants is independent of $s$ for a given gauge choice $f^{(1)}_j$, then the same must hold for any other choice $f^{(2)}_j$, related through $f^{(2)}_j=f^{(1)}_j+F_{j+1}-F_j$ for some suitable function $F_j$. Defining $\Hat{U} = e^{\frac{\tilde{s}}{2} \sum_{j=1}^{L} F_j c^{\dagger}_j c_j}$, as in Appendix~\ref{sec:proofGT}, and denoting by $d\mathcal{L}_s^{(1)}$ and $d\mathcal{L}_s^{(2)}$ the generators of the stochastic dynamics associated with the gauges $f^{(1)}_j$ and $f^{(2)}_j$, respectively, one can show that
\begin{equation}
    G_s^{(2)}(t):=\frac{\Tr\left(\mathcal{T}e^{\int_0^t d\mathcal{L}_s^{(2)}}\left(\rho_s(0)\right)c^{\dagger}_k c_j\right)}{\Tr\left(\mathcal{T}e^{\int_0^t d\mathcal{L}_s^{(2)}}\left(\rho_s(0)\right)\right)}=\frac{\Tr\left(\hat{U}\mathcal{T}e^{\int_0^t d\mathcal{L}_s^{(1)}}\left(\hat{U}^{-1}\rho_s(0)\hat{U}^{-1}\right)\hat{U}c^{\dagger}_k c_j\right)}{\Tr\left(\hat{U}\mathcal{T}e^{\int_0^t d\mathcal{L}_s^{(1)}}\left(\hat{U}^{-1}\rho_s(0)\hat{U}^{-1}\right)\hat{U}\right)}=_{t\to \infty}\frac{\Tr\left(\hat{U}\mathcal{T}e^{\int_0^t d\mathcal{L}_s^{(1)}}\left(\rho_s(0)\right)\hat{U}c^{\dagger}_k c_j\right)}{\Tr\left(\hat{U}\mathcal{T}e^{\int_0^t d\mathcal{L}_s^{(1)}}\left(\rho_s(0)\right)\hat{U}\right)},
\end{equation}
where the last equality is to be understood within expectation values and in the limit $t\to\infty$.
Using the formalism developed in Appendix~\ref{sec:ProofGaussianity} for Gaussian density matrices, and defining $G_s^{(1)}$ analogously to $G_s^{(2)}$ (with $(2)$ replaced by $(1)$), one finds that the above relation reduces to
\begin{equation}
    \mathbb{1}-G^{(2)}= U\left(\mathbb{1}-G^{(1)}\right)\left(\mathbb{1}+\left(U^{-2}-\mathbb{1}\right)\left(\mathbb{1}-G^{(1)}\right)\right)^{-1}U, \text{ where } U_{j,k} = e^{\frac{s F_k}{2}}\delta_{j,k}.
\end{equation}
This identity shows that all matrix elements of $G^{(2)}$ appearing inside a correlator can be expressed as functions of $G^{(1)}$. In particular, by expanding the inverse—whose series has a nonzero radius of convergence around $s=0$—one obtains, for the matrix element $\mathbb{1}_{j,k}-G^{(2)}_{j,k}$,
\begin{equation}
    e^{-s\left(F_j+F_k\right)/2} \left(\mathbb{1}-G^{(2)}\right)_{j,k}  =  \left(\mathbb{1}-G^{(1)}\right)_{j,k} +  \sum_{m=1}^{\infty}\sum_{l_1,...,l_m=1}^L \left(\mathbb{1}-G^{(1)}\right)_{j,l_1}   \prod_{i=1}^{m}   \left(1-e^{-sF_{l_i}}\right)\left(\mathbb{1}-G^{(1)}\right)_{l_i,l_{i+1}}, \text{ for } l_{m+1}:=k.
\end{equation}
Roughly, each term in this expansion effectively introduces an additional matrix element of $G_s$ into the correlator, thereby reducing its order by $\sim L^{-1}$ according to the scaling rules of~\cite{PhysRevLett.123.080601}. This reduction is compensated by the summation over the newly introduced free index, which increases the order by $\sim L$. These two effects cancel, and the $L$-scaling of the correlator is therefore preserved, even though the limiting rescaled function changes. This step can be carried out in a fully rigorous manner, although we do not present the detailed derivation here for the sake of clarity and brevity. This concludes our numerically supported argument for the validity of Eq.~(\ref{eq:assumptions}).

\subsection{Detailed analysis}

We now turn to a more systematic argument to establish the validity of the assumptions in Eq.~(\ref{eq:assumptions}). These assumptions are obviously true at $s=0$, see~\cite{PhysRevLett.123.080601}. It is thus natural to consider a taylor expansion of $\mathbb{E}^c_{\infty}\left[G_s^{\otimes n}\right]\eqqcolon \mathbb{G}_n(s)$ around $s=0$, $ \mathbb{G}_n(s) = \sum_{r=0}^\infty \frac{1}{r!}\mathbb{G}_n^{(r)} s^r$. For sufficiently well-behaved functions, this taylor expansion has a non-zero radius of convergence and thus validity of assumptions Eq.~(\ref{eq:assumptions}) follows from $\mathbb{G}_n^{(r)}$ having the same $L$-scaling for any $r$.
To establish this, we shall follow closely the work in~\cite{PhysRevLett.123.080601} to solve the (perturbative) steady state equations. 

The first important remark is that, even in the presence of the counting field $s$, the property extensively discussed in Appendix~\ref{sec:snl} and originally in~\cite{PhysRevLett.123.080601}—namely, that the set of left indices of $\mathbb{G}_n(s)$ must coincide with the set of right indices for the correlator to be non-vanishing—remains valid. As a consequence, we introduce a more compact notation for the indices of $\mathbb{G}_n(s)$. Specifically, we define
$\mathbf{i}_{\sigma_n}\coloneqq ((i_1,i_{\sigma(1)}),\ldots,(i_n,i_{\sigma(n)}))$,
where $\sigma_n$ is a permutation of $n$ elements.

Owing to the structure of Eq.~(\ref{eq:completeGs}), the steady-state equation for $\left(\mathbb{G}_n(s)\right)_{\mathbf{i}_{\sigma_n}}$ can be decomposed as
\begin{equation}\label{eq:HnFn}
    0=\left(d\mathbb{G}_n(s)\right)_{\mathbf{i}_{\sigma_n}}=\mathcal{H}_{\mathbf{i}_{\sigma_n}}\left[\left\{\mathbb{G}_n(s)\right\}_{n'\leq n}\right]+\mathcal{F}_{\mathbf{i}_{\sigma_n}}\left[\left\{\mathbb{G}_n(s)\right\}_{n'\leq n+2};s\right],
\end{equation}
where the functional $\mathcal{H}_{\mathbf{i}_{\sigma_n}}\left[\cdot\right]$ is independent of $s$, while $\mathcal{F}_{\mathbf{i}_{\sigma_n}}\left[\cdot;s\right]$ satisfies $\lim_{s\to 0}\mathcal{F}_{\mathbf{i}_{\sigma_n}}\left[\cdot;s\right]=0$. By construction, the condition $\mathcal{H}_{\mathbf{i}_{\sigma_n}}\left[\left\{\mathbb{G}_n(s=0)\right\}_{n'\leq n}\right]=0$ reproduces the steady-state equations derived in~\cite{PhysRevLett.123.080601}. The functional $\mathcal{F}_{\mathbf{i}_{\sigma_n}}\left[\left\{\mathbb{G}_n(s)\right\}_{n'\leq n+2};s\right]$, instead, collects all contributions arising from higher-order cumulants of $G_s$, as can be seen from Eq.~(\ref{eq:completeGs}).

For instance, at $n=1$, the functional  $\mathcal{H}_j[\mathbb{G}_1(s)]$ corresponds to the first line on the right-hand side of Eq.~(\ref{Eq:Gevs}), while $\mathcal{F}_j[\mathbb{G}_1(s);s]$ corresponds to the second and third lines. Expanding Eq.~(\ref{eq:HnFn}) perturbatively in $s$ yields
\begin{equation}\label{eq:H+R_order_r}
    0=\left(d\mathbb{G}_n(s)\right)^{(r)}_{\mathbf{i}_{\sigma_n}}=\mathcal{H}_{\mathbf{i}_{\sigma_n}}^{(r)}\left[\left\{\mathbb{G}_{n'}^{(r')}\right\}_{\substack{n'\leq n,\\r'\leq r}}\right]+\mathcal{F}_{\mathbf{i}_{\sigma_n}}^{(r)}\left[\left\{\mathbb{G}_{n'}^{(r')}\right\}_{\substack{n'\leq n+2,\\r'< r}}\right],
\end{equation}
where the superscript $r$ indicates that all terms are evaluated at order $r$ in the expansion in $s$. Importantly, the second term only depends on the $r'$ strictly smaller than $r$.

In close analogy with~\cite{PhysRevLett.123.080601}, it is convenient to decompose the functionals $\mathcal{H}_{\mathbf{i}_{\sigma_n}}^{(r)}$ and $\mathcal{F}_{\mathbf{i}_{\sigma_n}}^{(r)}$ into three distinct contributions,
\begin{equation}
    \mathcal{B}_{\mathbf{i}_{\sigma_n}}^{(r)}\left[\cdot\right]=\left(\mathcal{B}_{bndry}\right)_{\mathbf{i}_{\sigma_n}}^{(r)}\left[\cdot\right]+\left(\mathcal{B}_{contact}\right)_{\mathbf{i}_{\sigma_n}}^{(r)}\left[\cdot\right]+\left(\mathcal{B}_{bulk}\right)_{\mathbf{i}_{\sigma_n}}^{(r)}\left[\cdot\right], \; \mathcal{B}\in\{\mathcal{H},\mathcal{F}\},
\end{equation}
where $\mathcal{B}_{bndry}$ accounts for contributions arising from the steady-state equations at the boundaries (i.e., when at least one of the indices in $\mathbf{i}$ is equal to $1$ or $L$), $\mathcal{B}_{contact}$ collects contributions associated with contact points (when two or more indices in $\mathbf{i}$ coincide, $\mathbf{i}_j=\mathbf{i}_k$, or are adjacent, $\mathbf{i}_j=\mathbf{i}_k\pm 1$), and $\mathcal{B}_{bulk}$ corresponds to the remaining configurations, in which all indices in $\mathbf{i}$ are well separated from one another and from the boundaries. Although extending the arguments of \cite{PhysRevLett.123.080601} which leads to exact formulas for the connected cumulants to finite $s$ is hard, here we only need to focus on the scaling with the system size and we can keep the discussion more schematic; thus, we introduce
\begin{equation}
    \Theta_L(f_L(\mathbf{x})):=\lim_{L \to \infty} \log\left(f_L(\mathbf{x})\right)/\log(L),
\end{equation}
that gives the relevant scaling with $L$ of a quantity $f_L(\mathbf{x})$. We will now argue that
\begin{equation}
    \Theta_L(\mathbb{G}_n^{(r)}) = \Theta_L(\mathbb{G}_n^{(r=0)}) \;, \quad \forall r > 0.
\end{equation}
To show that the scaling of $\mathbb{G}_{n}^{(r)}$ is independent of $r$—and therefore to justify the assumptions made in Eq.~(\ref{eq:assumptions}), as discussed at the beginning of this appendix—we proceed by induction. We assume that the scaling behaviour and continuity of the cumulants established in~\cite{PhysRevLett.123.080601} hold for all pairs $(n',r')$ such that $r'<r$ and $n'\leq n+2\,r$ (our induction hypothesis), and we then demonstrate that the same properties necessarily follow at order $r$ for the $n$-correlator (induction step). Under this assumption, we can use Eq.~\eqref{eq:H+R_order_r} to determine $G_{n}^{(r)}$. 
Then, as we noted,
the functional $\mathcal{F}_{\mathbf{i}_{\sigma_n}}^{(r)}\left[\left\{\mathbb{G}_{n'}^{(r')}\right\}_{\substack{n'\leq n+2,\\r'< r}}\right]$
only involves cumulants of strictly lower order in $s$ than $r$, which are therefore covered by the inductive hypothesis, and whose scaling is known from~\cite{PhysRevLett.123.080601}: thus, it
always provides inhomogeneous contributions to the equations. 
Following~\cite{PhysRevLett.123.080601}, or equivalently by direct inspection of Eq.~(\ref{eq:completeGs}), one finds that
\begin{equation}
    \left(\mathcal{H}_{bulk}\right)_{\mathbf{i}_{\sigma_n}}^{(r)}\left[\left\{\mathbb{G}_{n'}^{(r')}\right\}_{\substack{n'\leq n,\\r'\leq r}}\right] = \Delta_{\mathbf{i}}^{dis} \cdot  \left(\mathbb{G}_{n}^{(r)}\right)_{\mathbf{i}_{\sigma_n}}\Longrightarrow \Theta_L\left(\left(\mathcal{H}_{bulk}\right)_{\mathbf{x}_{\sigma_n}}^{(r)}\left[\left\{\mathbb{G}_{n'}^{(r')}\right\}_{\substack{n'\leq n,\\r'\leq r}}\right]\right) \leq \Theta_L\left(L^{-2}\left(\mathbb{G}_n^{(r)}\right)_{\mathbf{x}_{\sigma_n}}\right),
\end{equation}
for $\Delta_{\mathbf{i}}^{dis}$ the $n$-dimensional discrete Laplacian,
$\Delta_{\mathbf{i}}^{dis} f(i_1,...i_n) = \sum_k \left(f(...,i_k+1,...)+f(...,i_k-1,...)-2f(...,i_k,...)\right)$ and $\mathbf{x}_{\sigma_n} = \mathbf{i}_{\sigma_n}/L$.
It then follows directly from Eq.~(\ref{eq:H+R_order_r}) that, in the bulk,
\begin{equation}
     \Theta_L \left(L^2\,\left(\mathcal{F}_{bulk}\right)_{\mathbf{x}_{\sigma_n}}^{(r)}\left[\left\{\mathbb{G}_{n'}^{(r')}\right\}_{\substack{n'\leq n+2,\\r'< r}}\right]\right) \leq \Theta_L\left(\left(\mathbb{G}_n^{(r)}\right)_{\mathbf{x}_{\sigma_n}}\right) .
\end{equation}
Therefore, in order for $\left(\mathbb{G}_n^{(r)}\right)_{\mathbf{x}_{\sigma_n}}$ to exhibit the same $L$-scaling independently of $r$, it is necessary that
\begin{equation}\label{eq:scalingassumption1}
    \Theta_L \left(L^2\,\left(\mathcal{F}_{bulk}\right)_{\mathbf{x}_{\sigma_n}}^{(r)}\left[\left\{\mathbb{G}_{n'}^{(r')}\right\}_{\substack{n'\leq n+2,\\r'< r}}\right]\right) \leq \Theta_L\left(\left(\mathbb{G}_n^{(0)}\right)_{\mathbf{x}_{\sigma_n}}\right)\eqqcolon K_{\mathbf{x}_{\sigma_n}}.
\end{equation}
Additionally, as can be inferred from~\cite{PhysRevLett.123.080601}, the contact equations
\begin{equation}\label{eq:contact_G_r}
    \left(\mathcal{H}_{contact}\right)_{\mathbf{i}_{\sigma_n}}^{(r)}\left[\left\{\mathbb{G}_{n'}^{(r')}\right\}_{\substack{n'\leq n,\\r'\leq r}}\right]+\left(\mathcal{F}_{contact}\right)_{\mathbf{i}_{\sigma_n}}^{(r)}\left[\left\{\mathbb{G}_{n'}^{(r')}\right\}_{\substack{n'\leq n+2,\\r'< r}}\right]=0,
\end{equation}
can be combined to derive linear equations in the discrete derivatives $\nabla_{\mathbf{i}}^{dis} \left(\mathbb{G}_{n}^{(r)}\right)_{\mathbf{i}_{\sigma_n}}$ at the contact points. Since $\nabla_{\mathbf{i}}^{dis}\sim L^{-1}$, preserving the $L$-scaling of  $\left(\mathbb{G}_n^{(r)}\right)_{\mathbf{x}_{\sigma_n}}$ independently of $r$ requires that
\begin{equation}\label{eq:scalingassumption2}
    \Theta_L \left(L\,\left(\mathcal{F}_{contact}\right)_{\mathbf{x}_{\sigma_n}}^{(r)}\left[\left\{\mathbb{G}_{n'}^{(r')}\right\}_{\substack{n'\leq n+2,\\r'< r}}\right]\right) \leq \Theta_L\left(\left(\mathbb{G}_n^{(0)}\right)_{\mathbf{x}_{\sigma_n}}\right).
\end{equation}
Note that the continuity of the cumulants 
$\left(\mathbb{G}_n^{(r)}\right)_{\mathbf{x}_{\sigma_n}}$ at the contact points—established in~\cite{PhysRevLett.123.080601} through the analysis of $\mathcal{H}_{\mathrm{contact}}$—remains valid upon including 
$\left(\mathcal{F}_{\mathrm{contact}}\right)_{\mathbf{x}_{\sigma_n}}$, since, by the induction hypothesis, this additional contribution is itself continuous at the contact points. 
Similarly, the boundary conditions are easily seen to remain unchanged by the inclusion of 
$\left(\mathcal{F}_{\mathrm{boundary}}\right)_{\mathbf{x}_{\sigma_n}}$.

Before discussing the validity of Eqs.~(\ref{eq:scalingassumption1},\ref{eq:scalingassumption2}), we temporarily assume them and investigate their consequences. 
Indeed, under these assumptions, one may introduce the following functions in the continuum,
\begin{equation}
    g^{(r)}_{\sigma_n}(\mathbf{x})=\lim_{L \to \infty} L^{K_{\mathbf{x}_{\sigma_n}}} \left(\mathbb{G}_{n}^{(r)}\right)_{\mathbf{x}_{\sigma_n}},\; \mathfrak{f}^{(r)}_{\sigma_n}(\mathbf{x})=\lim_{L \to \infty} L^{2+K_{\mathbf{x}_{\sigma_n}}} \left(\mathcal{F}_{bulk}\right)_{\mathbf{x}_{\sigma_n}}^{(r)}\left[\left\{\mathbb{G}_{n'}^{(r')}\right\}_{\substack{n'\leq n+2,\\r'< r}}\right],
\end{equation}
which satisfy the bulk equation
\begin{equation}\label{eq:continuum_geq_r}
    \Delta \cdot g^{(r)}_{\sigma_n}(\mathbf{x}) + \mathfrak{f}^{(r)}_{\sigma_n}(\mathbf{x}) = 0.
\end{equation}
Within our inductive scheme, the function $\mathfrak{f}^{(r)}_{\sigma_n}(\mathbf{x})$ is known and continuous. Eq.~(\ref{eq:continuum_geq_r}) therefore constitutes a linear inhomogeneous partial differential equation to be solved for $g^{(r)}_{\sigma_n}(\mathbf{x})$. We emphasize that this equation is valid only away from the boundaries and contact points, which instead determine the appropriate boundary conditions for the PDE. At the contact points, and under the scaling prescribed in Eq.~(\ref{eq:scalingassumption2}), one obtains von Neumann–type boundary equations for $g^{(r)}_{\sigma_n}(\mathbf{x})$, which can be implicitly written in a compact form
\begin{equation}\label{eq:continuum_geq_contact}
    \exists\,j,k : \mathbf{x}_j=\mathbf{x}_k,\,\mathfrak{h}^c_{\mathbf{x}_{\sigma_n}}\left[\{g_{\sigma_{n'}}(s)\}_{n'\leq n}\right]+\mathfrak{f}^c_{\mathbf{x}_{\sigma_n}}(\{g_{\sigma_{n'}}(s)\}_{n'\leq n+2})=0,
\end{equation}
where
$\mathfrak{h}^c_{\mathbf{x}_{\sigma_n}}\left[\{g_{\sigma_{n'}}(s)\}_{n'\leq n}\right]$ and $\mathfrak{f}^c_{\mathbf{x}_{\sigma_n}}(\{g_{\sigma_{n'}}(s)\}_{n'\leq n+2})$ denote the scaling functions associated with the appropriate combinations of 
$\mathcal{H}_{\mathrm{contact}}$ and $\mathcal{F}_{\mathrm{contact}}$, respectively, 
which give rise to the continuum boundary conditions in Eq.~(\ref{eq:continuum_geq_contact}).
Finally, just as in~\cite{PhysRevLett.123.080601}, Dirichlet boundary conditions are imposed at the edges of the domain,
\begin{equation}\label{eq:continuum_geq_boundary}
    \exists\,j : \mathbf{x}_j\in\{0,1\},\, \begin{cases}
        g^{(r)}_{\sigma_n}(\mathbf{x}) = \rho_{\text{L}}, \text{ for } n=1,r=0, \mathbf{x}_j=0,\\
        g^{(r)}_{\sigma_n}(\mathbf{x}) = \rho_{\text{R}}, \text{ for } n=1,r=0, \mathbf{x}_j=1, \\
        g^{(r)}_{\sigma_n}(\mathbf{x}) = 0, \text{ otherwise.} 
    \end{cases}
\end{equation}
The equations~(\ref{eq:continuum_geq_r},\ref{eq:continuum_geq_contact},\ref{eq:continuum_geq_boundary}) form a well-posed boundary-value problem, consisting of a linear inhomogeneous partial differential equation supplemented by contact and boundary conditions. This problem admits a non-trivial continuous solution, thereby confirming the consistency of the $L$-scaling assumed in the induction hypothesis. This demonstrates that Eqs.~(\ref{eq:scalingassumption1},\ref{eq:scalingassumption2}) are not only necessary, but also sufficient to complete the induction step.
To complete the inductive proof, it therefore remains to establish the validity of these assumptions.

As discussed above, owing to the structure of the functional 
$\mathcal{F}_{\mathbf{x}_{\sigma_n}}\!\left[\left\{\mathbb{G}_{n'}(s)\right\}_{n'\leq n}; s\right]$, 
its perturbative expansion in $s$ at order $r$ depends only on cumulants 
$\mathbb{G}_{n'}^{(r')}$ with $r'<r$, which, by the induction hypothesis, exhibit the same $L$-scaling and continuity properties as 
$\mathbb{G}_{n'}^{(0)}$. 
Consequently, since these are the only properties that will be necessary to establish the identities in Eqs.~(\ref{eq:scalingassumption1},\ref{eq:scalingassumption2}), all cumulants 
$\mathbb{G}_{n'}^{(r')}$ may be replaced by $\mathbb{G}_{n'}^{(0)}$ when evaluating 
$\mathcal{F}_{\mathbf{x}_{\sigma_n}}^{(r)}\!\left[\left\{\mathbb{G}_{n'}^{(r')}\right\}_{\substack{n'\leq n+2,\\ r'< r}}\right]$ 
inside the scaling function $\Theta_L(\cdot)$.
This observation shows that establishing Eqs.~(\ref{eq:scalingassumption1},\ref{eq:scalingassumption2}) for all $r$ and $n$ is equivalent to showing that
\begin{equation}\label{eq:indstep}
    \Theta_L\left(\left(\mathcal{F}_{bulk}\right)_{\mathbf{x}_{\sigma_n}}\left[\left\{\mathbb{G}_{n'}(s=0)\right\}_{n'\leq n};s=0\right]\right)\leq K_{\mathbf{x}_{\sigma_n}}-2,\;\Theta_L\left(\left(\mathcal{F}_{contact}\right)_{\mathbf{x}_{\sigma_n}}\left[\left\{\mathbb{G}_{n'}(s=0)\right\}_{n'\leq n};s=0\right]\right)\leq K_{\mathbf{x}_{\sigma_n}}-1, \;\;\forall n.
\end{equation}
For $n=1$, this identity was already established in going from Eq.~(\ref{Eq:Gevs}) to Eq.~(\ref{eq:eqforgs}) using the assumptions in~Eq.~(\ref{eq:assumptions}). 
We now explicitly verify Eq.~(\ref{eq:indstep}) for $n=2$ and $\mathbf{x}_{p_2}\coloneqq((\mathbf{x}_1,\mathbf{x}_2),(\mathbf{x}_2,\mathbf{x}_1))$ with $\mathbf{x}_1=j/L \neq k/L=\mathbf{x}_2$, starting by introducing the notation $G\coloneqq G_{s=0}$ and writing
{\small
\begin{multline}
\mathcal{F}_{\mathbf{x}_{p_2}}\left[\left\{\mathbb{G}_{n'}(0)\right\}_{n'\leq 4};0\right] =
    4\sum_{l} \sinh^2{\left(\frac{\tilde{s} f_l}{2}\right)}\left( \overline{\left(G_{j,l}G_{l,k} G_{k,j}+G_{j,k}G_{k,l} G_{l,j}\right) G_{l+1,l+1} } + \overline{\left(G_{j,l+1}G_{l+1,k} G_{k,j}+G_{j,k}G_{k,l+1} G_{l+1,j}\right) G_{l,l} }\right)\\-2\sum_{l} \left(e^{\tilde{s} f_l} + e^{-\tilde{s} f_{l-1}} -2\right) \left(  \overline{G_{j,l}G_{l,k} G_{k,j}}+ \overline{G_{j,k}G_{k,l} G_{l,j}}\right)-2\left(e^{\tilde{s} f_{j-1}}-1\right) \overline{G_{j,k}G_{k,j}G_{j-1,j-1}}-2\left(e^{-\tilde{s} f_{j}}-1\right) \overline{G_{j,k}G_{k,j}G_{j+1,j+1}}
    \\
    -2\left(e^{\tilde{s} f_{k-1}}-1\right)\overline{G_{j,k}G_{k,j}G_{k-1,k-1}}-2\left(e^{-\tilde{s} f_{k}}-1\right) \overline{G_{j,k}G_{k,j}G_{k+1,k+1}}+ \left(e^{\tilde{s} f_{j-1}}-1\right) \overline{G_{j-1,k}G_{k,j-1}}+\left(e^{-\tilde{s} f_{j}}-1\right) \overline{G_{j+1,k}G_{k,j+1}}\\
    +\left(e^{\tilde{s} f_{k-1}}-1\right) \overline{G_{j,k-1}G_{k-1,j}}
    +\left(e^{-\tilde{s} f_{k}}-1\right) \overline{G_{j,k+1}G_{k+1,j}}
    + 4 \sum_{l=1}^{L-1} \sinh^2{\left(\frac{\tilde{s} f_l}{2}\right)}\left(\overline{G_{j,l+1}G_{l+1,j}G_{l,k}G_{k,l}}+\overline{G_{j,l}G_{l,j}G_{k,l+1}G_{l+1,k}}\right).
\end{multline}
}
Analysing each of these terms individually yields the correct scaling described in Eq.~(\ref{eq:indstep}). We remark that the terms of the type $\overline{G_{j,l}G_{l,k} G_{k,j} G_{l+1,l+1}}$, corresponding to the insertion of a new vertex $l$ to the loop $j-k$ and of $G_{l+1,l+1}$, were already present at order $n=1$. From the scaling described in~\cite{PhysRevLett.123.080601}, one concludes that
\begin{equation}
\Theta_L \left(\sinh^2{\left(\frac{\tilde{s} f_l}{2}\right)}\overline{G_{j,l}G_{l,k} G_{k,j} G_{l+1,l+1}}\right)=
     \begin{cases}
         K_{\mathbf{x}_{p_2}}-2\text{ if }l \in \{j,k\},\\
         K_{\mathbf{x}_{p_2}}-3\text{ else},
     \end{cases}\hspace{-4mm}\Longrightarrow \Theta_L \left(\sum_{l}\sinh^2{\left(\frac{\tilde{s} f_l}{2}\right)}\overline{G_{j,l}G_{l,k} G_{k,j} G_{l+1,l+1}}\right)=K_{\mathbf{x}_{p_2}}-2.
\end{equation}
Additionally, continuity of the cumulants and of the function $f$ leads to 
\begin{align}
    &\Theta_L\left(2\left(e^{\tilde{s} f_{j-1}}-1\right) \overline{G_{j,k}G_{k,j}G_{j-1,j-1}}+2\left(e^{-\tilde{s} f_{j}}-1\right) \overline{G_{j,k}G_{k,j}G_{j+1,j+1}}\right)=K_{\mathbf{x}_{p_2}}-2,\\
    &\Theta_L\left( \sum_{\sigma\in\{-1,1\}}\left(\left(e^{-\sigma \tilde{s} f_{j+\frac{\sigma-1}{2}}}-1\right) \overline{G_{j+\sigma,k}G_{k,j+\sigma}}+\left(e^{-\sigma \tilde{s} f_{k+\frac{\sigma-1}{2}}}-1\right)\overline{G_{j,k+\sigma}G_{k+\sigma,j}}\right)\right) = \begin{cases}
        K_{\mathbf{x}_{p_2}}-2 \text{ if } j\neq k\pm 1,\\
        K_{\mathbf{x}_{p_2}}-1 \text{ if } j= k\pm 1.
    \end{cases}
\end{align}
Finally, the last additional terms that come from the contribution of  $\overline{\left(dG_s\right)_{j,k} \left(dG_s\right)_{k,j}}$ to the steady state equations, which of course was not present at $n=1$, also lead to
\begin{equation}
\begin{aligned}
&\text{If  } j \neq k+ 1,\\ 
     &\Theta_L \left(\sinh^2{\left(\frac{\tilde{s} f_l}{2}\right)}\overline{G_{j,l+1}G_{l+1,j}G_{l,k}G_{k,l}}\right)=\begin{cases}
         K_{\mathbf{x}_{p_2}}-2\text{ if }l \in \{j-1,k\},\vspace{1mm}\\ 
         K_{\mathbf{x}_{p_2}}-3\text{ else},
     \end{cases}\hspace{-10mm}\Longrightarrow \Theta_L \left(\sum_{l}\sinh^2{\left(\frac{\tilde{s} f_l}{2}\right)}\overline{G_{j,l+1}G_{l+1,j}G_{l,k}G_{k,l}}\right)=K_{\mathbf{x}_{p_2}}-2,\\
&\text{If  } j = k+1,\\ 
     &\Theta_L \left(\sinh^2{\left(\frac{\tilde{s} f_l}{2}\right)}\overline{G_{j,l+1}G_{l+1,j}G_{l,k}G_{k,l}}\right)=\begin{cases}
         K_{\mathbf{x}_{p_2}}-1\text{ if }l = k,\\
         K_{\mathbf{x}_{p_2}}-3\text{ else},
     \end{cases}\hspace{-4mm}\Longrightarrow \Theta_L \left(\sum_{l}\sinh^2{\left(\frac{\tilde{s} f_l}{2}\right)}\overline{G_{j,l+1}G_{l+1,j}G_{l,k}G_{k,l}}\right)=K_{\mathbf{x}_{p_2}}-1. 
\end{aligned}
\end{equation}
This completes the proof of Eq.~(\ref{eq:indstep}) for the case $n=2$ and the permutation $p_2$. 
For larger values of $n$ and for different permutations $\sigma$, the action of the functional 
$\mathcal{F}$ on higher-order correlators follows the same structural pattern, leading to analogous combinations of correlators. 
This stems from the structure of the dynamical equations and, more specifically, from It\^o calculus, which restricts $\mathcal{F}$ to act on at most two $G_s$'s within each term. 
We expect a proof for general $n$ to proceed exactly as in the present case. However, treating generic $n$-correlators is complex and we leave it to future work.
Based on these strong arguments, we expect that Eq.~(\ref{eq:indstep}) holds for all $n$.

\section{On the order of limits: $L \to \infty$ vs. $\gamma \to \infty$ then $L \to \infty$}\label{sec:EquivalenceOfLimits}
In this appendix, we address the equivalence between the limits $L \to \infty$ and $\gamma \to \infty$ followed by $L \to \infty$. In Appendix~\ref{sec:snl}, our focus was strictly on the limit $\gamma \to \infty$, and therefore we expanded the equations of motion only to leading order in $\gamma^{-1}$. Here, instead, we consider a finite but large $\gamma$, take the limit $L \to \infty$, and compare the resulting behaviour with that obtained by first retaining only the leading-order terms in $\gamma^{-1}$ and then sending $L \to \infty$. To carry out this comparison consistently, we must consider the full expansion of the equations of motion in powers of $\gamma^{-1}$.
To simplify the analysis, we restrict ourselves to a closed system with the same bulk dynamics, omitting both boundary terms and the counting field. Since our conclusions rely solely on bulk properties and their scaling in the thermodynamic limit, these additional contributions do not modify the result. A more detailed treatment including them leads to the same conclusions.

As in appendix~\ref{sec:snl}, we represent $\overline{G^{\otimes n}}$ as a vector $\ket{g^{(n)}}$ in an $L^{2n}$-dimensional vector space, whose time evolution is governed by the dynamics of the \qnoise model. The noise contribution is encoded in the operator $\Hat{F}$, defined in Eq.~(\ref{Eq:Dn}), which possesses a non-trivial kernel. This naturally leads us to decompose $\ket{g^{(n)}}$ into its projection onto the kernel of $\Hat{F}$, denoted by $\ket{g_{\parallel}^{(n)}}$, and its orthogonal complement, $\ket{g_{\perp}^{(n)}}$, so that $\ket{g^{(n)}} = \ket{g_{\parallel}^{(n)}} + \ket{g_{\perp}^{(n)}}$. With this decomposition, the full time evolution can be written as
\begin{equation}
        \frac{d}{dt} \begin{bmatrix}
        \ket{g_{\parallel}^{(n)}}\\
        \ket{g_{\perp}^{(n)}}
    \end{bmatrix} = - \begin{bmatrix}
        0 &   \Hat{B}_{\parallel,\perp} \\
         \Hat{B}_{\perp,\parallel} &  \gamma \Hat{F} + \Hat{B}_{\perp,\perp}
    \end{bmatrix} \begin{bmatrix}
        \ket{g_{\parallel}^{(n)}}\\
        \ket{g_{\perp}^{(n)}}
    \end{bmatrix}, 
    \label{eq:NtensorEQclosed}
\end{equation}
where $\Hat{B}_{\alpha,\beta}$ denotes the Hamiltonian contribution to the time evolution, as defined in Eq.~(\ref{Eq:ABdef}). The term $\Hat{B}_{\perp,\perp}$ was omitted in Eq.~(\ref{eq:NtensorEQ}) because there we focused exclusively on the leading-order contribution in $\gamma^{-1}$, for which it does not play a role. In order to perform the full expansion in $\gamma^{-1}$, however, this term must be retained.
After the time rescaling $t\to \gamma t$ and $\ket{g_{\perp}^{(n)}} \to \gamma^{-1} \ket{g_{\perp}^{(n)}}$, Eq.~(\ref{eq:NtensorEQclosed}) becomes
\begin{equation}
        \frac{d}{dt} \begin{bmatrix}
        \ket{g_{\parallel}^{(n)}}\\
        \ket{g_{\perp}^{(n)}}
    \end{bmatrix} = - \begin{bmatrix}
        0 &   \Hat{B}_{\parallel,\perp} \\
         \gamma^2\Hat{B}_{\perp,\parallel} &  \gamma^2 \Hat{F} + \gamma \Hat{B}_{\perp,\perp}
    \end{bmatrix} \begin{bmatrix}
        \ket{g_{\parallel}^{(n)}}\\
        \ket{g_{\perp}^{(n)}}
    \end{bmatrix}.
    \label{eq:NtensorEQclosedRescaled}
\end{equation}
From this appropriately rescaled dynamical equation, we can now analyze the time evolution of each component in the Taylor expansion
$\ket{g_{\perp/\parallel}^{(n)}}=\sum_{r=0}^{\infty} \gamma^{-r} \ket{g_{\perp/\parallel}^{(n,r)}}$.
Substituting this expansion into Eq.~(\ref{eq:NtensorEQclosedRescaled}) yields the recursive set of equations
\begin{equation}\label{eq:dyneqpert0}
\partial_t\ket{g_{\parallel}^{(n,r)}}= -\Hat{B}_{\parallel,\perp} \ket{g_{\perp}^{(n,r)}},\;\ket{g_{\perp}^{(n,r)}}=
\begin{cases}
  -\Hat{F}^{-1} \Hat{B}_{\perp,\parallel} \ket{g_{\parallel}^{(n,r)}}-\Hat{F}^{-1} \Hat{B}_{\perp,\perp} \ket{g_{\perp}^{(n,r-1)}}-\Hat{F}^{-1} \partial_t \ket{g_{\perp}^{(n,r-2)}}, \text{ for } r\geq0,\\
  0, \text{ for } r\leq-1.
\end{cases}
\end{equation}
The term $\Hat{F}^{-1} \Hat{B}_{\perp,\perp} \ket{g_{\perp}^{(n,r-1)}}$ can be systematically rewritten by recursively applying the same relation at lower orders. This leads to
\begin{equation}\label{eq:dyneqpert1}
\partial_t\ket{g_{\parallel}^{(n,r)}}= -\Hat{B}_{\parallel,\perp} \ket{g_{\perp}^{(n,r)}},\;\ket{g_{\perp}^{(n,r)}}=
\begin{cases}
  -\sum_{k=0}^r \left(-\Hat{F}^{-1} \Hat{B}_{\perp,\perp}\right)^k\Hat{F}^{-1}\left( \Hat{B}_{\perp,\parallel} \ket{g_{\parallel}^{(n,r-k)}}+\partial_t\ket{g_{\perp}^{(n,r-2-k)}}\right), \text{ for } r\geq0,\\
  0, \text{ for } r\leq-1.
\end{cases}
\end{equation}
Summing over all orders, we obtain
\begin{multline}
    \partial_t\ket{g_{\parallel}^{(n)}} =  -\Hat{B}_{\parallel,\perp} \ket{g_{\perp}^{(n)}},\,\ket{g_{\perp}^{(n)}} = - \sum_{k=0}^{\infty} \left(-\gamma^{-1}\Hat{F}^{-1} \Hat{B}_{\perp,\perp}\right)^k \Hat{F}^{-1} \left(\Hat{B}_{\perp,\parallel} \ket{g_{\parallel}^{(n)}}+\gamma^{-2}\partial_t\ket{g_{\perp}^{(n)}}\right) \Longrightarrow\\
    \Longrightarrow \partial_t\ket{g_{\parallel}^{(n)}} =\sum_{k=0}^{\infty} \gamma^{-k} \Hat{B}_{\parallel,\perp}\left(-\Hat{F}^{-1} \Hat{B}_{\perp,\perp}\right)^k \Hat{F}^{-1} \Hat{B}_{\perp,\parallel} \ket{g_{\parallel}^{(n)}} -\gamma^{-2}\sum_{k,k'=0}^{\infty} \gamma^{-k-k'} \Hat{B}_{\parallel,\perp}\left(-\Hat{F}^{-1} \Hat{B}_{\perp,\perp}\right)^{k+k'} \Hat{F}^{-1} \Hat{B}_{\perp,\parallel} \partial_t \ket{g_{\parallel}^{(n)}} + \mathcal{O}(\partial_t^2),
\end{multline}
where $\mathcal{O}(\partial_t^2)$ collects all terms containing higher order time derivatives. Operators of the form $\hat{B}_{\parallel,\perp} \left(-\Hat{F}^{-1} \Hat{B}_{\perp,\perp}\right)^k \Hat{F}^{-1} \Hat{B}_{\perp,\parallel}$ always implement discrete second (or higher) derivatives, and thus their scaling with $L$ is at most of order $\sim L^{-2}$. This identifies a diffusive regime at timescales $\tau=t/L^2$. In this regime $\partial_t^n = L^{-2n}\partial_\tau^n$ and all terms beyond the first on the right-hand side of the previous equation are subleading in $L$. We therefore obtain
\begin{equation}
    \partial_\tau\ket{g_{\parallel}^{(n)}} = L^2\sum_{k=0}^{\infty} \gamma^{-k} \Hat{B}_{\parallel,\perp}\left(-\Hat{F}^{-1} \Hat{B}_{\perp,\perp}\right)^k \Hat{F}^{-1} \Hat{B}_{\perp,\parallel} \ket{g_{\parallel}^{(n)}} + \mathcal{O}\left(L^{-2}\right).
\end{equation}
To make this structure explicit, we focus on the case $n=1$, where $\Hat{F}=\mathbb{1}_\perp$ (for $\mathbb{1}_\perp$ the identity on the $\perp$-subspace). The extension to $n>1$ follows along the same lines, up to minor technical considerations concerning the spectrum of $\Hat{F}$ and possible coincidences of pairs of indices (contact points), which do not modify the conclusion. Indeed, for $n=1$, one finds
\begin{equation}
    \partial_\tau\ket{g_{\parallel}^{(1)}} = L^2\sum_{k=0}^{\infty} \gamma^{-k} \Hat{B}_{\parallel,\perp}\left(-\Hat{B}_{\perp,\perp}\right)^k  \Hat{B}_{\perp,\parallel} \ket{g_{\parallel}^{(1)}} + \mathcal{O}\left(L^{-2}\right).
\end{equation}
A direct, though somewhat lengthy, computation shows that, using the notation introduced at the beginning of appendix~\ref{sec:snl},
\begin{equation}
    \Hat{B}_{\parallel,\perp} \Hat{B}_{\perp,\perp}^{2k}  \Hat{B}_{\perp,\parallel} \propto \Hat{D}_{2(k+1)}, \, \forall k\in \mathbb{N}_0, \text{ where }  \Hat{D}_{2k} =\sum_j \sum_{m=-k}^k (-1)^{m+k}\binom{2k}{m+k} \ket{j;j}\bra{j+m;j+m}
\end{equation}
is the operator implementing the discrete $2k$-th derivative. Since $\mathcal{O}\left(\Hat{D}_{2k}\right)\sim L^{-2k}$, only the $k=0$ contribution survives at leading order in the thermodynamic limit. We finally obtain
\begin{equation}
    \partial_\tau\ket{g_{\parallel}^{(1)}} = L^2 \Hat{B}_{\parallel,\perp}  \Hat{B}_{\perp,\parallel} \ket{g_{\parallel}^{(1)}} + \mathcal{O}\left(L^{-2}\right).
\end{equation}
As emphasized above, the same reasoning applies for arbitrary $n$: the operators generated at higher orders in $\gamma^{-1}$ correspond to higher-order discrete derivatives and are therefore increasingly suppressed in the thermodynamic limit. We thus conclude that the limit $L\to\infty$ isolates precisely the contribution arising from the leading-order term in the $\gamma^{-1}$ expansion. This establishes that taking $L \to \infty$ is equivalent to first sending $\gamma \to \infty$ and subsequently taking $L \to \infty$.

\section{QSSEP in higher dimensions}\label{sec:higherdimensions}
In this appendix, we discuss how the Gauge Trick used in one dimension extends to the computation of the QSSEP's CGF in arbitrary dimension $d$. We take the system to occupy a regular bounded domain $\Lambda$, which we regularize by imposing a UV cutoff through the lattice
$\Lambda_a = \Lambda \cap \left(a \mathbb{Z}\right)^d$, with one fermionic degree of freedom at each lattice site. The QSSEP dynamics is generated by stochastic hopping between nearest-neighbour sites,
\begin{equation}
    d\hat{H} = \sum_{\mathbf{j}\in \Lambda_a,\mathbf{e}\in\mathcal{E}_+} \left(d\xi_{\mathbf{j},\mathbf{e}}\, \hat{L}_{\mathbf{j},\mathbf{e}} + \text{H.c}\right),\; \hat{L}_{\mathbf{j},\mathbf{e}} = c^{\dagger}_{\mathbf{j}+\mathbf{e}} c_{\mathbf{j}},
\end{equation}
where the $d\xi_{\mathbf{j},\mathbf{e}}$ are complex independent Wiener increments satisfying
$\overline{d\xi_{\mathbf{j},\mathbf{e}} d\xi_{\mathbf{k},\mathbf{e}'}^*} = dt \delta_{\mathbf{j},\mathbf{k}} \delta_{\mathbf{e},\mathbf{e}'}$, and
$\mathcal{E}_+ = \{(1,0,\ldots,0),\ldots,(0,\ldots,0,1)\}$ is the set of positive lattice directions. As in the main text, we couple the system to Markovian reservoirs that exchange particles through two disjoint boundary regions, denoted by $\partial \text{L}$ and $\partial \text{R}$. The time evolution is then given by Eq.~(\ref{Eq:TimeEvGen}), with the bath contribution replaced by
\begin{equation}
    [d\rho]_{\rm bath} := \sum_{\substack{\mathbf{j}\in \{\partial\text{L}, \partial \text{R}\},\\ \sigma\in\{-1,1\}} } \left(\hat{L}_{\mathbf{j},\sigma} \rho \hat{L}_{\mathbf{j},\sigma}^{\dagger} - \frac{1}{2} \left\{\hat{L}_{\mathbf{j},\sigma}^{\dagger}\hat{L}_{\mathbf{j},\sigma},\rho\right\}\right)dt,\;\text{ where } \hat{L}_{\mathbf{j},1} = \sqrt{\Gamma_{\mathbf{j},1}}\,c^{\dagger}_{\mathbf{j}},\;\hat{L}_{\mathbf{j},-1} = \sqrt{\Gamma_{\mathbf{j},-1}}\,c_{\mathbf{j}}.
\end{equation} 
A counting field can be introduced exactly as in Eq.~(\ref{Eq:TimeEvGenCGF}), leading to
\begin{equation}
    d\rho_s = [d\rho_s]_{\rm uni} + 
    [d\rho_s]_{\rm bath}+
    \sum_{\substack{\mathbf{j}\in \partial \text{R}, \sigma\in\{-1,1\}}} \left(e^{-\sigma s} - 1\right) \Hat{L}_{\mathbf{j},\sigma} \rho_s \Hat{L}_{\mathbf{j},\sigma}^{\dagger} dt \;.
    \label{Eq:TimeEvGenCGFd-dim}
\end{equation}
A gauge transformation of the form Eq.~(\ref{eq:gaugetransf}) can then be used to distribute the counting field through the bulk,
\begin{multline}
    d\rho_s = \sum_{\substack{\mathbf{j}\in \Lambda_a,\\\mathbf{e}\in\mathcal{E}_+}} \left( e^{s \mathbf{f}_\mathbf{j} \cdot \mathbf{e}} \hat{L}_{\mathbf{j},\mathbf{e}} \rho_s \hat{L}^{\dagger}_{\mathbf{j},\mathbf{e}} + e^{-s \mathbf{f}_\mathbf{j} \cdot \mathbf{e}} \hat{L}^{\dagger}_{\mathbf{j},\mathbf{e}} \rho_s \hat{L}_{\mathbf{j},\mathbf{e}} - \frac{1}{2} \{\hat{L}^{\dagger}_{\mathbf{j},\mathbf{e}} \hat{L}_{\mathbf{j},\mathbf{e}} + \hat{L}_{\mathbf{j},\mathbf{e}} \hat{L}^{\dagger}_{\mathbf{j},\mathbf{e}}, \rho_s \} \right)  + \sum_{\substack{\mathbf{j}\in \{\partial\text{L}, \partial \text{R}\},\\ \sigma\in\{-1,1\}}} \left( \hat{L}_{\mathbf{j},\sigma} \rho \hat{L}_{\mathbf{j},\sigma}^{\dagger} - \frac{1}{2} \left\{\hat{L}_{\mathbf{j},\sigma}^{\dagger}\hat{L}_{\mathbf{j},\sigma},\rho\right\}\right)-  \\ 
    - i \sum_{\mathbf{j}\in \Lambda_a,\mathbf{e}\in\mathcal{E}_+} \left( e^{s \mathbf{f}_\mathbf{j}\cdot \mathbf{e}/2}\hat{L}_{\mathbf{j},\mathbf{e}} \rho_s d\xi_{\mathbf{j},\mathbf{e}} + e^{-s \mathbf{f}_\mathbf{j}\cdot \mathbf{e}/2}\hat{L}^{\dagger}_{\mathbf{j},\mathbf{e}} \rho_s d\xi^*_{\mathbf{j},\mathbf{e}} - e^{-s \mathbf{f}_\mathbf{j}\cdot \mathbf{e}/2} d\xi_{\mathbf{j},\mathbf{e}}  \rho_s \hat{L}_{\mathbf{j},\mathbf{e}} - e^{s \mathbf{f}_\mathbf{j}\cdot \mathbf{e}/2} d\xi^*_{\mathbf{j},\mathbf{e}} \rho_s \hat{L}^{\dagger}_{\mathbf{j},\mathbf{e}}\right).
\label{eq:BulkCFmaind-dim}
\end{multline} 
Here the vector field
$\mathbf{f}_{\mathbf{j}} = \sum_{\mathbf{e}\in \mathcal{E}_+} (F_{\mathbf{j}+\mathbf{e}}-F_{\mathbf{j}})\;\mathbf{e}$
is the discrete gradient of the gauge function $F$ appearing in Eq.~(\ref{eq:gaugetransf}). We choose $F$ so that the counting field is removed from the boundary terms, namely
$F_{\mathbf{j}}=1$ for $\mathbf{j}\in \partial \text{R}$ and $F_{\mathbf{j}}=0$ for $\mathbf{j}\in \partial \text{L}$. As in one dimension, the remaining freedom in $F$ will be fixed below by a convenient gauge choice.

The same computation of $d\log\Tr(\rho_s)$ that led to Eq.~(\ref{eq:cgfexact}) gives, in the present lattice approximation and with the choice of $F$ described,
\begin{equation}
    \lambda_a(s)=
    \sum_{\substack{\mathbf{j}\in \Lambda_a,\\\mathbf{e}\in\mathcal{E}_+}}
    \left(e^{s\mathbf{f}_{\mathbf{j}}\cdot\mathbf{e}}-1\right)
    \mathbb{E}_{\infty}\left[
    \left(G_s\right)_{\mathbf{j},\mathbf{j}}
    \left(1-\left(G_s\right)_{\mathbf{j}+\mathbf{e},\mathbf{j}+\mathbf{e}}\right)
    \right]
    +
    \sum_{\substack{\mathbf{j}\in \Lambda_a,\\\mathbf{e}\in\mathcal{E}_+}}
    \left(e^{-s\mathbf{f}_{\mathbf{j}}\cdot\mathbf{e}}-1\right)
    \mathbb{E}_{\infty}\left[
    \left(G_s\right)_{\mathbf{j}+\mathbf{e},\mathbf{j}+\mathbf{e}}
    \left(1-\left(G_s\right)_{\mathbf{j},\mathbf{j}}\right)
    \right].
    \label{eq:cgfexactd-dim}
\end{equation}
Writing Eq.~(\ref{eq:BulkCFmaind-dim}) for the diagonal element of the correlation matrix and averaging over the noise gives the direct $d$-dimensional analogue of Eq.~(\ref{Eq:Gevs}):
\begin{multline}\label{Eq:Gevsd-dim}
    \frac{d}{dt} \overline{\left(G_s\right)}_{\mathbf{j},\mathbf{j}} =
    \sum_{\mathbf{e}\in\mathcal{E}_+}\left(
    \overline{\left(G_s\right)}_{\mathbf{j}+\mathbf{e},\mathbf{j}+\mathbf{e}}
    +\overline{\left(G_s\right)}_{\mathbf{j}-\mathbf{e},\mathbf{j}-\mathbf{e}}
    -2\overline{\left(G_s\right)}_{\mathbf{j},\mathbf{j}}\right)
    + \sum_{\substack{\mathbf{j}\in \{\partial\text{L}, \partial \text{R}\},\\ \sigma\in\{-1,1\}}} \left(\delta_{\sigma,1}-\overline{\left(G_s\right)}_{\mathbf{j},\mathbf{j}}\right)\\
    - \sum_{\mathbf{k}\in\Lambda_a,\,\mathbf{e}\in\mathcal{E}_+}
    \left(e^{s\mathbf{f}_{\mathbf{k}}\cdot\mathbf{e}}-1\right)
    \left(
    \overline{\left(G_s\right)_{\mathbf{k},\mathbf{j}}\left(G_s\right)_{\mathbf{j},\mathbf{k}}
    \left(1-\left(G_s\right)_{\mathbf{k}+\mathbf{e},\mathbf{k}+\mathbf{e}}\right)}
    -\overline{\left(G_s\right)_{\mathbf{k},\mathbf{k}}
    \left(\delta_{\mathbf{j},\mathbf{k}+\mathbf{e}}-\left(G_s\right)_{\mathbf{j},\mathbf{k}+\mathbf{e}}\right)
    \left(\delta_{\mathbf{j},\mathbf{k}+\mathbf{e}}-\left(G_s\right)_{\mathbf{k}+\mathbf{e},\mathbf{j}}\right)}
    \right)\\
    - \sum_{\mathbf{k}\in\Lambda_a,\,\mathbf{e}\in\mathcal{E}_+}
    \left(e^{-s\mathbf{f}_{\mathbf{k}}\cdot\mathbf{e}}-1\right)
    \left(
    \overline{\left(G_s\right)_{\mathbf{j},\mathbf{k}+\mathbf{e}}
    \left(G_s\right)_{\mathbf{k}+\mathbf{e},\mathbf{j}}
    \left(1-\left(G_s\right)_{\mathbf{k},\mathbf{k}}\right)}
    -\overline{\left(G_s\right)_{\mathbf{k}+\mathbf{e},\mathbf{k}+\mathbf{e}}
    \left(\delta_{\mathbf{j},\mathbf{k}}-\left(G_s\right)_{\mathbf{j},\mathbf{k}}\right)
    \left(\delta_{\mathbf{j},\mathbf{k}}-\left(G_s\right)_{\mathbf{k},\mathbf{j}}\right)}
    \right).
\end{multline}
To pass to the continuum, we set
$g_s(\mathbf{x})=\lim_{a\to0}\mathbb{E}_{\infty}[(G_s)_{\mathbf{x}/a,\mathbf{x}/a}]$, with $\mathbf{x}=\mathbf{j}\,a$,
and assume the higher-dimensional analogue of Eq.~(\ref{eq:assumptions}). Namely, for $\mathbf{x}\neq\mathbf{y}$,
\begin{equation}
    \lim_{a\to0} a^{-d}\,
    \mathbb{E}_{\infty}^c\left[
    \left(G_s\right)_{\mathbf{x}/a,\mathbf{y}/a}
    \left(G_s\right)_{\mathbf{y}/a,\mathbf{x}/a}\right]
    = H_s(\mathbf{x},\mathbf{y}),
    \qquad
    \lim_{a\to0}
    \frac{\mathbb{E}_{\infty}^c\left[
    \left(G_s\right)_{\mathbf{x}/a,\mathbf{x}/a}\mathcal{F}(G_s)\right]}
    {\mathbb{E}_{\infty}\left[\left(G_s\right)_{\mathbf{x}/a,\mathbf{x}/a}\right]
    \mathbb{E}_{\infty}\left[\mathcal{F}(G_s)\right]}=0.
    \label{eq:assumptionsd-dim}
\end{equation}
We assume that these scalings hold for the tilted problem at arbitrary $s$. A full proof should proceed in two steps. First, one should establish the statement at $s=0$; this is natural to expect because, after averaging over the noise, the QSSEP reduces to the $d$-dimensional SSEP, for which the corresponding continuum scaling has been shown. Second, one should extend the statement to finite non-zero $s$. We do not prove this extension here, but the required adaptation of the perturbative argument of Appendix~\ref{sec:CGFassumptions} to the present geometry should be straightforward.

We next distinguish the microscopic CGF from its continuum scaling limit. If $\lambda_a(s)$ denotes the CGF on the lattice with spacing $a$, the finite continuum quantity is
\begin{equation}
    \tilde{\lambda}_d(s)=\lim_{a\to0} a^{d-2}\lambda_{d,a}(s),
    \qquad\text{or equivalently}\qquad
    \lambda_{d,a}(s)=a^{2-d}\tilde{\lambda}_d(s)+o\left(a^{2-d}\right).
    \label{eq:lambdascalingd-dim}
\end{equation}
For $d=1$ and $a\simeq L^{-1}$, this reduces to the usual 1d QSSEP scaling $\lambda_{1\mathrm{d},a}(s)\simeq L^{-1}\tilde{\lambda}_{1\mathrm{d}}(s)$. With this convention and the assumptions in Eq.~(\ref{eq:assumptionsd-dim}), the discrete formula in Eq.~(\ref{eq:cgfexactd-dim}) has the continuum limit
\begin{equation}
    \tilde{\lambda}_d(s)
    =
    -s\int_{\Lambda}d^d\mathbf{x}\,\mathbf{f}(\mathbf{x})\cdot\nabla g_s(\mathbf{x})
    +s^2\int_{\Lambda}d^d\mathbf{x}\,
    \left(\mathbf{f}(\mathbf{x})\cdot\mathbf{f}(\mathbf{x})\right)
    g_s(\mathbf{x})\left(1-g_s(\mathbf{x})\right).
    \label{eq:cgfcontinuumd-dim}
\end{equation}
The same continuum limit applied to the steady-state version of Eq.~(\ref{Eq:Gevsd-dim}) gives the bulk equation
\begin{multline}
    \nabla^2 g_s(\mathbf{x})
    - s\left[
    2\left(\mathbf{f}(\mathbf{x})\cdot\nabla g_s(\mathbf{x})\right)
    \left(1-2g_s(\mathbf{x})\right)
    +\left(\nabla\cdot\mathbf{f}(\mathbf{x})\right)g_s(\mathbf{x})
    \left(1-g_s(\mathbf{x})\right)\right]\\
    +s^2\left(\mathbf{f}(\mathbf{x})\cdot\mathbf{f}(\mathbf{x})\right)
    g_s(\mathbf{x})\left(1-g_s(\mathbf{x})\right)
    \left(1-2g_s(\mathbf{x})\right)
    =
    -\int_{\Lambda} d^d\mathbf{y}\,
    \left[
    s^2\left(\mathbf{f}(\mathbf{y})\cdot\mathbf{f}(\mathbf{y})\right)
    \left(2g_s(\mathbf{y})-1\right)
    -s\nabla\cdot\mathbf{f}(\mathbf{y})
    \right]H_s(\mathbf{x},\mathbf{y}).
    \label{eq:eqforgsd-dim}
\end{multline}
We now use the gauge freedom. We note that $\mathbf{f}$ is a gradient gauge field,
\begin{equation}
    \mathbf{f}(\mathbf{x})=\nabla F_s(\mathbf{x}),
    \qquad
    F_s=0 \text{ on } \partial\text{L},\qquad
    F_s=1 \text{ on } \partial\text{R}.
    \label{eq:gradientgauged-dim}
\end{equation}
As in one dimension, the remaining gauge freedom can be fixed so as to remove the dependence on the unknown two-point scaling function $H_s$. In terms of $F_s$, this amounts to imposing
\begin{equation}
    \nabla^2 F_s(\mathbf{x})
    =
    \nabla\cdot\mathbf{f}(\mathbf{x})
    =
    s\,\mathbf{f}(\mathbf{x})\cdot\mathbf{f}(\mathbf{x})
    \left(2g_s(\mathbf{x})-1\right).
    \label{eq:fconstrd-dim}
\end{equation}
With this choice, $g_s$ obeys the closed equation
\begin{equation}
    \nabla^2 g_s
    - s\left[
    2 \left(\mathbf{f}\cdot\nabla g_s\right) \left(1-2 g_s\right)
    +\left(\nabla\cdot \mathbf{f}\right)g_s(1-g_s)\right]
    +s^2 \left(\mathbf{f}\cdot \mathbf{f}\right)g_s(1-g_s)(1-2g_s) =0.
    \label{eq:closedg-dim}
\end{equation}
These equations are supplemented by the reservoir boundary conditions
$g_s=\rho_{\text{L}}$ on $\partial\text{L}$ and
$g_s=\rho_{\text{R}}$ on $\partial\text{R}$. On the remaining, open part of the boundary
$\partial\Lambda_{\mathrm{open}}\coloneqq \partial\Lambda\setminus(\partial\text{L}\cup\partial\text{R})$, no particles are exchanged with reservoirs, and the continuum limit gives the reflecting boundary conditions
\begin{equation}
    \nabla F_s(\mathbf{x})\cdot\mathbf{n}=\mathbf{f}(\mathbf{x})\cdot\mathbf{n}=0,\qquad \nabla g_s\cdot\mathbf{n}=0,
    \qquad \mathbf{x}\in\partial\Lambda_{\mathrm{open}},
\end{equation}
where $\mathbf{n}$ is the outward normal. Together with $F_s=0$ on $\partial\text{L}$ and $F_s=1$ on $\partial\text{R}$, these conditions fix the normalization of the gauge potential.

This system can be reduced exactly to the one-dimensional equations by choosing a harmonic coordinate $u(\mathbf{x})$, in the spirit of the $d$-dimensional SSEP reduction of Ref.~\cite{Akkermans2013}, satisfying
\begin{equation}
    \nabla^2 u=0,\qquad
    u=0 \text{ on } \partial\text{L},\qquad
    u=1 \text{ on } \partial\text{R},\qquad
    \nabla u\cdot\mathbf{n}=0 \text{ on } \partial\Lambda_{\mathrm{open}}.
\end{equation}
The standard existence and uniqueness theorem for the mixed Dirichlet--Neumann problem ensures that such a harmonic coordinate exists for regular domains, while the strong maximum principle and Hopf boundary-point lemma imply that its extrema are attained only on the Dirichlet parts of the boundary~\cite{GilbargTrudinger,EvansPDE}. Thus the left reservoir boundary sits at $u=0$ and the right reservoir boundary at $u=1$, just as in one dimension.
We then consider fields that depend only on $u$,
\begin{equation}
    g_s(\mathbf{x})=\phi_s(u(\mathbf{x})),\qquad
    \mathbf{f}(\mathbf{x})=f_s(u(\mathbf{x}))\,\nabla u(\mathbf{x}).
\end{equation}
Equivalently, $\mathbf{f}=\nabla F_s$ with $F_s(\mathbf{x})=\Phi_s(u(\mathbf{x}))$ and $\Phi_s'(u)=f_s(u)$. These fields explicitly satisfy the boundary conditions stated above: on $\partial\Lambda_{\mathrm{open}}$, $\nabla u\cdot\mathbf{n}=0$ implies $\mathbf{f}\cdot\mathbf{n}=0$ and $\nabla g_s\cdot\mathbf{n}=\phi_s'(u)\nabla u\cdot\mathbf{n}=0$, while the Dirichlet conditions on the reservoirs reduce to $\phi_s(0)=\rho_{\text{L}}$, $\phi_s(1)=\rho_{\text{R}}$, $\Phi_s(0)=0$, and $\Phi_s(1)=1$.
Since $\nabla^2 u=0$, one has
\begin{equation}
    \nabla^2 g_s=\phi_s''(u)\left|\nabla u\right|^2,\qquad
    \mathbf{f}\cdot\nabla g_s=f_s(u)\phi_s'(u)\left|\nabla u\right|^2,\qquad
    \nabla\cdot\mathbf{f}=f_s'(u)\left|\nabla u\right|^2.
\end{equation}
After dividing by the common factor $\left|\nabla u\right|^2$, Eqs.~(\ref{eq:fconstrd-dim},\ref{eq:closedg-dim}) become
\begin{equation}
    f_s'(u)=s f_s^2(u)\left(2\phi_s(u)-1\right),
\end{equation}
\begin{equation}
    \phi_s''(u)
    -s\left[
    2f_s(u)\phi_s'(u)\left(1-2\phi_s(u)\right)
    +f_s'(u)\phi_s(u)\left(1-\phi_s(u)\right)\right]
    +s^2 f_s^2(u)\phi_s(u)\left(1-\phi_s(u)\right)\left(1-2\phi_s(u)\right)=0,
\end{equation}
which are exactly the one-dimensional equations written in the coordinate $u$. In this sense, the harmonic coordinate collapses the higher-dimensional problem onto the same one-dimensional Gauge Trick construction.
Inserting this form in Eq.~(\ref{eq:cgfcontinuumd-dim}) factorizes the CGF into the one-dimensional functional times the conductance of the harmonic coordinate, so that
\begin{equation}
    \tilde{\lambda}_{d}(s)
    = \mathcal{C}_{\Lambda}\,\tilde{\lambda}_{1\mathrm{d}}(s),\qquad
    \mathcal{C}_{\Lambda}=\int_{\Lambda} d^d\mathbf{x}\,\left|\nabla u(\mathbf{x})\right|^2,
    \qquad
    \tilde{\lambda}_{1\mathrm{d}}(s)
    = \Theta(w_s-1)\left(\operatorname{arccosh} w_s\right)^2
    -\Theta(1-w_s)\left(\arccos w_s\right)^2,
\end{equation}
with $w_s=\sqrt{\left(1+\left(e^s-1\right)\rho_{\text{L}}\right)\left(1+\left(e^{-s}-1\right)\rho_{\text{R}}\right)}$, as in Eq.~(\ref{eq:CGFSSEP}). We thus conclude that, to leading order in the continuum limit, the $d$-dimensional QSSEP between two reservoirs has the same current CGF as the corresponding $d$-dimensional SSEP, with the geometry entering only through the harmonic conductance $\mathcal{C}_{\Lambda}$.


\begin{thebibliography}{91}%
\makeatletter
\providecommand \@ifxundefined [1]{%
 \@ifx{#1\undefined}
}%
\providecommand \@ifnum [1]{%
 \ifnum #1\expandafter \@firstoftwo
 \else \expandafter \@secondoftwo
 \fi
}%
\providecommand \@ifx [1]{%
 \ifx #1\expandafter \@firstoftwo
 \else \expandafter \@secondoftwo
 \fi
}%
\providecommand \natexlab [1]{#1}%
\providecommand \enquote  [1]{``#1''}%
\providecommand \bibnamefont  [1]{#1}%
\providecommand \bibfnamefont [1]{#1}%
\providecommand \citenamefont [1]{#1}%
\providecommand \href@noop [0]{\@secondoftwo}%
\providecommand \href [0]{\begingroup \@sanitize@url \@href}%
\providecommand \@href[1]{\@@startlink{#1}\@@href}%
\providecommand \@@href[1]{\endgroup#1\@@endlink}%
\providecommand \@sanitize@url [0]{\catcode `\\12\catcode `\$12\catcode
  `\&12\catcode `\#12\catcode `\^12\catcode `\_12\catcode `\%12\relax}%
\providecommand \@@startlink[1]{}%
\providecommand \@@endlink[0]{}%
\providecommand \url  [0]{\begingroup\@sanitize@url \@url }%
\providecommand \@url [1]{\endgroup\@href {#1}{\urlprefix }}%
\providecommand \urlprefix  [0]{URL }%
\providecommand \Eprint [0]{\href }%
\providecommand \doibase [0]{https://doi.org/}%
\providecommand \selectlanguage [0]{\@gobble}%
\providecommand \bibinfo  [0]{\@secondoftwo}%
\providecommand \bibfield  [0]{\@secondoftwo}%
\providecommand \translation [1]{[#1]}%
\providecommand \BibitemOpen [0]{}%
\providecommand \bibitemStop [0]{}%
\providecommand \bibitemNoStop [0]{.\EOS\space}%
\providecommand \EOS [0]{\spacefactor3000\relax}%
\providecommand \BibitemShut  [1]{\csname bibitem#1\endcsname}%
\let\auto@bib@innerbib\@empty
\bibitem [{\citenamefont {Eisert}\ \emph {et~al.}(2015)\citenamefont {Eisert},
  \citenamefont {Friesdorf},\ and\ \citenamefont
  {Gogolin}}]{eisert2015quantum}%
  \BibitemOpen
  \bibfield  {author} {\bibinfo {author} {\bibfnamefont {J.}~\bibnamefont
  {Eisert}}, \bibinfo {author} {\bibfnamefont {M.}~\bibnamefont {Friesdorf}},\
  and\ \bibinfo {author} {\bibfnamefont {C.}~\bibnamefont {Gogolin}},\
  }\bibfield  {title} {\bibinfo {title} {Quantum many-body systems out of
  equilibrium},\ }\href@noop {} {\bibfield  {journal} {\bibinfo  {journal}
  {Nature Physics}\ }\textbf {\bibinfo {volume} {11}},\ \bibinfo {pages} {124}
  (\bibinfo {year} {2015})}\BibitemShut {NoStop}%
\bibitem [{\citenamefont {D'Alessio}\ \emph {et~al.}(2016)\citenamefont
  {D'Alessio}, \citenamefont {Kafri}, \citenamefont {Polkovnikov},\ and\
  \citenamefont {Rigol}}]{d2016quantum}%
  \BibitemOpen
  \bibfield  {author} {\bibinfo {author} {\bibfnamefont {L.}~\bibnamefont
  {D'Alessio}}, \bibinfo {author} {\bibfnamefont {Y.}~\bibnamefont {Kafri}},
  \bibinfo {author} {\bibfnamefont {A.}~\bibnamefont {Polkovnikov}},\ and\
  \bibinfo {author} {\bibfnamefont {M.}~\bibnamefont {Rigol}},\ }\bibfield
  {title} {\bibinfo {title} {From quantum chaos and eigenstate thermalization
  to statistical mechanics and thermodynamics},\ }\href@noop {} {\bibfield
  {journal} {\bibinfo  {journal} {Advances in Physics}\ }\textbf {\bibinfo
  {volume} {65}},\ \bibinfo {pages} {239} (\bibinfo {year} {2016})}\BibitemShut
  {NoStop}%
\bibitem [{\citenamefont {Abanin}\ \emph {et~al.}(2019)\citenamefont {Abanin},
  \citenamefont {Altman}, \citenamefont {Bloch},\ and\ \citenamefont
  {Serbyn}}]{abanin2019colloquium}%
  \BibitemOpen
  \bibfield  {author} {\bibinfo {author} {\bibfnamefont {D.~A.}\ \bibnamefont
  {Abanin}}, \bibinfo {author} {\bibfnamefont {E.}~\bibnamefont {Altman}},
  \bibinfo {author} {\bibfnamefont {I.}~\bibnamefont {Bloch}},\ and\ \bibinfo
  {author} {\bibfnamefont {M.}~\bibnamefont {Serbyn}},\ }\bibfield  {title}
  {\bibinfo {title} {Colloquium: Many-body localization, thermalization, and
  entanglement},\ }\href@noop {} {\bibfield  {journal} {\bibinfo  {journal}
  {Reviews of Modern Physics}\ }\textbf {\bibinfo {volume} {91}},\ \bibinfo
  {pages} {021001} (\bibinfo {year} {2019})}\BibitemShut {NoStop}%
\bibitem [{\citenamefont {Tang}\ \emph {et~al.}(2018)\citenamefont {Tang},
  \citenamefont {Kao}, \citenamefont {Li}, \citenamefont {Seo}, \citenamefont
  {Mallayya}, \citenamefont {Rigol}, \citenamefont {Gopalakrishnan},\ and\
  \citenamefont {Lev}}]{tang2018thermalization}%
  \BibitemOpen
  \bibfield  {author} {\bibinfo {author} {\bibfnamefont {Y.}~\bibnamefont
  {Tang}}, \bibinfo {author} {\bibfnamefont {W.}~\bibnamefont {Kao}}, \bibinfo
  {author} {\bibfnamefont {K.-Y.}\ \bibnamefont {Li}}, \bibinfo {author}
  {\bibfnamefont {S.}~\bibnamefont {Seo}}, \bibinfo {author} {\bibfnamefont
  {K.}~\bibnamefont {Mallayya}}, \bibinfo {author} {\bibfnamefont
  {M.}~\bibnamefont {Rigol}}, \bibinfo {author} {\bibfnamefont
  {S.}~\bibnamefont {Gopalakrishnan}},\ and\ \bibinfo {author} {\bibfnamefont
  {B.~L.}\ \bibnamefont {Lev}},\ }\bibfield  {title} {\bibinfo {title}
  {Thermalization near integrability in a dipolar quantum newton’s cradle},\
  }\href@noop {} {\bibfield  {journal} {\bibinfo  {journal} {Physical Review
  X}\ }\textbf {\bibinfo {volume} {8}},\ \bibinfo {pages} {021030} (\bibinfo
  {year} {2018})}\BibitemShut {NoStop}%
\bibitem [{\citenamefont {Schneider}\ \emph {et~al.}(2012)\citenamefont
  {Schneider}, \citenamefont {Hackermüller}, \citenamefont {Ronzheimer},
  \citenamefont {Will}, \citenamefont {Braun}, \citenamefont {Best},
  \citenamefont {Bloch}, \citenamefont {Demler}, \citenamefont {Mandt},
  \citenamefont {Rasch},\ and\ \citenamefont {Rosch}}]{Schneider2012}%
  \BibitemOpen
  \bibfield  {author} {\bibinfo {author} {\bibfnamefont {U.}~\bibnamefont
  {Schneider}}, \bibinfo {author} {\bibfnamefont {L.}~\bibnamefont
  {Hackermüller}}, \bibinfo {author} {\bibfnamefont {J.~P.}\ \bibnamefont
  {Ronzheimer}}, \bibinfo {author} {\bibfnamefont {S.}~\bibnamefont {Will}},
  \bibinfo {author} {\bibfnamefont {S.}~\bibnamefont {Braun}}, \bibinfo
  {author} {\bibfnamefont {T.}~\bibnamefont {Best}}, \bibinfo {author}
  {\bibfnamefont {I.}~\bibnamefont {Bloch}}, \bibinfo {author} {\bibfnamefont
  {E.}~\bibnamefont {Demler}}, \bibinfo {author} {\bibfnamefont
  {S.}~\bibnamefont {Mandt}}, \bibinfo {author} {\bibfnamefont
  {D.}~\bibnamefont {Rasch}},\ and\ \bibinfo {author} {\bibfnamefont
  {A.}~\bibnamefont {Rosch}},\ }\bibfield  {title} {\bibinfo {title} {Fermionic
  transport and out-of-equilibrium dynamics in a homogeneous hubbard model with
  ultracold atoms},\ }\href {https://doi.org/10.1038/nphys2205} {\bibfield
  {journal} {\bibinfo  {journal} {Nature Physics}\ }\textbf {\bibinfo {volume}
  {8}},\ \bibinfo {pages} {213} (\bibinfo {year} {2012})}\BibitemShut {NoStop}%
\bibitem [{\citenamefont {Lebrat}\ \emph {et~al.}(2018)\citenamefont {Lebrat},
  \citenamefont {Grišins}, \citenamefont {Husmann}, \citenamefont {Häusler},
  \citenamefont {Corman}, \citenamefont {Giamarchi}, \citenamefont {Brantut},\
  and\ \citenamefont {Esslinger}}]{Lebrat2018}%
  \BibitemOpen
  \bibfield  {author} {\bibinfo {author} {\bibfnamefont {M.}~\bibnamefont
  {Lebrat}}, \bibinfo {author} {\bibfnamefont {P.}~\bibnamefont {Grišins}},
  \bibinfo {author} {\bibfnamefont {D.}~\bibnamefont {Husmann}}, \bibinfo
  {author} {\bibfnamefont {S.}~\bibnamefont {Häusler}}, \bibinfo {author}
  {\bibfnamefont {L.}~\bibnamefont {Corman}}, \bibinfo {author} {\bibfnamefont
  {T.}~\bibnamefont {Giamarchi}}, \bibinfo {author} {\bibfnamefont {J.-P.}\
  \bibnamefont {Brantut}},\ and\ \bibinfo {author} {\bibfnamefont
  {T.}~\bibnamefont {Esslinger}},\ }\bibfield  {title} {\bibinfo {title} {Band
  and correlated insulators of cold fermions in a mesoscopic lattice},\ }\href
  {https://doi.org/10.1103/PhysRevX.8.011053} {\bibfield  {journal} {\bibinfo
  {journal} {Physical Review X}\ }\textbf {\bibinfo {volume} {8}},\ \bibinfo
  {pages} {011053} (\bibinfo {year} {2018})}\BibitemShut {NoStop}%
\bibitem [{\citenamefont {Rauer}\ \emph {et~al.}(2018)\citenamefont {Rauer},
  \citenamefont {Erne}, \citenamefont {Schweigler}, \citenamefont {Cataldini},
  \citenamefont {Tajik},\ and\ \citenamefont
  {Schmiedmayer}}]{rauer2018recurrences}%
  \BibitemOpen
  \bibfield  {author} {\bibinfo {author} {\bibfnamefont {B.}~\bibnamefont
  {Rauer}}, \bibinfo {author} {\bibfnamefont {S.}~\bibnamefont {Erne}},
  \bibinfo {author} {\bibfnamefont {T.}~\bibnamefont {Schweigler}}, \bibinfo
  {author} {\bibfnamefont {F.}~\bibnamefont {Cataldini}}, \bibinfo {author}
  {\bibfnamefont {M.}~\bibnamefont {Tajik}},\ and\ \bibinfo {author}
  {\bibfnamefont {J.}~\bibnamefont {Schmiedmayer}},\ }\bibfield  {title}
  {\bibinfo {title} {Recurrences in an isolated quantum many-body system},\
  }\href@noop {} {\bibfield  {journal} {\bibinfo  {journal} {Science}\ }\textbf
  {\bibinfo {volume} {360}},\ \bibinfo {pages} {307} (\bibinfo {year}
  {2018})}\BibitemShut {NoStop}%
\bibitem [{\citenamefont {Bloch}\ \emph {et~al.}(2008)\citenamefont {Bloch},
  \citenamefont {Dalibard},\ and\ \citenamefont {Zwerger}}]{bloch2008many}%
  \BibitemOpen
  \bibfield  {author} {\bibinfo {author} {\bibfnamefont {I.}~\bibnamefont
  {Bloch}}, \bibinfo {author} {\bibfnamefont {J.}~\bibnamefont {Dalibard}},\
  and\ \bibinfo {author} {\bibfnamefont {W.}~\bibnamefont {Zwerger}},\
  }\bibfield  {title} {\bibinfo {title} {Many-body physics with ultracold
  gases},\ }\href@noop {} {\bibfield  {journal} {\bibinfo  {journal} {Reviews
  of modern physics}\ }\textbf {\bibinfo {volume} {80}},\ \bibinfo {pages}
  {885} (\bibinfo {year} {2008})}\BibitemShut {NoStop}%
\bibitem [{\citenamefont {Brown}\ \emph {et~al.}(2015)\citenamefont {Brown},
  \citenamefont {Wyllie}, \citenamefont {Koller}, \citenamefont {Goldschmidt},
  \citenamefont {Foss-Feig},\ and\ \citenamefont {Porto}}]{Brown2015}%
  \BibitemOpen
  \bibfield  {author} {\bibinfo {author} {\bibfnamefont {R.~C.}\ \bibnamefont
  {Brown}}, \bibinfo {author} {\bibfnamefont {R.}~\bibnamefont {Wyllie}},
  \bibinfo {author} {\bibfnamefont {S.~B.}\ \bibnamefont {Koller}}, \bibinfo
  {author} {\bibfnamefont {E.~A.}\ \bibnamefont {Goldschmidt}}, \bibinfo
  {author} {\bibfnamefont {M.}~\bibnamefont {Foss-Feig}},\ and\ \bibinfo
  {author} {\bibfnamefont {J.~V.}\ \bibnamefont {Porto}},\ }\bibfield  {title}
  {\bibinfo {title} {Two-dimensional superexchange-mediated magnetization
  dynamics in an optical lattice},\ }\href
  {https://doi.org/10.1126/science.aaa1385} {\bibfield  {journal} {\bibinfo
  {journal} {Science}\ }\textbf {\bibinfo {volume} {348}},\ \bibinfo {pages}
  {540} (\bibinfo {year} {2015})},\ \Eprint
  {https://arxiv.org/abs/https://www.science.org/doi/pdf/10.1126/science.aaa1385}
  {https://www.science.org/doi/pdf/10.1126/science.aaa1385} \BibitemShut
  {NoStop}%
\bibitem [{\citenamefont {Choi}\ \emph {et~al.}(2016)\citenamefont {Choi},
  \citenamefont {Hild}, \citenamefont {Zeiher}, \citenamefont {Schauß},
  \citenamefont {Rubio-Abadal}, \citenamefont {Yefsah}, \citenamefont
  {Khemani}, \citenamefont {Huse}, \citenamefont {Bloch},\ and\ \citenamefont
  {Gross}}]{Choi2016}%
  \BibitemOpen
  \bibfield  {author} {\bibinfo {author} {\bibfnamefont {J.-y.}\ \bibnamefont
  {Choi}}, \bibinfo {author} {\bibfnamefont {S.}~\bibnamefont {Hild}}, \bibinfo
  {author} {\bibfnamefont {J.}~\bibnamefont {Zeiher}}, \bibinfo {author}
  {\bibfnamefont {P.}~\bibnamefont {Schauß}}, \bibinfo {author} {\bibfnamefont
  {A.}~\bibnamefont {Rubio-Abadal}}, \bibinfo {author} {\bibfnamefont
  {T.}~\bibnamefont {Yefsah}}, \bibinfo {author} {\bibfnamefont
  {V.}~\bibnamefont {Khemani}}, \bibinfo {author} {\bibfnamefont {D.~A.}\
  \bibnamefont {Huse}}, \bibinfo {author} {\bibfnamefont {I.}~\bibnamefont
  {Bloch}},\ and\ \bibinfo {author} {\bibfnamefont {C.}~\bibnamefont {Gross}},\
  }\bibfield  {title} {\bibinfo {title} {Exploring the many-body localization
  transition in two dimensions},\ }\href
  {https://doi.org/10.1126/science.aaf8834} {\bibfield  {journal} {\bibinfo
  {journal} {Science}\ }\textbf {\bibinfo {volume} {352}},\ \bibinfo {pages}
  {1547} (\bibinfo {year} {2016})}\BibitemShut {NoStop}%
\bibitem [{\citenamefont {Boll}\ \emph {et~al.}(2016)\citenamefont {Boll},
  \citenamefont {Hilker}, \citenamefont {Salomon}, \citenamefont {Omran},
  \citenamefont {Nespolo}, \citenamefont {Pollet}, \citenamefont {Bloch},\ and\
  \citenamefont {Gross}}]{boll2016spin}%
  \BibitemOpen
  \bibfield  {author} {\bibinfo {author} {\bibfnamefont {M.}~\bibnamefont
  {Boll}}, \bibinfo {author} {\bibfnamefont {T.~A.}\ \bibnamefont {Hilker}},
  \bibinfo {author} {\bibfnamefont {G.}~\bibnamefont {Salomon}}, \bibinfo
  {author} {\bibfnamefont {A.}~\bibnamefont {Omran}}, \bibinfo {author}
  {\bibfnamefont {J.}~\bibnamefont {Nespolo}}, \bibinfo {author} {\bibfnamefont
  {L.}~\bibnamefont {Pollet}}, \bibinfo {author} {\bibfnamefont
  {I.}~\bibnamefont {Bloch}},\ and\ \bibinfo {author} {\bibfnamefont
  {C.}~\bibnamefont {Gross}},\ }\bibfield  {title} {\bibinfo {title} {Spin-and
  density-resolved microscopy of antiferromagnetic correlations in
  fermi-hubbard chains},\ }\href@noop {} {\bibfield  {journal} {\bibinfo
  {journal} {Science}\ }\textbf {\bibinfo {volume} {353}},\ \bibinfo {pages}
  {1257} (\bibinfo {year} {2016})}\BibitemShut {NoStop}%
\bibitem [{\citenamefont {Hild}\ \emph {et~al.}(2014)\citenamefont {Hild},
  \citenamefont {Fukuhara}, \citenamefont {Schau{\ss}}, \citenamefont {Zeiher},
  \citenamefont {Knap}, \citenamefont {Demler}, \citenamefont {Bloch},\ and\
  \citenamefont {Gross}}]{hild2014far}%
  \BibitemOpen
  \bibfield  {author} {\bibinfo {author} {\bibfnamefont {S.}~\bibnamefont
  {Hild}}, \bibinfo {author} {\bibfnamefont {T.}~\bibnamefont {Fukuhara}},
  \bibinfo {author} {\bibfnamefont {P.}~\bibnamefont {Schau{\ss}}}, \bibinfo
  {author} {\bibfnamefont {J.}~\bibnamefont {Zeiher}}, \bibinfo {author}
  {\bibfnamefont {M.}~\bibnamefont {Knap}}, \bibinfo {author} {\bibfnamefont
  {E.}~\bibnamefont {Demler}}, \bibinfo {author} {\bibfnamefont
  {I.}~\bibnamefont {Bloch}},\ and\ \bibinfo {author} {\bibfnamefont
  {C.}~\bibnamefont {Gross}},\ }\bibfield  {title} {\bibinfo {title}
  {Far-from-equilibrium spin transport in heisenberg quantum magnets},\
  }\href@noop {} {\bibfield  {journal} {\bibinfo  {journal} {Physical review
  letters}\ }\textbf {\bibinfo {volume} {113}},\ \bibinfo {pages} {147205}
  (\bibinfo {year} {2014})}\BibitemShut {NoStop}%
\bibitem [{\citenamefont {Scheie}\ \emph {et~al.}(2021)\citenamefont {Scheie},
  \citenamefont {Sherman}, \citenamefont {Dupont}, \citenamefont {Nagler},
  \citenamefont {Stone}, \citenamefont {Granroth}, \citenamefont {Moore},\ and\
  \citenamefont {Tennant}}]{scheie2021detection}%
  \BibitemOpen
  \bibfield  {author} {\bibinfo {author} {\bibfnamefont {A.}~\bibnamefont
  {Scheie}}, \bibinfo {author} {\bibfnamefont {N.}~\bibnamefont {Sherman}},
  \bibinfo {author} {\bibfnamefont {M.}~\bibnamefont {Dupont}}, \bibinfo
  {author} {\bibfnamefont {S.}~\bibnamefont {Nagler}}, \bibinfo {author}
  {\bibfnamefont {M.}~\bibnamefont {Stone}}, \bibinfo {author} {\bibfnamefont
  {G.}~\bibnamefont {Granroth}}, \bibinfo {author} {\bibfnamefont
  {J.}~\bibnamefont {Moore}},\ and\ \bibinfo {author} {\bibfnamefont
  {D.}~\bibnamefont {Tennant}},\ }\bibfield  {title} {\bibinfo {title}
  {Detection of kardar--parisi--zhang hydrodynamics in a quantum heisenberg
  spin-1/2 chain},\ }\href@noop {} {\bibfield  {journal} {\bibinfo  {journal}
  {Nature Physics}\ }\textbf {\bibinfo {volume} {17}},\ \bibinfo {pages} {726}
  (\bibinfo {year} {2021})}\BibitemShut {NoStop}%
\bibitem [{\citenamefont {Gustavsson}\ \emph {et~al.}(2006)\citenamefont
  {Gustavsson}, \citenamefont {Leturcq}, \citenamefont {Simovič},
  \citenamefont {Schleser}, \citenamefont {Ihn}, \citenamefont {Studerus},
  \citenamefont {Ensslin}, \citenamefont {Driscoll},\ and\ \citenamefont
  {Gossard}}]{Gustavsson2006}%
  \BibitemOpen
  \bibfield  {author} {\bibinfo {author} {\bibfnamefont {S.}~\bibnamefont
  {Gustavsson}}, \bibinfo {author} {\bibfnamefont {R.}~\bibnamefont {Leturcq}},
  \bibinfo {author} {\bibfnamefont {B.}~\bibnamefont {Simovič}}, \bibinfo
  {author} {\bibfnamefont {R.}~\bibnamefont {Schleser}}, \bibinfo {author}
  {\bibfnamefont {T.}~\bibnamefont {Ihn}}, \bibinfo {author} {\bibfnamefont
  {P.}~\bibnamefont {Studerus}}, \bibinfo {author} {\bibfnamefont
  {K.}~\bibnamefont {Ensslin}}, \bibinfo {author} {\bibfnamefont {D.~C.}\
  \bibnamefont {Driscoll}},\ and\ \bibinfo {author} {\bibfnamefont {A.~C.}\
  \bibnamefont {Gossard}},\ }\bibfield  {title} {\bibinfo {title} {Counting
  statistics of single electron transport in a quantum dot},\ }\href
  {https://doi.org/10.1103/PhysRevLett.96.076605} {\bibfield  {journal}
  {\bibinfo  {journal} {Physical Review Letters}\ }\textbf {\bibinfo {volume}
  {96}},\ \bibinfo {pages} {076605} (\bibinfo {year} {2006})}\BibitemShut
  {NoStop}%
\bibitem [{\citenamefont {Fujisawa}\ \emph {et~al.}(2006)\citenamefont
  {Fujisawa}, \citenamefont {Hayashi}, \citenamefont {Tomita},\ and\
  \citenamefont {Hirayama}}]{Fujisawa2006}%
  \BibitemOpen
  \bibfield  {author} {\bibinfo {author} {\bibfnamefont {T.}~\bibnamefont
  {Fujisawa}}, \bibinfo {author} {\bibfnamefont {T.}~\bibnamefont {Hayashi}},
  \bibinfo {author} {\bibfnamefont {R.}~\bibnamefont {Tomita}},\ and\ \bibinfo
  {author} {\bibfnamefont {Y.}~\bibnamefont {Hirayama}},\ }\bibfield  {title}
  {\bibinfo {title} {Bidirectional counting of single electrons},\ }\href
  {https://doi.org/10.1126/science.1126788} {\bibfield  {journal} {\bibinfo
  {journal} {Science}\ }\textbf {\bibinfo {volume} {312}},\ \bibinfo {pages}
  {1634} (\bibinfo {year} {2006})}\BibitemShut {NoStop}%
\bibitem [{\citenamefont {Flindt}\ \emph {et~al.}(2009)\citenamefont {Flindt},
  \citenamefont {Fricke}, \citenamefont {Hohls}, \citenamefont {Novotný},
  \citenamefont {Netočný}, \citenamefont {Brandes},\ and\ \citenamefont
  {Haug}}]{Flindt2009}%
  \BibitemOpen
  \bibfield  {author} {\bibinfo {author} {\bibfnamefont {C.}~\bibnamefont
  {Flindt}}, \bibinfo {author} {\bibfnamefont {C.}~\bibnamefont {Fricke}},
  \bibinfo {author} {\bibfnamefont {F.}~\bibnamefont {Hohls}}, \bibinfo
  {author} {\bibfnamefont {T.}~\bibnamefont {Novotný}}, \bibinfo {author}
  {\bibfnamefont {K.}~\bibnamefont {Netočný}}, \bibinfo {author}
  {\bibfnamefont {T.}~\bibnamefont {Brandes}},\ and\ \bibinfo {author}
  {\bibfnamefont {R.~J.}\ \bibnamefont {Haug}},\ }\bibfield  {title} {\bibinfo
  {title} {Universal oscillations in counting statistics},\ }\href
  {https://doi.org/10.1073/pnas.0901002106} {\bibfield  {journal} {\bibinfo
  {journal} {Proceedings of the National Academy of Sciences}\ }\textbf
  {\bibinfo {volume} {106}},\ \bibinfo {pages} {10116} (\bibinfo {year}
  {2009})}\BibitemShut {NoStop}%
\bibitem [{\citenamefont {Bertini}\ \emph {et~al.}(2016)\citenamefont
  {Bertini}, \citenamefont {Collura}, \citenamefont {De~Nardis},\ and\
  \citenamefont {Fagotti}}]{bertini_transport_2016}%
  \BibitemOpen
  \bibfield  {author} {\bibinfo {author} {\bibfnamefont {B.}~\bibnamefont
  {Bertini}}, \bibinfo {author} {\bibfnamefont {M.}~\bibnamefont {Collura}},
  \bibinfo {author} {\bibfnamefont {J.}~\bibnamefont {De~Nardis}},\ and\
  \bibinfo {author} {\bibfnamefont {M.}~\bibnamefont {Fagotti}},\ }\bibfield
  {title} {\bibinfo {title} {Transport in {Out}-of-{Equilibrium} \${XXZ}\$
  {Chains}: {Exact} {Profiles} of {Charges} and {Currents}},\ }\href
  {https://doi.org/10.1103/PhysRevLett.117.207201} {\bibfield  {journal}
  {\bibinfo  {journal} {Physical Review Letters}\ }\textbf {\bibinfo {volume}
  {117}},\ \bibinfo {pages} {207201} (\bibinfo {year} {2016})}\BibitemShut
  {NoStop}%
\bibitem [{\citenamefont {Castro-Alvaredo}\ \emph {et~al.}(2016)\citenamefont
  {Castro-Alvaredo}, \citenamefont {Doyon},\ and\ \citenamefont
  {Yoshimura}}]{PhysRevX.6.041065}%
  \BibitemOpen
  \bibfield  {author} {\bibinfo {author} {\bibfnamefont {O.~A.}\ \bibnamefont
  {Castro-Alvaredo}}, \bibinfo {author} {\bibfnamefont {B.}~\bibnamefont
  {Doyon}},\ and\ \bibinfo {author} {\bibfnamefont {T.}~\bibnamefont
  {Yoshimura}},\ }\bibfield  {title} {\bibinfo {title} {Emergent hydrodynamics
  in integrable quantum systems out of equilibrium},\ }\href
  {https://doi.org/10.1103/PhysRevX.6.041065} {\bibfield  {journal} {\bibinfo
  {journal} {Phys. Rev. X}\ }\textbf {\bibinfo {volume} {6}},\ \bibinfo {pages}
  {041065} (\bibinfo {year} {2016})}\BibitemShut {NoStop}%
\bibitem [{\citenamefont {De~Nardis}\ \emph {et~al.}(2018)\citenamefont
  {De~Nardis}, \citenamefont {Bernard},\ and\ \citenamefont
  {Doyon}}]{PhysRevLett.121.160603}%
  \BibitemOpen
  \bibfield  {author} {\bibinfo {author} {\bibfnamefont {J.}~\bibnamefont
  {De~Nardis}}, \bibinfo {author} {\bibfnamefont {D.}~\bibnamefont {Bernard}},\
  and\ \bibinfo {author} {\bibfnamefont {B.}~\bibnamefont {Doyon}},\ }\bibfield
   {title} {\bibinfo {title} {Hydrodynamic diffusion in integrable systems},\
  }\href {https://doi.org/10.1103/PhysRevLett.121.160603} {\bibfield  {journal}
  {\bibinfo  {journal} {Phys. Rev. Lett.}\ }\textbf {\bibinfo {volume} {121}},\
  \bibinfo {pages} {160603} (\bibinfo {year} {2018})}\BibitemShut {NoStop}%
\bibitem [{\citenamefont {Schemmer}\ \emph {et~al.}(2019)\citenamefont
  {Schemmer}, \citenamefont {Bouchoule}, \citenamefont {Doyon},\ and\
  \citenamefont {Dubail}}]{schemmer2019generalized}%
  \BibitemOpen
  \bibfield  {author} {\bibinfo {author} {\bibfnamefont {M.}~\bibnamefont
  {Schemmer}}, \bibinfo {author} {\bibfnamefont {I.}~\bibnamefont {Bouchoule}},
  \bibinfo {author} {\bibfnamefont {B.}~\bibnamefont {Doyon}},\ and\ \bibinfo
  {author} {\bibfnamefont {J.}~\bibnamefont {Dubail}},\ }\bibfield  {title}
  {\bibinfo {title} {Generalized hydrodynamics on an atom chip},\ }\href@noop
  {} {\bibfield  {journal} {\bibinfo  {journal} {Physical review letters}\
  }\textbf {\bibinfo {volume} {122}},\ \bibinfo {pages} {090601} (\bibinfo
  {year} {2019})}\BibitemShut {NoStop}%
\bibitem [{\citenamefont {Bastianello}\ \emph {et~al.}(2020)\citenamefont
  {Bastianello}, \citenamefont {De~Luca}, \citenamefont {Doyon},\ and\
  \citenamefont {De~Nardis}}]{bastianello2020thermalization}%
  \BibitemOpen
  \bibfield  {author} {\bibinfo {author} {\bibfnamefont {A.}~\bibnamefont
  {Bastianello}}, \bibinfo {author} {\bibfnamefont {A.}~\bibnamefont
  {De~Luca}}, \bibinfo {author} {\bibfnamefont {B.}~\bibnamefont {Doyon}},\
  and\ \bibinfo {author} {\bibfnamefont {J.}~\bibnamefont {De~Nardis}},\
  }\bibfield  {title} {\bibinfo {title} {Thermalization of a trapped
  one-dimensional bose gas via diffusion},\ }\href@noop {} {\bibfield
  {journal} {\bibinfo  {journal} {Physical Review Letters}\ }\textbf {\bibinfo
  {volume} {125}},\ \bibinfo {pages} {240604} (\bibinfo {year}
  {2020})}\BibitemShut {NoStop}%
\bibitem [{\citenamefont {Doyon}\ \emph {et~al.}(2025)\citenamefont {Doyon},
  \citenamefont {Gopalakrishnan}, \citenamefont {M{\o}ller}, \citenamefont
  {Schmiedmayer},\ and\ \citenamefont {Vasseur}}]{doyon2025generalized}%
  \BibitemOpen
  \bibfield  {author} {\bibinfo {author} {\bibfnamefont {B.}~\bibnamefont
  {Doyon}}, \bibinfo {author} {\bibfnamefont {S.}~\bibnamefont
  {Gopalakrishnan}}, \bibinfo {author} {\bibfnamefont {F.}~\bibnamefont
  {M{\o}ller}}, \bibinfo {author} {\bibfnamefont {J.}~\bibnamefont
  {Schmiedmayer}},\ and\ \bibinfo {author} {\bibfnamefont {R.}~\bibnamefont
  {Vasseur}},\ }\bibfield  {title} {\bibinfo {title} {Generalized
  hydrodynamics: a perspective},\ }\href@noop {} {\bibfield  {journal}
  {\bibinfo  {journal} {Physical Review X}\ }\textbf {\bibinfo {volume} {15}},\
  \bibinfo {pages} {010501} (\bibinfo {year} {2025})}\BibitemShut {NoStop}%
\bibitem [{\citenamefont {Gallavotti}\ and\ \citenamefont
  {Cohen}(1995)}]{gallavotti1995dynamical}%
  \BibitemOpen
  \bibfield  {author} {\bibinfo {author} {\bibfnamefont {G.}~\bibnamefont
  {Gallavotti}}\ and\ \bibinfo {author} {\bibfnamefont {E.~G.~D.}\ \bibnamefont
  {Cohen}},\ }\bibfield  {title} {\bibinfo {title} {Dynamical ensembles in
  nonequilibrium statistical mechanics},\ }\href@noop {} {\bibfield  {journal}
  {\bibinfo  {journal} {Physical review letters}\ }\textbf {\bibinfo {volume}
  {74}},\ \bibinfo {pages} {2694} (\bibinfo {year} {1995})}\BibitemShut
  {NoStop}%
\bibitem [{\citenamefont {Crooks}(1999)}]{crooks1999entropy}%
  \BibitemOpen
  \bibfield  {author} {\bibinfo {author} {\bibfnamefont {G.~E.}\ \bibnamefont
  {Crooks}},\ }\bibfield  {title} {\bibinfo {title} {Entropy production
  fluctuation theorem and the nonequilibrium work relation for free energy
  differences},\ }\href@noop {} {\bibfield  {journal} {\bibinfo  {journal}
  {Physical Review E}\ }\textbf {\bibinfo {volume} {60}},\ \bibinfo {pages}
  {2721} (\bibinfo {year} {1999})}\BibitemShut {NoStop}%
\bibitem [{\citenamefont {Maes}(1999)}]{maes1999fluctuation}%
  \BibitemOpen
  \bibfield  {author} {\bibinfo {author} {\bibfnamefont {C.}~\bibnamefont
  {Maes}},\ }\bibfield  {title} {\bibinfo {title} {The fluctuation theorem as a
  gibbs property},\ }\href@noop {} {\bibfield  {journal} {\bibinfo  {journal}
  {Journal of statistical physics}\ }\textbf {\bibinfo {volume} {95}},\
  \bibinfo {pages} {367} (\bibinfo {year} {1999})}\BibitemShut {NoStop}%
\bibitem [{\citenamefont {Krajnik}\ \emph {et~al.}(2022)\citenamefont
  {Krajnik}, \citenamefont {Schmidt}, \citenamefont {Pasquier}, \citenamefont
  {Ilievski},\ and\ \citenamefont {Prosen}}]{krajnik2022exact}%
  \BibitemOpen
  \bibfield  {author} {\bibinfo {author} {\bibfnamefont {{\v{Z}}.}~\bibnamefont
  {Krajnik}}, \bibinfo {author} {\bibfnamefont {J.}~\bibnamefont {Schmidt}},
  \bibinfo {author} {\bibfnamefont {V.}~\bibnamefont {Pasquier}}, \bibinfo
  {author} {\bibfnamefont {E.}~\bibnamefont {Ilievski}},\ and\ \bibinfo
  {author} {\bibfnamefont {T.}~\bibnamefont {Prosen}},\ }\bibfield  {title}
  {\bibinfo {title} {Exact anomalous current fluctuations in a deterministic
  interacting model},\ }\href@noop {} {\bibfield  {journal} {\bibinfo
  {journal} {Physical Review Letters}\ }\textbf {\bibinfo {volume} {128}},\
  \bibinfo {pages} {160601} (\bibinfo {year} {2022})}\BibitemShut {NoStop}%
\bibitem [{\citenamefont {Gopalakrishnan}\ \emph {et~al.}(2024)\citenamefont
  {Gopalakrishnan}, \citenamefont {McCulloch},\ and\ \citenamefont
  {Vasseur}}]{gopalakrishnan2024non}%
  \BibitemOpen
  \bibfield  {author} {\bibinfo {author} {\bibfnamefont {S.}~\bibnamefont
  {Gopalakrishnan}}, \bibinfo {author} {\bibfnamefont {E.}~\bibnamefont
  {McCulloch}},\ and\ \bibinfo {author} {\bibfnamefont {R.}~\bibnamefont
  {Vasseur}},\ }\bibfield  {title} {\bibinfo {title} {Non-gaussian diffusive
  fluctuations in dirac fluids},\ }\href@noop {} {\bibfield  {journal}
  {\bibinfo  {journal} {Proceedings of the National Academy of Sciences}\
  }\textbf {\bibinfo {volume} {121}},\ \bibinfo {pages} {e2403327121} (\bibinfo
  {year} {2024})}\BibitemShut {NoStop}%
\bibitem [{\citenamefont {Gopalakrishnan}\ and\ \citenamefont
  {Vasseur}(2019)}]{gopalakrishnan2019kinetic}%
  \BibitemOpen
  \bibfield  {author} {\bibinfo {author} {\bibfnamefont {S.}~\bibnamefont
  {Gopalakrishnan}}\ and\ \bibinfo {author} {\bibfnamefont {R.}~\bibnamefont
  {Vasseur}},\ }\bibfield  {title} {\bibinfo {title} {Kinetic theory of spin
  diffusion and superdiffusion in xxz spin chains},\ }\href@noop {} {\bibfield
  {journal} {\bibinfo  {journal} {Physical review letters}\ }\textbf {\bibinfo
  {volume} {122}},\ \bibinfo {pages} {127202} (\bibinfo {year}
  {2019})}\BibitemShut {NoStop}%
\bibitem [{\citenamefont {Gardiner}\ and\ \citenamefont
  {Zoller}(2017)}]{gardiner2017quantum}%
  \BibitemOpen
  \bibfield  {author} {\bibinfo {author} {\bibfnamefont {C.~W.}\ \bibnamefont
  {Gardiner}}\ and\ \bibinfo {author} {\bibfnamefont {P.}~\bibnamefont
  {Zoller}},\ }\href@noop {} {\emph {\bibinfo {title} {Quantum World Of
  Ultra-cold Atoms And Light, The-Book Iii: Ultra-cold Atoms}}},\ Vol.~\bibinfo
  {volume} {5}\ (\bibinfo  {publisher} {World Scientific},\ \bibinfo {year}
  {2017})\BibitemShut {NoStop}%
\bibitem [{\citenamefont {{\v Z}nidari{\v
  c}}(2010)}]{znidaric_dephasing-induced_2010}%
  \BibitemOpen
  \bibfield  {author} {\bibinfo {author} {\bibfnamefont {M.}~\bibnamefont {{\v
  Z}nidari{\v c}}},\ }\bibfield  {title} {{\selectlanguage {en}\bibinfo {title}
  {Dephasing-induced diffusive transport in the anisotropic {Heisenberg}
  model}},\ }\href {https://doi.org/10.1088/1367-2630/12/4/043001} {\bibfield
  {journal} {\bibinfo  {journal} {New Journal of Physics}\ }\textbf {\bibinfo
  {volume} {12}},\ \bibinfo {pages} {043001} (\bibinfo {year} {2010})},\
  \bibinfo {note} {publisher: IOP Publishing}\BibitemShut {NoStop}%
\bibitem [{\citenamefont {Carollo}\ \emph {et~al.}(2017)\citenamefont
  {Carollo}, \citenamefont {Garrahan}, \citenamefont {Lesanovsky},\ and\
  \citenamefont {P\'erez-Espigares}}]{PhysRevE.96.052118}%
  \BibitemOpen
  \bibfield  {author} {\bibinfo {author} {\bibfnamefont {F.}~\bibnamefont
  {Carollo}}, \bibinfo {author} {\bibfnamefont {J.~P.}\ \bibnamefont
  {Garrahan}}, \bibinfo {author} {\bibfnamefont {I.}~\bibnamefont
  {Lesanovsky}},\ and\ \bibinfo {author} {\bibfnamefont {C.}~\bibnamefont
  {P\'erez-Espigares}},\ }\bibfield  {title} {\bibinfo {title} {Fluctuating
  hydrodynamics, current fluctuations, and hyperuniformity in boundary-driven
  open quantum chains},\ }\href {https://doi.org/10.1103/PhysRevE.96.052118}
  {\bibfield  {journal} {\bibinfo  {journal} {Phys. Rev. E}\ }\textbf {\bibinfo
  {volume} {96}},\ \bibinfo {pages} {052118} (\bibinfo {year}
  {2017})}\BibitemShut {NoStop}%
\bibitem [{\citenamefont {Knap}(2018)}]{knap2018entanglement}%
  \BibitemOpen
  \bibfield  {author} {\bibinfo {author} {\bibfnamefont {M.}~\bibnamefont
  {Knap}},\ }\bibfield  {title} {\bibinfo {title} {Entanglement production and
  information scrambling in a noisy spin system},\ }\href@noop {} {\bibfield
  {journal} {\bibinfo  {journal} {Physical Review B}\ }\textbf {\bibinfo
  {volume} {98}},\ \bibinfo {pages} {184416} (\bibinfo {year}
  {2018})}\BibitemShut {NoStop}%
\bibitem [{\citenamefont {Fisher}\ \emph {et~al.}(2023)\citenamefont {Fisher},
  \citenamefont {Khemani}, \citenamefont {Nahum},\ and\ \citenamefont
  {Vijay}}]{annurev:/content/journals/10.1146/annurev-conmatphys-031720-030658}%
  \BibitemOpen
  \bibfield  {author} {\bibinfo {author} {\bibfnamefont {M.~P.}\ \bibnamefont
  {Fisher}}, \bibinfo {author} {\bibfnamefont {V.}~\bibnamefont {Khemani}},
  \bibinfo {author} {\bibfnamefont {A.}~\bibnamefont {Nahum}},\ and\ \bibinfo
  {author} {\bibfnamefont {S.}~\bibnamefont {Vijay}},\ }\bibfield  {title}
  {\bibinfo {title} {Random quantum circuits},\ }\href
  {https://doi.org/https://doi.org/10.1146/annurev-conmatphys-031720-030658}
  {\bibfield  {journal} {\bibinfo  {journal} {Annual Review of Condensed Matter
  Physics}\ }\textbf {\bibinfo {volume} {14}},\ \bibinfo {pages} {335}
  (\bibinfo {year} {2023})}\BibitemShut {NoStop}%
\bibitem [{\citenamefont {Bauer}\ \emph {et~al.}(2017)\citenamefont {Bauer},
  \citenamefont {Bernard},\ and\ \citenamefont
  {Jin}}]{10.21468/SciPostPhys.3.5.033}%
  \BibitemOpen
  \bibfield  {author} {\bibinfo {author} {\bibfnamefont {M.}~\bibnamefont
  {Bauer}}, \bibinfo {author} {\bibfnamefont {D.}~\bibnamefont {Bernard}},\
  and\ \bibinfo {author} {\bibfnamefont {T.}~\bibnamefont {Jin}},\ }\bibfield
  {title} {\bibinfo {title} {{Stochastic dissipative quantum spin chains (I) :
  Quantum fluctuating discrete hydrodynamics}},\ }\href
  {https://doi.org/10.21468/SciPostPhys.3.5.033} {\bibfield  {journal}
  {\bibinfo  {journal} {SciPost Phys.}\ }\textbf {\bibinfo {volume} {3}},\
  \bibinfo {pages} {033} (\bibinfo {year} {2017})}\BibitemShut {NoStop}%
\bibitem [{\citenamefont {Rowlands}\ and\ \citenamefont
  {Lamacraft}(2018)}]{rowlands2018}%
  \BibitemOpen
  \bibfield  {author} {\bibinfo {author} {\bibfnamefont {D.~A.}\ \bibnamefont
  {Rowlands}}\ and\ \bibinfo {author} {\bibfnamefont {A.}~\bibnamefont
  {Lamacraft}},\ }\bibfield  {title} {\bibinfo {title} {Noisy spins and the
  {Richardson-Gaudin} model},\ }\href
  {http://dx.doi.org/10.1103/PhysRevLett.120.090401} {\bibfield  {journal}
  {\bibinfo  {journal} {Phys. Rev. Lett.}\ }\textbf {\bibinfo {volume} {120}},\
  \bibinfo {pages} {090401} (\bibinfo {year} {2018})}\BibitemShut {NoStop}%
\bibitem [{\citenamefont {Christopoulos}\ \emph {et~al.}(2023)\citenamefont
  {Christopoulos}, \citenamefont {Le~Doussal}, \citenamefont {Bernard},\ and\
  \citenamefont {De~Luca}}]{christopoulos2023universal}%
  \BibitemOpen
  \bibfield  {author} {\bibinfo {author} {\bibfnamefont {A.}~\bibnamefont
  {Christopoulos}}, \bibinfo {author} {\bibfnamefont {P.}~\bibnamefont
  {Le~Doussal}}, \bibinfo {author} {\bibfnamefont {D.}~\bibnamefont
  {Bernard}},\ and\ \bibinfo {author} {\bibfnamefont {A.}~\bibnamefont
  {De~Luca}},\ }\bibfield  {title} {\bibinfo {title} {Universal
  out-of-equilibrium dynamics of 1d critical quantum systems perturbed by noise
  coupled to energy},\ }\href@noop {} {\bibfield  {journal} {\bibinfo
  {journal} {Physical Review X}\ }\textbf {\bibinfo {volume} {13}},\ \bibinfo
  {pages} {011043} (\bibinfo {year} {2023})}\BibitemShut {NoStop}%
\bibitem [{\citenamefont {Swann}\ \emph {et~al.}(2023)\citenamefont {Swann},
  \citenamefont {Bernard},\ and\ \citenamefont {Nahum}}]{swann2023spacetime}%
  \BibitemOpen
  \bibfield  {author} {\bibinfo {author} {\bibfnamefont {T.}~\bibnamefont
  {Swann}}, \bibinfo {author} {\bibfnamefont {D.}~\bibnamefont {Bernard}},\
  and\ \bibinfo {author} {\bibfnamefont {A.}~\bibnamefont {Nahum}},\ }\bibfield
   {title} {\bibinfo {title} {Spacetime picture for entanglement generation in
  noisy fermion chains},\ }\href@noop {} {\bibfield  {journal} {\bibinfo
  {journal} {arXiv preprint arXiv:2302.12212}\ } (\bibinfo {year}
  {2023})}\BibitemShut {NoStop}%
\bibitem [{\citenamefont {Nahum}\ \emph {et~al.}(2017)\citenamefont {Nahum},
  \citenamefont {Ruhman}, \citenamefont {Vijay},\ and\ \citenamefont
  {Haah}}]{nahum}%
  \BibitemOpen
  \bibfield  {author} {\bibinfo {author} {\bibfnamefont {A.}~\bibnamefont
  {Nahum}}, \bibinfo {author} {\bibfnamefont {J.}~\bibnamefont {Ruhman}},
  \bibinfo {author} {\bibfnamefont {S.}~\bibnamefont {Vijay}},\ and\ \bibinfo
  {author} {\bibfnamefont {J.}~\bibnamefont {Haah}},\ }\bibfield  {title}
  {\bibinfo {title} {Quantum entanglement growth under random unitary
  dynamics},\ }\href@noop {} {\bibfield  {journal} {\bibinfo  {journal}
  {Physical Review X}\ }\textbf {\bibinfo {volume} {7}},\ \bibinfo {pages}
  {031016} (\bibinfo {year} {2017})}\BibitemShut {NoStop}%
\bibitem [{\citenamefont {Nahum}\ \emph {et~al.}(2018)\citenamefont {Nahum},
  \citenamefont {Vijay},\ and\ \citenamefont {Haah}}]{nahum2018operator}%
  \BibitemOpen
  \bibfield  {author} {\bibinfo {author} {\bibfnamefont {A.}~\bibnamefont
  {Nahum}}, \bibinfo {author} {\bibfnamefont {S.}~\bibnamefont {Vijay}},\ and\
  \bibinfo {author} {\bibfnamefont {J.}~\bibnamefont {Haah}},\ }\bibfield
  {title} {\bibinfo {title} {Operator spreading in random unitary circuits},\
  }\href@noop {} {\bibfield  {journal} {\bibinfo  {journal} {Physical Review
  X}\ }\textbf {\bibinfo {volume} {8}},\ \bibinfo {pages} {021014} (\bibinfo
  {year} {2018})}\BibitemShut {NoStop}%
\bibitem [{\citenamefont {Chan}\ \emph
  {et~al.}(2018{\natexlab{a}})\citenamefont {Chan}, \citenamefont {De~Luca},\
  and\ \citenamefont {Chalker}}]{chan2018solution}%
  \BibitemOpen
  \bibfield  {author} {\bibinfo {author} {\bibfnamefont {A.}~\bibnamefont
  {Chan}}, \bibinfo {author} {\bibfnamefont {A.}~\bibnamefont {De~Luca}},\ and\
  \bibinfo {author} {\bibfnamefont {J.~T.}\ \bibnamefont {Chalker}},\
  }\bibfield  {title} {\bibinfo {title} {Solution of a minimal model for
  many-body quantum chaos},\ }\href@noop {} {\bibfield  {journal} {\bibinfo
  {journal} {Physical Review X}\ }\textbf {\bibinfo {volume} {8}},\ \bibinfo
  {pages} {041019} (\bibinfo {year} {2018}{\natexlab{a}})}\BibitemShut
  {NoStop}%
\bibitem [{\citenamefont {Chan}\ \emph
  {et~al.}(2018{\natexlab{b}})\citenamefont {Chan}, \citenamefont {De~Luca},\
  and\ \citenamefont {Chalker}}]{chan2018spectral}%
  \BibitemOpen
  \bibfield  {author} {\bibinfo {author} {\bibfnamefont {A.}~\bibnamefont
  {Chan}}, \bibinfo {author} {\bibfnamefont {A.}~\bibnamefont {De~Luca}},\ and\
  \bibinfo {author} {\bibfnamefont {J.}~\bibnamefont {Chalker}},\ }\bibfield
  {title} {\bibinfo {title} {Spectral statistics in spatially extended chaotic
  quantum many-body systems},\ }\href@noop {} {\bibfield  {journal} {\bibinfo
  {journal} {Physical review letters}\ }\textbf {\bibinfo {volume} {121}},\
  \bibinfo {pages} {060601} (\bibinfo {year} {2018}{\natexlab{b}})}\BibitemShut
  {NoStop}%
\bibitem [{\citenamefont {Friedman}\ \emph {et~al.}(2019)\citenamefont
  {Friedman}, \citenamefont {Chan}, \citenamefont {De~Luca},\ and\
  \citenamefont {Chalker}}]{friedman2019spectral}%
  \BibitemOpen
  \bibfield  {author} {\bibinfo {author} {\bibfnamefont {A.~J.}\ \bibnamefont
  {Friedman}}, \bibinfo {author} {\bibfnamefont {A.}~\bibnamefont {Chan}},
  \bibinfo {author} {\bibfnamefont {A.}~\bibnamefont {De~Luca}},\ and\ \bibinfo
  {author} {\bibfnamefont {J.}~\bibnamefont {Chalker}},\ }\bibfield  {title}
  {\bibinfo {title} {Spectral statistics and many-body quantum chaos with
  conserved charge},\ }\href@noop {} {\bibfield  {journal} {\bibinfo  {journal}
  {Physical Review Letters}\ }\textbf {\bibinfo {volume} {123}},\ \bibinfo
  {pages} {210603} (\bibinfo {year} {2019})}\BibitemShut {NoStop}%
\bibitem [{\citenamefont {Shivam}\ \emph {et~al.}(2023)\citenamefont {Shivam},
  \citenamefont {De~Luca}, \citenamefont {Huse},\ and\ \citenamefont
  {Chan}}]{shivam2023many}%
  \BibitemOpen
  \bibfield  {author} {\bibinfo {author} {\bibfnamefont {S.}~\bibnamefont
  {Shivam}}, \bibinfo {author} {\bibfnamefont {A.}~\bibnamefont {De~Luca}},
  \bibinfo {author} {\bibfnamefont {D.~A.}\ \bibnamefont {Huse}},\ and\
  \bibinfo {author} {\bibfnamefont {A.}~\bibnamefont {Chan}},\ }\bibfield
  {title} {\bibinfo {title} {Many-body quantum chaos and emergence of ginibre
  ensemble},\ }\href@noop {} {\bibfield  {journal} {\bibinfo  {journal}
  {Physical review letters}\ }\textbf {\bibinfo {volume} {130}},\ \bibinfo
  {pages} {140403} (\bibinfo {year} {2023})}\BibitemShut {NoStop}%
\bibitem [{\citenamefont {Chan}\ \emph {et~al.}(2022)\citenamefont {Chan},
  \citenamefont {Shivam}, \citenamefont {Huse},\ and\ \citenamefont
  {De~Luca}}]{chan2022many}%
  \BibitemOpen
  \bibfield  {author} {\bibinfo {author} {\bibfnamefont {A.}~\bibnamefont
  {Chan}}, \bibinfo {author} {\bibfnamefont {S.}~\bibnamefont {Shivam}},
  \bibinfo {author} {\bibfnamefont {D.~A.}\ \bibnamefont {Huse}},\ and\
  \bibinfo {author} {\bibfnamefont {A.}~\bibnamefont {De~Luca}},\ }\bibfield
  {title} {\bibinfo {title} {Many-body quantum chaos and space-time
  translational invariance},\ }\href@noop {} {\bibfield  {journal} {\bibinfo
  {journal} {Nature communications}\ }\textbf {\bibinfo {volume} {13}},\
  \bibinfo {pages} {7484} (\bibinfo {year} {2022})}\BibitemShut {NoStop}%
\bibitem [{\citenamefont {Bertini}\ \emph {et~al.}(2018)\citenamefont
  {Bertini}, \citenamefont {Kos},\ and\ \citenamefont
  {Prosen}}]{bertini2018exact}%
  \BibitemOpen
  \bibfield  {author} {\bibinfo {author} {\bibfnamefont {B.}~\bibnamefont
  {Bertini}}, \bibinfo {author} {\bibfnamefont {P.}~\bibnamefont {Kos}},\ and\
  \bibinfo {author} {\bibfnamefont {T.}~\bibnamefont {Prosen}},\ }\bibfield
  {title} {\bibinfo {title} {Exact spectral form factor in a minimal model of
  many-body quantum chaos},\ }\href@noop {} {\bibfield  {journal} {\bibinfo
  {journal} {Physical review letters}\ }\textbf {\bibinfo {volume} {121}},\
  \bibinfo {pages} {264101} (\bibinfo {year} {2018})}\BibitemShut {NoStop}%
\bibitem [{\citenamefont {Bertini}\ \emph {et~al.}(2019)\citenamefont
  {Bertini}, \citenamefont {Kos},\ and\ \citenamefont
  {Prosen}}]{bertini2019entanglement}%
  \BibitemOpen
  \bibfield  {author} {\bibinfo {author} {\bibfnamefont {B.}~\bibnamefont
  {Bertini}}, \bibinfo {author} {\bibfnamefont {P.}~\bibnamefont {Kos}},\ and\
  \bibinfo {author} {\bibfnamefont {T.}~\bibnamefont {Prosen}},\ }\bibfield
  {title} {\bibinfo {title} {Entanglement spreading in a minimal model of
  maximal many-body quantum chaos},\ }\href@noop {} {\bibfield  {journal}
  {\bibinfo  {journal} {Physical Review X}\ }\textbf {\bibinfo {volume} {9}},\
  \bibinfo {pages} {021033} (\bibinfo {year} {2019})}\BibitemShut {NoStop}%
\bibitem [{\citenamefont {Mezei}(2018)}]{mezei2018membrane}%
  \BibitemOpen
  \bibfield  {author} {\bibinfo {author} {\bibfnamefont {M.}~\bibnamefont
  {Mezei}},\ }\bibfield  {title} {\bibinfo {title} {Membrane theory of
  entanglement dynamics from holography},\ }\href@noop {} {\bibfield  {journal}
  {\bibinfo  {journal} {Physical Review D}\ }\textbf {\bibinfo {volume} {98}},\
  \bibinfo {pages} {106025} (\bibinfo {year} {2018})}\BibitemShut {NoStop}%
\bibitem [{\citenamefont {Zhou}\ and\ \citenamefont
  {Nahum}(2019)}]{zhou2019emergent}%
  \BibitemOpen
  \bibfield  {author} {\bibinfo {author} {\bibfnamefont {T.}~\bibnamefont
  {Zhou}}\ and\ \bibinfo {author} {\bibfnamefont {A.}~\bibnamefont {Nahum}},\
  }\bibfield  {title} {\bibinfo {title} {Emergent statistical mechanics of
  entanglement in random unitary circuits},\ }\href@noop {} {\bibfield
  {journal} {\bibinfo  {journal} {Physical Review B}\ }\textbf {\bibinfo
  {volume} {99}},\ \bibinfo {pages} {174205} (\bibinfo {year}
  {2019})}\BibitemShut {NoStop}%
\bibitem [{\citenamefont {Zhou}\ and\ \citenamefont
  {Nahum}(2020)}]{zhou2020entanglement}%
  \BibitemOpen
  \bibfield  {author} {\bibinfo {author} {\bibfnamefont {T.}~\bibnamefont
  {Zhou}}\ and\ \bibinfo {author} {\bibfnamefont {A.}~\bibnamefont {Nahum}},\
  }\bibfield  {title} {\bibinfo {title} {Entanglement membrane in chaotic
  many-body systems},\ }\href@noop {} {\bibfield  {journal} {\bibinfo
  {journal} {Physical Review X}\ }\textbf {\bibinfo {volume} {10}},\ \bibinfo
  {pages} {031066} (\bibinfo {year} {2020})}\BibitemShut {NoStop}%
\bibitem [{\citenamefont {Gopalakrishnan}\ \emph {et~al.}(2018)\citenamefont
  {Gopalakrishnan}, \citenamefont {Huse}, \citenamefont {Khemani},\ and\
  \citenamefont {Vasseur}}]{gopalakrishnan2018hydrodynamics}%
  \BibitemOpen
  \bibfield  {author} {\bibinfo {author} {\bibfnamefont {S.}~\bibnamefont
  {Gopalakrishnan}}, \bibinfo {author} {\bibfnamefont {D.~A.}\ \bibnamefont
  {Huse}}, \bibinfo {author} {\bibfnamefont {V.}~\bibnamefont {Khemani}},\ and\
  \bibinfo {author} {\bibfnamefont {R.}~\bibnamefont {Vasseur}},\ }\bibfield
  {title} {\bibinfo {title} {Hydrodynamics of operator spreading and
  quasiparticle diffusion in interacting integrable systems},\ }\href@noop {}
  {\bibfield  {journal} {\bibinfo  {journal} {Physical Review B}\ }\textbf
  {\bibinfo {volume} {98}},\ \bibinfo {pages} {220303} (\bibinfo {year}
  {2018})}\BibitemShut {NoStop}%
\bibitem [{\citenamefont {Chan}\ \emph {et~al.}(2019)\citenamefont {Chan},
  \citenamefont {De~Luca},\ and\ \citenamefont {Chalker}}]{chan2019eigenstate}%
  \BibitemOpen
  \bibfield  {author} {\bibinfo {author} {\bibfnamefont {A.}~\bibnamefont
  {Chan}}, \bibinfo {author} {\bibfnamefont {A.}~\bibnamefont {De~Luca}},\ and\
  \bibinfo {author} {\bibfnamefont {J.}~\bibnamefont {Chalker}},\ }\bibfield
  {title} {\bibinfo {title} {Eigenstate correlations, thermalization, and the
  butterfly effect},\ }\href@noop {} {\bibfield  {journal} {\bibinfo  {journal}
  {Physical Review Letters}\ }\textbf {\bibinfo {volume} {122}},\ \bibinfo
  {pages} {220601} (\bibinfo {year} {2019})}\BibitemShut {NoStop}%
\bibitem [{\citenamefont {Xu}\ and\ \citenamefont
  {Swingle}(2019)}]{xu2019locality}%
  \BibitemOpen
  \bibfield  {author} {\bibinfo {author} {\bibfnamefont {S.}~\bibnamefont
  {Xu}}\ and\ \bibinfo {author} {\bibfnamefont {B.}~\bibnamefont {Swingle}},\
  }\bibfield  {title} {\bibinfo {title} {Locality, quantum fluctuations, and
  scrambling},\ }\href@noop {} {\bibfield  {journal} {\bibinfo  {journal}
  {Physical Review X}\ }\textbf {\bibinfo {volume} {9}},\ \bibinfo {pages}
  {031048} (\bibinfo {year} {2019})}\BibitemShut {NoStop}%
\bibitem [{\citenamefont {Bernard}\ and\ \citenamefont
  {Doyon}(2016)}]{Bernard_2016}%
  \BibitemOpen
  \bibfield  {author} {\bibinfo {author} {\bibfnamefont {D.}~\bibnamefont
  {Bernard}}\ and\ \bibinfo {author} {\bibfnamefont {B.}~\bibnamefont
  {Doyon}},\ }\bibfield  {title} {\bibinfo {title} {Conformal field theory out
  of equilibrium: a review},\ }\href
  {https://doi.org/10.1088/1742-5468/2016/06/064005} {\bibfield  {journal}
  {\bibinfo  {journal} {Journal of Statistical Mechanics: Theory and
  Experiment}\ }\textbf {\bibinfo {volume} {2016}},\ \bibinfo {pages} {064005}
  (\bibinfo {year} {2016})}\BibitemShut {NoStop}%
\bibitem [{\citenamefont {Bernard}(2021)}]{Bernard_2021}%
  \BibitemOpen
  \bibfield  {author} {\bibinfo {author} {\bibfnamefont {D.}~\bibnamefont
  {Bernard}},\ }\bibfield  {title} {\bibinfo {title} {Can the macroscopic
  fluctuation theory be quantized?},\ }\href
  {https://doi.org/10.1088/1751-8121/ac2597} {\bibfield  {journal} {\bibinfo
  {journal} {Journal of Physics A: Mathematical and Theoretical}\ }\textbf
  {\bibinfo {volume} {54}},\ \bibinfo {pages} {433001} (\bibinfo {year}
  {2021})}\BibitemShut {NoStop}%
\bibitem [{\citenamefont {Gullans}\ and\ \citenamefont
  {Huse}(2019)}]{gullans2019entanglement}%
  \BibitemOpen
  \bibfield  {author} {\bibinfo {author} {\bibfnamefont {M.~J.}\ \bibnamefont
  {Gullans}}\ and\ \bibinfo {author} {\bibfnamefont {D.~A.}\ \bibnamefont
  {Huse}},\ }\bibfield  {title} {\bibinfo {title} {Entanglement structure of
  current-driven diffusive fermion systems},\ }\href@noop {} {\bibfield
  {journal} {\bibinfo  {journal} {Physical Review X}\ }\textbf {\bibinfo
  {volume} {9}},\ \bibinfo {pages} {021007} (\bibinfo {year}
  {2019})}\BibitemShut {NoStop}%
\bibitem [{\citenamefont {Cao}\ \emph {et~al.}(2018)\citenamefont {Cao},
  \citenamefont {Tilloy},\ and\ \citenamefont {De~Luca}}]{DeLuca}%
  \BibitemOpen
  \bibfield  {author} {\bibinfo {author} {\bibfnamefont {X.}~\bibnamefont
  {Cao}}, \bibinfo {author} {\bibfnamefont {A.}~\bibnamefont {Tilloy}},\ and\
  \bibinfo {author} {\bibfnamefont {A.}~\bibnamefont {De~Luca}},\ }\bibfield
  {title} {\bibinfo {title} {Entanglement in a fermion chain under continuous
  monitoring},\ }\href@noop {} {\bibfield  {journal} {\bibinfo  {journal}
  {arXiv preprint arXiv:1804.04638}\ } (\bibinfo {year} {2018})}\BibitemShut
  {NoStop}%
\bibitem [{\citenamefont {Bernard}\ and\ \citenamefont
  {Jin}(2019)}]{PhysRevLett.123.080601}%
  \BibitemOpen
  \bibfield  {author} {\bibinfo {author} {\bibfnamefont {D.}~\bibnamefont
  {Bernard}}\ and\ \bibinfo {author} {\bibfnamefont {T.}~\bibnamefont {Jin}},\
  }\bibfield  {title} {\bibinfo {title} {Open quantum symmetric simple
  exclusion process},\ }\href {https://doi.org/10.1103/PhysRevLett.123.080601}
  {\bibfield  {journal} {\bibinfo  {journal} {Phys. Rev. Lett.}\ }\textbf
  {\bibinfo {volume} {123}},\ \bibinfo {pages} {080601} (\bibinfo {year}
  {2019})}\BibitemShut {NoStop}%
\bibitem [{\citenamefont {Bauer}\ \emph {et~al.}(2019)\citenamefont {Bauer},
  \citenamefont {Bernard},\ and\ \citenamefont {Jin}}]{bauer2019equilibrium}%
  \BibitemOpen
  \bibfield  {author} {\bibinfo {author} {\bibfnamefont {M.}~\bibnamefont
  {Bauer}}, \bibinfo {author} {\bibfnamefont {D.}~\bibnamefont {Bernard}},\
  and\ \bibinfo {author} {\bibfnamefont {T.}~\bibnamefont {Jin}},\ }\bibfield
  {title} {\bibinfo {title} {Equilibrium fluctuations in maximally noisy
  extended quantum systems},\ }\href@noop {} {\bibfield  {journal} {\bibinfo
  {journal} {SciPost Physics}\ }\textbf {\bibinfo {volume} {6}},\ \bibinfo
  {pages} {045} (\bibinfo {year} {2019})}\BibitemShut {NoStop}%
\bibitem [{\citenamefont {Eisler}(2011)}]{eisler_crossover_2011}%
  \BibitemOpen
  \bibfield  {author} {\bibinfo {author} {\bibfnamefont {V.}~\bibnamefont
  {Eisler}},\ }\bibfield  {title} {{\selectlanguage {en}\bibinfo {title}
  {Crossover between ballistic and diffusive transport: the quantum exclusion
  process}},\ }\href {https://doi.org/10.1088/1742-5468/2011/06/P06007}
  {\bibfield  {journal} {\bibinfo  {journal} {Journal of Statistical Mechanics:
  Theory and Experiment}\ }\textbf {\bibinfo {volume} {2011}},\ \bibinfo
  {pages} {P06007} (\bibinfo {year} {2011})},\ \bibinfo {note} {publisher: IOP
  Publishing}\BibitemShut {NoStop}%
\bibitem [{\citenamefont {Temme}\ \emph {et~al.}(2012)\citenamefont {Temme},
  \citenamefont {Wolf},\ and\ \citenamefont {Verstraete}}]{Temme_2012}%
  \BibitemOpen
  \bibfield  {author} {\bibinfo {author} {\bibfnamefont {K.}~\bibnamefont
  {Temme}}, \bibinfo {author} {\bibfnamefont {M.~M.}\ \bibnamefont {Wolf}},\
  and\ \bibinfo {author} {\bibfnamefont {F.}~\bibnamefont {Verstraete}},\
  }\bibfield  {title} {\bibinfo {title} {Stochastic exclusion processes versus
  coherent transport},\ }\href {https://doi.org/10.1088/1367-2630/14/7/075004}
  {\bibfield  {journal} {\bibinfo  {journal} {New Journal of Physics}\ }\textbf
  {\bibinfo {volume} {14}},\ \bibinfo {pages} {075004} (\bibinfo {year}
  {2012})}\BibitemShut {NoStop}%
\bibitem [{\citenamefont {Bernard}\ and\ \citenamefont
  {Jin}(2021)}]{Bernard2021-vx}%
  \BibitemOpen
  \bibfield  {author} {\bibinfo {author} {\bibfnamefont {D.}~\bibnamefont
  {Bernard}}\ and\ \bibinfo {author} {\bibfnamefont {T.}~\bibnamefont {Jin}},\
  }\bibfield  {title} {\bibinfo {title} {Solution to the quantum symmetric
  simple exclusion process: The continuous case},\ }\href@noop {} {\bibfield
  {journal} {\bibinfo  {journal} {Communications in Mathematical Physics}\
  }\textbf {\bibinfo {volume} {384}},\ \bibinfo {pages} {1141} (\bibinfo {year}
  {2021})}\BibitemShut {NoStop}%
\bibitem [{\citenamefont {Hruza}\ and\ \citenamefont
  {Bernard}(2023)}]{PhysRevX.13.011045}%
  \BibitemOpen
  \bibfield  {author} {\bibinfo {author} {\bibfnamefont {L.}~\bibnamefont
  {Hruza}}\ and\ \bibinfo {author} {\bibfnamefont {D.}~\bibnamefont
  {Bernard}},\ }\bibfield  {title} {\bibinfo {title} {Coherent fluctuations in
  noisy mesoscopic systems, the open quantum ssep, and free probability},\
  }\href {https://doi.org/10.1103/PhysRevX.13.011045} {\bibfield  {journal}
  {\bibinfo  {journal} {Phys. Rev. X}\ }\textbf {\bibinfo {volume} {13}},\
  \bibinfo {pages} {011045} (\bibinfo {year} {2023})}\BibitemShut {NoStop}%
\bibitem [{\citenamefont {Biane}(2023)}]{biane2023combinatorics}%
  \BibitemOpen
  \bibfield  {author} {\bibinfo {author} {\bibfnamefont {P.}~\bibnamefont
  {Biane}},\ }\bibfield  {title} {\bibinfo {title} {Combinatorics of the
  quantum symmetric simple exclusion process, associahedra and free
  cumulants},\ }\href@noop {} {\bibfield  {journal} {\bibinfo  {journal}
  {Annales de l’Institut Henri Poincar{\'e} D}\ } (\bibinfo {year}
  {2023})}\BibitemShut {NoStop}%
\bibitem [{\citenamefont {Derrida}\ and\ \citenamefont
  {Gerschenfeld}(2009)}]{Derrida2009-nc}%
  \BibitemOpen
  \bibfield  {author} {\bibinfo {author} {\bibfnamefont {B.}~\bibnamefont
  {Derrida}}\ and\ \bibinfo {author} {\bibfnamefont {A.}~\bibnamefont
  {Gerschenfeld}},\ }\bibfield  {title} {\bibinfo {title} {Current fluctuations
  of the one dimensional symmetric simple exclusion process with step initial
  condition},\ }\href@noop {} {\bibfield  {journal} {\bibinfo  {journal}
  {Journal of Statistical Physics}\ }\textbf {\bibinfo {volume} {136}},\
  \bibinfo {pages} {1} (\bibinfo {year} {2009})}\BibitemShut {NoStop}%
\bibitem [{\citenamefont {Eyink}\ \emph {et~al.}(1990)\citenamefont {Eyink},
  \citenamefont {Lebowitz},\ and\ \citenamefont {Spohn}}]{Eyink1990}%
  \BibitemOpen
  \bibfield  {author} {\bibinfo {author} {\bibfnamefont {G.}~\bibnamefont
  {Eyink}}, \bibinfo {author} {\bibfnamefont {J.~L.}\ \bibnamefont
  {Lebowitz}},\ and\ \bibinfo {author} {\bibfnamefont {H.}~\bibnamefont
  {Spohn}},\ }\bibfield  {title} {\bibinfo {title} {Hydrodynamics of stationary
  non-equilibrium states for some stochastic lattice gas models},\ }\href
  {https://doi.org/10.1007/BF02278011} {\bibfield  {journal} {\bibinfo
  {journal} {Communications in Mathematical Physics}\ }\textbf {\bibinfo
  {volume} {132}},\ \bibinfo {pages} {253} (\bibinfo {year}
  {1990})}\BibitemShut {NoStop}%
\bibitem [{\citenamefont {Bertini}\ \emph {et~al.}(2015)\citenamefont
  {Bertini}, \citenamefont {De~Sole}, \citenamefont {Gabrielli}, \citenamefont
  {Jona-Lasinio},\ and\ \citenamefont {Landim}}]{RevModPhys.87.593}%
  \BibitemOpen
  \bibfield  {author} {\bibinfo {author} {\bibfnamefont {L.}~\bibnamefont
  {Bertini}}, \bibinfo {author} {\bibfnamefont {A.}~\bibnamefont {De~Sole}},
  \bibinfo {author} {\bibfnamefont {D.}~\bibnamefont {Gabrielli}}, \bibinfo
  {author} {\bibfnamefont {G.}~\bibnamefont {Jona-Lasinio}},\ and\ \bibinfo
  {author} {\bibfnamefont {C.}~\bibnamefont {Landim}},\ }\bibfield  {title}
  {\bibinfo {title} {Macroscopic fluctuation theory},\ }\href
  {https://doi.org/10.1103/RevModPhys.87.593} {\bibfield  {journal} {\bibinfo
  {journal} {Rev. Mod. Phys.}\ }\textbf {\bibinfo {volume} {87}},\ \bibinfo
  {pages} {593} (\bibinfo {year} {2015})}\BibitemShut {NoStop}%
\bibitem [{\citenamefont {Kipnis}\ and\ \citenamefont
  {Landim}(2013)}]{kipnis2013scaling}%
  \BibitemOpen
  \bibfield  {author} {\bibinfo {author} {\bibfnamefont {C.}~\bibnamefont
  {Kipnis}}\ and\ \bibinfo {author} {\bibfnamefont {C.}~\bibnamefont
  {Landim}},\ }\href@noop {} {\emph {\bibinfo {title} {Scaling limits of
  interacting particle systems}}},\ Vol.\ \bibinfo {volume} {320}\ (\bibinfo
  {publisher} {Springer Science \& Business Media},\ \bibinfo {year}
  {2013})\BibitemShut {NoStop}%
\bibitem [{\citenamefont {Spohn}(2012)}]{spohn2012large}%
  \BibitemOpen
  \bibfield  {author} {\bibinfo {author} {\bibfnamefont {H.}~\bibnamefont
  {Spohn}},\ }\href@noop {} {\emph {\bibinfo {title} {Large scale dynamics of
  interacting particles}}}\ (\bibinfo  {publisher} {Springer Science \&
  Business Media},\ \bibinfo {year} {2012})\BibitemShut {NoStop}%
\bibitem [{\citenamefont {Kipnis}\ \emph {et~al.}(1989)\citenamefont {Kipnis},
  \citenamefont {Olla},\ and\ \citenamefont
  {Varadhan}}]{kipnis1989hydrodynamics}%
  \BibitemOpen
  \bibfield  {author} {\bibinfo {author} {\bibfnamefont {C.}~\bibnamefont
  {Kipnis}}, \bibinfo {author} {\bibfnamefont {S.}~\bibnamefont {Olla}},\ and\
  \bibinfo {author} {\bibfnamefont {S.~S.}\ \bibnamefont {Varadhan}},\
  }\bibfield  {title} {\bibinfo {title} {Hydrodynamics and large deviation for
  simple exclusion processes},\ }\href@noop {} {\bibfield  {journal} {\bibinfo
  {journal} {Communications on Pure and Applied Mathematics}\ }\textbf
  {\bibinfo {volume} {42}},\ \bibinfo {pages} {115} (\bibinfo {year}
  {1989})}\BibitemShut {NoStop}%
\bibitem [{\citenamefont {Bertini}\ \emph {et~al.}(2005)\citenamefont
  {Bertini}, \citenamefont {De~Sole}, \citenamefont {Gabrielli}, \citenamefont
  {Jona-Lasinio},\ and\ \citenamefont {Landim}}]{PhysRevLett.94.030601}%
  \BibitemOpen
  \bibfield  {author} {\bibinfo {author} {\bibfnamefont {L.}~\bibnamefont
  {Bertini}}, \bibinfo {author} {\bibfnamefont {A.}~\bibnamefont {De~Sole}},
  \bibinfo {author} {\bibfnamefont {D.}~\bibnamefont {Gabrielli}}, \bibinfo
  {author} {\bibfnamefont {G.}~\bibnamefont {Jona-Lasinio}},\ and\ \bibinfo
  {author} {\bibfnamefont {C.}~\bibnamefont {Landim}},\ }\bibfield  {title}
  {\bibinfo {title} {Current fluctuations in stochastic lattice gases},\ }\href
  {https://doi.org/10.1103/PhysRevLett.94.030601} {\bibfield  {journal}
  {\bibinfo  {journal} {Phys. Rev. Lett.}\ }\textbf {\bibinfo {volume} {94}},\
  \bibinfo {pages} {030601} (\bibinfo {year} {2005})}\BibitemShut {NoStop}%
\bibitem [{\citenamefont {McCulloch}\ \emph {et~al.}(2023)\citenamefont
  {McCulloch}, \citenamefont {De~Nardis}, \citenamefont {Gopalakrishnan},\ and\
  \citenamefont {Vasseur}}]{PhysRevLett.131.210402}%
  \BibitemOpen
  \bibfield  {author} {\bibinfo {author} {\bibfnamefont {E.}~\bibnamefont
  {McCulloch}}, \bibinfo {author} {\bibfnamefont {J.}~\bibnamefont
  {De~Nardis}}, \bibinfo {author} {\bibfnamefont {S.}~\bibnamefont
  {Gopalakrishnan}},\ and\ \bibinfo {author} {\bibfnamefont {R.}~\bibnamefont
  {Vasseur}},\ }\bibfield  {title} {\bibinfo {title} {Full counting statistics
  of charge in chaotic many-body quantum systems},\ }\href
  {https://doi.org/10.1103/PhysRevLett.131.210402} {\bibfield  {journal}
  {\bibinfo  {journal} {Phys. Rev. Lett.}\ }\textbf {\bibinfo {volume} {131}},\
  \bibinfo {pages} {210402} (\bibinfo {year} {2023})}\BibitemShut {NoStop}%
\bibitem [{\citenamefont {Hruza}\ and\ \citenamefont
  {Jin}(2024)}]{Tony3dAnderson}%
  \BibitemOpen
  \bibfield  {author} {\bibinfo {author} {\bibfnamefont {L.}~\bibnamefont
  {Hruza}}\ and\ \bibinfo {author} {\bibfnamefont {T.}~\bibnamefont {Jin}},\
  }\bibfield  {title} {\bibinfo {title} {Fluctuations of quantum coherences in
  the anderson model are described by the quantum symmetric simple exclusion
  process},\ }\href {https://doi.org/10.1103/PhysRevB.110.L220202} {\bibfield
  {journal} {\bibinfo  {journal} {Phys. Rev. B}\ }\textbf {\bibinfo {volume}
  {110}},\ \bibinfo {pages} {L220202} (\bibinfo {year} {2024})}\BibitemShut
  {NoStop}%
\bibitem [{Note1()}]{Note1}%
  \BibitemOpen
  \bibinfo {note} {See supplemental material for extra details.}\BibitemShut
  {Stop}%
\bibitem [{\citenamefont {Jin}\ \emph {et~al.}(2022)\citenamefont {Jin},
  \citenamefont {Ferreira}, \citenamefont {Filippone},\ and\ \citenamefont
  {Giamarchi}}]{TonyQResistors}%
  \BibitemOpen
  \bibfield  {author} {\bibinfo {author} {\bibfnamefont {T.}~\bibnamefont
  {Jin}}, \bibinfo {author} {\bibfnamefont {J.~a.~S.}\ \bibnamefont
  {Ferreira}}, \bibinfo {author} {\bibfnamefont {M.}~\bibnamefont
  {Filippone}},\ and\ \bibinfo {author} {\bibfnamefont {T.}~\bibnamefont
  {Giamarchi}},\ }\bibfield  {title} {\bibinfo {title} {Exact description of
  quantum stochastic models as quantum resistors},\ }\href
  {https://doi.org/10.1103/PhysRevResearch.4.013109} {\bibfield  {journal}
  {\bibinfo  {journal} {Phys. Rev. Res.}\ }\textbf {\bibinfo {volume} {4}},\
  \bibinfo {pages} {013109} (\bibinfo {year} {2022})}\BibitemShut {NoStop}%
\bibitem [{\citenamefont {Medvedyeva}\ \emph {et~al.}(2016)\citenamefont
  {Medvedyeva}, \citenamefont {Essler},\ and\ \citenamefont
  {Prosen}}]{medvedyeva2016}%
  \BibitemOpen
  \bibfield  {author} {\bibinfo {author} {\bibfnamefont {M.~V.}\ \bibnamefont
  {Medvedyeva}}, \bibinfo {author} {\bibfnamefont {F.~H.~L.}\ \bibnamefont
  {Essler}},\ and\ \bibinfo {author} {\bibfnamefont {T.}~\bibnamefont
  {Prosen}},\ }\bibfield  {title} {\bibinfo {title} {Exact {Bethe} ansatz
  spectrum of a tight-binding chain with dephasing noise},\ }\href
  {http://dx.doi.org/10.1103/PhysRevLett.117.137202} {\bibfield  {journal}
  {\bibinfo  {journal} {Phys. Rev. Lett.}\ }\textbf {\bibinfo {volume} {117}},\
  \bibinfo {pages} {137202} (\bibinfo {year} {2016})}\BibitemShut {NoStop}%
\bibitem [{\citenamefont
  {{\v{Z}}nidari{\v{c}}}(2010{\natexlab{a}})}]{znidaric2010}%
  \BibitemOpen
  \bibfield  {author} {\bibinfo {author} {\bibfnamefont {M.}~\bibnamefont
  {{\v{Z}}nidari{\v{c}}}},\ }\bibfield  {title} {\bibinfo {title} {Exact
  solution for a diffusive nonequilibrium steady state of an open quantum
  chain},\ }\href {https://doi.org/10.1088/1742-5468/2010/05/L05002} {\bibfield
   {journal} {\bibinfo  {journal} {J. Stat. Mech.: Theory Exp.}\ }\textbf
  {\bibinfo {volume} {2010}}\bibinfo  {number} { (05)},\ \bibinfo {pages}
  {L05002}}\BibitemShut {NoStop}%
\bibitem [{\citenamefont
  {{\v{Z}}nidari{\v{c}}}(2010{\natexlab{b}})}]{znidaric2010jphysa}%
  \BibitemOpen
\bibfield  {number} {  }\bibfield  {author} {\bibinfo {author} {\bibfnamefont
  {M.}~\bibnamefont {{\v{Z}}nidari{\v{c}}}},\ }\bibfield  {title} {\bibinfo
  {title} {A matrix product solution for a nonequilibrium steady state of an xx
  chain},\ }\href {https://doi.org/10.1088/1751-8113/43/41/415004} {\bibfield
  {journal} {\bibinfo  {journal} {J. Phys. A: Math. Theor.}\ }\textbf {\bibinfo
  {volume} {43}},\ \bibinfo {pages} {415004} (\bibinfo {year}
  {2010}{\natexlab{b}})}\BibitemShut {NoStop}%
\bibitem [{\citenamefont {{\v{Z}}nidari{\v{c}}}(2014)}]{LDdeph}%
  \BibitemOpen
  \bibfield  {author} {\bibinfo {author} {\bibfnamefont {M.}~\bibnamefont
  {{\v{Z}}nidari{\v{c}}}},\ }\bibfield  {title} {\bibinfo {title}
  {Large-deviation statistics of a diffusive quantum spin chain and the
  additivity principle},\ }\bibfield  {journal} {\bibinfo  {journal} {Physical
  Review E}\ }\textbf {\bibinfo {volume} {89}},\ \href
  {https://doi.org/10.1103/physreve.89.042140} {10.1103/physreve.89.042140}
  (\bibinfo {year} {2014})\BibitemShut {NoStop}%
\bibitem [{Note2()}]{Note2}%
  \BibitemOpen
  \bibinfo {note} {Even though in this paper the authors work with a spin
  chain, their results can be directly mapped to spinless fermions through a
  Jordan-Wigner transform}\BibitemShut {NoStop}%
\bibitem [{\citenamefont {Touchette}(2009)}]{LDtouchette}%
  \BibitemOpen
  \bibfield  {author} {\bibinfo {author} {\bibfnamefont {H.}~\bibnamefont
  {Touchette}},\ }\bibfield  {title} {\bibinfo {title} {The large deviation
  approach to statistical mechanics},\ }\href
  {https://doi.org/https://doi.org/10.1016/j.physrep.2009.05.002} {\bibfield
  {journal} {\bibinfo  {journal} {Physics Reports}\ }\textbf {\bibinfo {volume}
  {478}},\ \bibinfo {pages} {1} (\bibinfo {year} {2009})}\BibitemShut {NoStop}%
\bibitem [{\citenamefont {Bodineau}\ and\ \citenamefont
  {Derrida}(2004)}]{DerridaAdditivityPrinciple}%
  \BibitemOpen
  \bibfield  {author} {\bibinfo {author} {\bibfnamefont {T.}~\bibnamefont
  {Bodineau}}\ and\ \bibinfo {author} {\bibfnamefont {B.}~\bibnamefont
  {Derrida}},\ }\bibfield  {title} {\bibinfo {title} {Current fluctuations in
  nonequilibrium diffusive systems: An additivity principle},\ }\href
  {https://doi.org/10.1103/PhysRevLett.92.180601} {\bibfield  {journal}
  {\bibinfo  {journal} {Phys. Rev. Lett.}\ }\textbf {\bibinfo {volume} {92}},\
  \bibinfo {pages} {180601} (\bibinfo {year} {2004})}\BibitemShut {NoStop}%
\bibitem [{\citenamefont {Derrida}(2007)}]{Derrida_2007}%
  \BibitemOpen
  \bibfield  {author} {\bibinfo {author} {\bibfnamefont {B.}~\bibnamefont
  {Derrida}},\ }\bibfield  {title} {\bibinfo {title} {Non-equilibrium steady
  states: fluctuations and large deviations of the density and of the
  current},\ }\href {https://doi.org/10.1088/1742-5468/2007/07/P07023}
  {\bibfield  {journal} {\bibinfo  {journal} {Journal of Statistical Mechanics:
  Theory and Experiment}\ }\textbf {\bibinfo {volume} {2007}},\ \bibinfo
  {pages} {P07023} (\bibinfo {year} {2007})}\BibitemShut {NoStop}%
\bibitem [{\citenamefont {Derrida}(2011)}]{derrida2011microscopic}%
  \BibitemOpen
  \bibfield  {author} {\bibinfo {author} {\bibfnamefont {B.}~\bibnamefont
  {Derrida}},\ }\bibfield  {title} {\bibinfo {title} {Microscopic versus
  macroscopic approaches to non-equilibrium systems},\ }\href@noop {}
  {\bibfield  {journal} {\bibinfo  {journal} {Journal of Statistical Mechanics:
  Theory and Experiment}\ }\textbf {\bibinfo {volume} {2011}},\ \bibinfo
  {pages} {P01030} (\bibinfo {year} {2011})}\BibitemShut {NoStop}%
\bibitem [{\citenamefont {{Derrida}}\ \emph {et~al.}(2004)\citenamefont
  {{Derrida}}, \citenamefont {{Dou{\c{c}}ot}},\ and\ \citenamefont
  {{Roche}}}]{Derrida2004}%
  \BibitemOpen
  \bibfield  {author} {\bibinfo {author} {\bibfnamefont {B.}~\bibnamefont
  {{Derrida}}}, \bibinfo {author} {\bibfnamefont {B.}~\bibnamefont
  {{Dou{\c{c}}ot}}},\ and\ \bibinfo {author} {\bibfnamefont {P.~E.}\
  \bibnamefont {{Roche}}},\ }\bibfield  {title} {\bibinfo {title} {{Current
  Fluctuations in the One-Dimensional Symmetric Exclusion Process with Open
  Boundaries}},\ }\href {https://doi.org/10.1023/B:JOSS.0000022379.95508.b2}
  {\bibfield  {journal} {\bibinfo  {journal} {Journal of Statistical Physics}\
  }\textbf {\bibinfo {volume} {115}},\ \bibinfo {pages} {717} (\bibinfo {year}
  {2004})},\ \Eprint {https://arxiv.org/abs/cond-mat/0310453}
  {arXiv:cond-mat/0310453 [cond-mat.dis-nn]} \BibitemShut {NoStop}%
\bibitem [{\citenamefont {Mallick}(2015)}]{mallick2015exclusion}%
  \BibitemOpen
  \bibfield  {author} {\bibinfo {author} {\bibfnamefont {K.}~\bibnamefont
  {Mallick}},\ }\bibfield  {title} {\bibinfo {title} {The exclusion process: A
  paradigm for non-equilibrium behaviour},\ }\href@noop {} {\bibfield
  {journal} {\bibinfo  {journal} {Physica A: Statistical Mechanics and its
  Applications}\ }\textbf {\bibinfo {volume} {418}},\ \bibinfo {pages} {17}
  (\bibinfo {year} {2015})}\BibitemShut {NoStop}%
\bibitem [{\citenamefont {Derrida}\ \emph {et~al.}(1993)\citenamefont
  {Derrida}, \citenamefont {Evans}, \citenamefont {Hakim},\ and\ \citenamefont
  {Pasquier}}]{derrida1993exact}%
  \BibitemOpen
  \bibfield  {author} {\bibinfo {author} {\bibfnamefont {B.}~\bibnamefont
  {Derrida}}, \bibinfo {author} {\bibfnamefont {M.~R.}\ \bibnamefont {Evans}},
  \bibinfo {author} {\bibfnamefont {V.}~\bibnamefont {Hakim}},\ and\ \bibinfo
  {author} {\bibfnamefont {V.}~\bibnamefont {Pasquier}},\ }\bibfield  {title}
  {\bibinfo {title} {Exact solution of a 1d asymmetric exclusion model using a
  matrix formulation},\ }\href@noop {} {\bibfield  {journal} {\bibinfo
  {journal} {Journal of Physics A: Mathematical and General}\ }\textbf
  {\bibinfo {volume} {26}},\ \bibinfo {pages} {1493} (\bibinfo {year}
  {1993})}\BibitemShut {NoStop}%
\bibitem [{\citenamefont {Furstenberg}\ and\ \citenamefont
  {Kesten}(1960)}]{Furstenberg-Kesten}%
  \BibitemOpen
  \bibfield  {author} {\bibinfo {author} {\bibfnamefont {H.}~\bibnamefont
  {Furstenberg}}\ and\ \bibinfo {author} {\bibfnamefont {H.}~\bibnamefont
  {Kesten}},\ }\bibfield  {title} {\bibinfo {title} {Products of random
  matrices},\ }\href {http://www.jstor.org/stable/2237962} {\bibfield
  {journal} {\bibinfo  {journal} {The Annals of Mathematical Statistics}\
  }\textbf {\bibinfo {volume} {31}},\ \bibinfo {pages} {457} (\bibinfo {year}
  {1960})}\BibitemShut {NoStop}%
\bibitem [{\citenamefont {Oseledets}(1968)}]{oseledets1968multiplicative}%
  \BibitemOpen
  \bibfield  {author} {\bibinfo {author} {\bibfnamefont {V.}~\bibnamefont
  {Oseledets}},\ }\bibfield  {title} {\bibinfo {title} {A multiplicative
  ergodic theorem. lyapunov characteristic numbers for dynamical systems},\
  }\href@noop {} {\bibfield  {journal} {\bibinfo  {journal} {Transactions of
  the Moscow Mathematical Society}\ }\textbf {\bibinfo {volume} {19}},\
  \bibinfo {pages} {197} (\bibinfo {year} {1968})}\BibitemShut {NoStop}%
\bibitem [{\citenamefont {Akkermans}\ \emph {et~al.}(2013)\citenamefont
  {Akkermans}, \citenamefont {Bodineau}, \citenamefont {Derrida},\ and\
  \citenamefont {Shpielberg}}]{Akkermans2013}%
  \BibitemOpen
  \bibfield  {author} {\bibinfo {author} {\bibfnamefont {E.}~\bibnamefont
  {Akkermans}}, \bibinfo {author} {\bibfnamefont {T.}~\bibnamefont {Bodineau}},
  \bibinfo {author} {\bibfnamefont {B.}~\bibnamefont {Derrida}},\ and\ \bibinfo
  {author} {\bibfnamefont {O.}~\bibnamefont {Shpielberg}},\ }\bibfield  {title}
  {\bibinfo {title} {Universal current fluctuations in the symmetric exclusion
  process and other diffusive systems},\ }\href
  {https://doi.org/10.1209/0295-5075/103/20001} {\bibfield  {journal} {\bibinfo
   {journal} {EPL}\ }\textbf {\bibinfo {volume} {103}},\ \bibinfo {pages}
  {20001} (\bibinfo {year} {2013})}\BibitemShut {NoStop}%
\bibitem [{\citenamefont {Gilbarg}\ and\ \citenamefont
  {Trudinger}(2001)}]{GilbargTrudinger}%
  \BibitemOpen
  \bibfield  {author} {\bibinfo {author} {\bibfnamefont {D.}~\bibnamefont
  {Gilbarg}}\ and\ \bibinfo {author} {\bibfnamefont {N.~S.}\ \bibnamefont
  {Trudinger}},\ }\href@noop {} {\emph {\bibinfo {title} {Elliptic Partial
  Differential Equations of Second Order}}},\ \bibinfo {edition} {reprint of
  the 1998 edition}\ ed.\ (\bibinfo  {publisher} {Springer},\ \bibinfo
  {address} {Berlin},\ \bibinfo {year} {2001})\BibitemShut {NoStop}%
\bibitem [{\citenamefont {Evans}(2010)}]{EvansPDE}%
  \BibitemOpen
  \bibfield  {author} {\bibinfo {author} {\bibfnamefont {L.~C.}\ \bibnamefont
  {Evans}},\ }\href@noop {} {\emph {\bibinfo {title} {Partial Differential
  Equations}}},\ \bibinfo {edition} {2nd}\ ed.\ (\bibinfo  {publisher}
  {American Mathematical Society},\ \bibinfo {address} {Providence, RI},\
  \bibinfo {year} {2010})\BibitemShut {NoStop}%
\end{thebibliography}
\end{document}